\documentclass[runningheads]{llncs}
\newif\ifdraft\draftfalse

\newif\iffull\fulltrue % with appendix
\usepackage{graphicx}
\usepackage{amsfonts} 
\usepackage{amsmath}
\usepackage{amssymb}
\usepackage{microtype}
\usepackage{algorithm}
\usepackage{algpseudocode}
\usepackage{todonotes}
\usepackage[framemethod=tikz]{mdframed}
\usepackage{stmaryrd}
\usepackage{cite}
\usepackage{booktabs}
\usepackage{colonequals}
\usepackage{subcaption}
\usepackage{pgfplots}
\usepgfplotslibrary{external}
\usepackage{wrapfig}
\usepackage{algorithm}
\usepackage{algpseudocode}
\algtext*{EndWhile}% Remove "end while" text
\algtext*{EndIf}% Remove "end if" text
\algtext*{EndFor}% Remove "end for" text
\usepackage{amsmath}
\usepackage{nicematrix}
\usepackage{csvsimple}
\usepackage{tikz}
\usetikzlibrary{calc}
\usetikzlibrary {arrows.meta,automata,positioning}
\usepackage{tikz-cd}
\usepackage[all]{xy} 
\tikzstyle{point}=[circle,  inner sep=2pt, fill]

\usepackage{hyperref} % "available" badge is a link to the actual DOI
\usepackage{cleveref}
\usepackage[firstpage]{draftwatermark} % free badge placement

\SetWatermarkAngle{0}
\SetWatermarkText{\raisebox{15.5cm}{%
\hspace{0.1cm}%
\href{https://doi.org/10.5281/zenodo.11002681}{\includegraphics{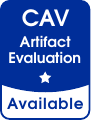}}%
\hspace{8cm}%
\includegraphics{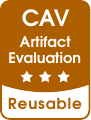}%
}}

\usepackage{comment}
\specialcomment{auxproof}
{\mbox{}\newline\textbf{BEGIN: AUX-PROOF}\dotfill\newline}
{\mbox{}\newline\textbf{END: AUX-PROOF}\dotfill\newline}
\excludecomment{auxproof}

\algdef{SE}[DOWHILE]{Do}{doWhile}{\algorithmicdo}[1]{\algorithmicwhile\ #1}%
\algnewcommand\algorithmicforeach{\textbf{for each}}
\algdef{S}[FOR]{ForEach}[1]{\algorithmicforeach\ #1\ \algorithmicdo}

\newcommand{\littletaller}{\mathchoice{\vphantom{\big|}}{}{}{}}

\newcommand\restr[2]{{% we make the whole thing an ordinary symbol
  \left.\kern-\nulldelimiterspace % automatically resize the bar with \right
  #1 % the function
  \littletaller % pretend it's a little taller at normal size
  \right|_{#2} % this is the delimiter
  }}

\usepackage{theorem}
\def\@thmcountersep{.}
\theorembodyfont{\itshape}
\newtheorem{mytheorem}{Theorem}[section]
\newtheorem{mylemma}[mytheorem]{Lemma}
\newtheorem{myproposition}[mytheorem]{Proposition}

\theorembodyfont{\rmfamily}

\newtheorem{myremark}[mytheorem]{Remark}
\newtheorem{myexample}[mytheorem]{Example}
\newtheorem{mydefinition}[mytheorem]{Definition}

\spnewtheorem*{myproof}{Proof}{\itshape}{\rmfamily}
\newcommand{\last}[1]{\mathrm{last}(#1)}

\newcommand{\defeq}{\colonequals}
\newcommand{\nat}{\mathbb{N}}
\newcommand{\dr}{\mathbf{r}}
\newcommand{\dl}{\mathbf{l}}
\newcommand{\seqcomp}{\fatsemi}

\newcommand{\interface}{\mathsf{IO}}

\newcommand{\nset}[1]{[#1]}
\newcommand{\real}{\mathbb R}

\newcommand{\dist}[1]{\mathcal{D}(#1)}

\newcommand{\leaf}[1]{\mathop{\mathrm{nCP}}(#1)}
\newcommand{\CP}[1]{\mathop{\mathrm{CP}}(#1)}
\newcommand{\constMDP}[1]{\mathsf{c}_{#1}}

\newcommand{\Reacha}[3]{\mathrm{RPr}^{#1,#3}(#2)}
\newcommand{\WReacha}[3]{\mathrm{WRPr}^{#1}(#2, #3)}
\newcommand{\MaxWReacha}[3]{\mathrm{WRPr}^{#1}_{\mathrm{max}}(#2, #3)}

\newcommand{\probinterval}{[0, 1]}

\newcommand{\sd}[1]{\mathbb{#1}}

\newcommand{\typemdp}[1]{\mathrm{arity}(#1)}
\newcommand{\semantics}[1]{\llbracket #1 \rrbracket}

\newcommand{\FinPaths}[1]{\mathsf{FPath}_{#1}}
\newcommand{\TFinPaths}[2]{\mathsf{FPath}_{#1}(#2)}

\newcommand{\Possched}[1]{\Sigma^{#1}}
\newcommand{\Dmsched}[1]{\Sigma_\text{d}^{#1}}

\newcommand{\point}{\mathbf{p}}
\newcommand{\achievable}[2]{\mathsf{Ach}_{#2}}

\newcommand{\achievablesched}[3]{\mathsf{Ach}^{#3}_{#2}}

\newcommand{\dwconvcl}[1]{\mathsf{DwConvCl}(#1)}

\newcommand{\mdp}[1]{\mathcal{#1}}
\newcommand{\ecmdp}[1]{\mathcal{C}(#1)}
\newcommand{\cmdp}[1]{\mathcal{C}(#1)}

\newcommand{\arr}[1]{#1_{\dr}}
\newcommand{\arl}[1]{#1_{\dl}}

\newcommand{\convw}[1]{\mathbf{#1}}
\newcommand{\en}{I}
\newcommand{\ex}{O}
\newcommand{\enLocal}[1]{\en_{\mathrm{lc}}(#1)}
\newcommand{\exLocal}[1]{\ex_{\mathrm{lc}}(#1)}
\newcommand{\enGlobal}[1]{\en^{\semantics{#1}}}
\newcommand{\exGlobal}[1]{\ex^{\semantics{#1}}}
\newcommand{\enr}{I_{\dr}}
\newcommand{\enrarg}[1]{i_{\dr, #1}}
\newcommand{\enl}{I_{\dl}}
\newcommand{\enlarg}[1]{i_{\dl, #1}}
\newcommand{\exr}{O_{\dr}}
\newcommand{\exrarg}[1]{o_{\dr, #1}}
\newcommand{\exl}{O_{\dl}}
\newcommand{\exlarg}[1]{o_{\dl, #1}}

\newcommand{\prmeas}[2]{\mathrm{Pr}^{#1}_{#2}}
\newcommand{\lfpoint}[1]{\mu#1}

\newcommand{\bellman}[2]{\Phi_{#1,#2}}
\newcommand{\wbellman}[3]{\Phi_{#1,#2;#3}}
\newcommand{\compBellman}[2]{\Psi_{#1;#2}}

\newcommand{\shortsubst}{\mathcal{C}}

\newcommand{\paretoAssign}{\mathbf{C}}

\renewcommand{\cref}[1]{\Cref{#1}}
\creflabelformat{enumi}{#2(#1)#3}
\crefname{mytheorem}{Thm.}{Thms}
\crefname{mydefinition}{Def.}{Defs}
\crefname{myproposition}{Prop.}{Props}
\crefname{myremark}{Rem.}{Remarks}
\crefname{mylemma}{Lem.}{Lemmas}
\crefname{myproof}{Proof.}{Proofs}
\crefname{appendix}{Appendix}{Appendixes}
\crefname{algorithm}{Alg.}{Algs}
\crefformat{section}{{\S}#2#1#3}
\crefname{figure}{Fig.}{Figs}
\Crefname{equation}{}{}

\newenvironment{proofs}{%
  \proof}{\endproof}

\newcommand{\myparagraph}[1]{\smallskip\noindent \textbf{#1}}

\newcommand{\scatterplotsize}[0]{0.98\textwidth}
\newcommand{\marksize}{1.8}

\newcommand{\scatterplot}[5]{%
	\begin{tikzpicture}
	\begin{axis}[
	width=\scatterplotsize,
	height=\scatterplotsize,
	axis equal image,
	xmin=1,
	ymin=1,
	ymax=3000,
	xmax=3000,
	xmode=log,
	ymode=log,
	axis x line=bottom,
	axis y line=left,
	xtick={1,9,90,900},
	xticklabels={1,9,90,900},
	extra x ticks = {2700},
	extra x tick labels = {OoR},
	extra x tick style = {grid = major},
	ytick={1,9,90,900},
	yticklabels={1,9,90,900},
	extra y ticks = {2700},
	extra y tick labels = {OoR},
	extra y tick style = {grid = major},
	xlabel={\scriptsize #3},
	xlabel style={yshift=0cm},
	ylabel={\scriptsize #5},
	ylabel style={yshift=-0.4cm},
	yticklabel style={font=\tiny},
	xticklabel style={rotate=290,anchor=west,font=\tiny},
 legend pos=outer north east,
	]

	\addplot[ % BiroomsBig
	scatter,
	only marks,
	mark=asterisk,
 mark size=\marksize,
 fill opacity=0.5,
	scatter/use mapped color={
    draw=black,
    fill=black,
	}
	]%
	table [col sep=comma,x=#2,y=#4] {#1_0.csv};

	\addplot[ % BiroomsSmall
	scatter,
	only marks,
	mark=triangle,
 mark size=\marksize,
 fill opacity=0.5,
	scatter/use mapped color={
    draw=black,
    fill=black,
	}
	]%
	table [col sep=comma,x=#2,y=#4] {#1_1.csv};

	\addplot[ % ChainsBig
	scatter,
	only marks,
	mark=square,
 mark size=\marksize,
 fill opacity=0.5,
	scatter/use mapped color={
    draw=black,
    fill=black,
	}
	]%
	table [col sep=comma,x=#2,y=#4] {#1_2.csv};

	\addplot[ % ChainsDice3
	scatter,
	only marks,
	mark=diamond,
 mark size=\marksize,
 fill opacity=0.5,
	scatter/use mapped color={
    draw=black,
    fill=black,
	}
	]%
	table [col sep=comma,x=#2,y=#4] {#1_3.csv};

	\addplot[ % ChainsDice4
	scatter,
	only marks,
	mark=-,
 mark size=\marksize,
 fill opacity=0.5,
	scatter/use mapped color={
    draw=black,
    fill=black,
	}
	]%
	table [col sep=comma,x=#2,y=#4] {#1_4.csv};

	\addplot[ % ChainsDice5
	scatter,
	only marks,
	mark=|,
 mark size=\marksize,
 fill opacity=0.5,
	scatter/use mapped color={
    draw=black,
    fill=black,
	}
	]%
	table [col sep=comma,x=#2,y=#4] {#1_5.csv};

	\addplot[ % ChainsDiceLoop3
	scatter,
	only marks,
	mark=Mercedes star,
 mark size=\marksize,
 fill opacity=0.5,
	scatter/use mapped color={
    draw=black,
    fill=black,
	}
	]%
	table [col sep=comma,x=#2,y=#4] {#1_6.csv};

	\addplot[ % ChainsDiceLoop4
	scatter,
	only marks,
	mark=Mercedes star flipped,
 mark size=\marksize,
 fill opacity=0.5,
	scatter/use mapped color={
    draw=black,
    fill=black,
	}
	]%
	table [col sep=comma,x=#2,y=#4] {#1_7.csv};

	\addplot[ % ChainsDiceLoop5
	scatter,
	only marks,
	mark=heart,
 mark size=\marksize,
 fill opacity=0.5,
	scatter/use mapped color={
    draw=black,
    fill=black,
	}
	]%
	table [col sep=comma,x=#2,y=#4] {#1_8.csv};

	\addplot[ % ChainsSmall
	scatter,
	only marks,
	mark=pentagon,
 mark size=\marksize,
 fill opacity=0.5,
	scatter/use mapped color={
    draw=black,
    fill=black,
	}
	]%
	table [col sep=comma,x=#2,y=#4] {#1_9.csv};

	\addplot[ % RoomsBig
	scatter,
	only marks,
	mark=oplus,
 mark size=\marksize,
 fill opacity=0.5,
	scatter/use mapped color={
    draw=black,
    fill=black,
	}
	]%
	table [col sep=comma,x=#2,y=#4] {#1_10.csv};

	\addplot[ % RoomsDice
	scatter,
	only marks,
	mark=otimes,
 mark size=\marksize,
 fill opacity=0.5,
	scatter/use mapped color={
    draw=black,
    fill=black,
	}
	]%
	table [col sep=comma,x=#2,y=#4] {#1_11.csv};

	\addplot[ % RoomsSmall
	scatter,
	only marks,
	mark=x,
 mark size=\marksize,
 fill opacity=0.5,
	scatter/use mapped color={
    draw=black,
    fill=black,
	}
	]%
	table [col sep=comma,x=#2,y=#4] {#1_12.csv};

	\addplot[no marks] coordinates {(0.01,0.01) (2700,2700) };
	\addplot[no marks, densely dotted] coordinates {(0.01,0.1) (270,2700)};
	\addplot[no marks, densely dotted] coordinates {(0.1,0.01) (2700,270)};
	\end{axis}
	\end{tikzpicture}
}

\newcommand{\algrule}[1][.2pt]{\par\vskip.5\baselineskip\hrule height #1\par\vskip.5\baselineskip}

\makeatletter
\newcommand{\printfnsymbol}[1]{%
  \textsuperscript{\@fnsymbol{#1}}%
}
\makeatother

\begin{document}
\title{Compositional Value Iteration\\ with Pareto Caching\thanks{K.W.\ and I.H.\ are supported by ERATO HASUO Metamathematics for Systems Design Project (No.\ JPMJER1603) and the ASPIRE grant No.\ JPMJAP2301, JST. K.W.\ is supported by the JST grants No.\ JPMJFS2136 and  JPMJAX23CU. S.J. is supported by the NWO Veni ProMiSe (222.147).}}
\author{Kazuki Watanabe\inst{1,2,}\thanks{Equal contribution.} \and Marck van der Vegt\inst{3,}\printfnsymbol{2}\and\\ Sebastian Junges\inst{3} \and Ichiro Hasuo\inst{1,2}}
\institute{National Institute of Informatics, Japan \and The Graduate University for Advanced Studies (SOKENDAI), Japan  \and Radboud University, Nijmegen, the Netherlands}
 \authorrunning{Watanabe et al.}
\maketitle              
\begin{abstract} 
The de-facto standard approach in MDP verification is based on value iteration (VI). 
We propose \emph{compositional VI}, a framework for model checking compositional MDPs, that addresses efficiency while maintaining soundness.
Concretely, compositional MDPs naturally arise from the combination of individual components, and their structure can be expressed using, e.g., string diagrams. 
Towards efficiency, we observe that compositional VI repeatedly verifies individual components. 
We propose a technique called \emph{Pareto caching} that allows to reuse verification results, even for previously unseen queries.
Towards soundness, we present two stopping criteria:  
one generalizes the optimistic value iteration paradigm and the other uses Pareto caches in conjunction with recent baseline algorithms. 
Our experimental evaluations shows the promise of the novel algorithm and its variations, and identifies challenges for future work.
\end{abstract}

\section{Introduction}\label{sec:intro}
\myparagraph{MDP Model Checking and Value Iteration}
\emph{Markov decision processes (MDPs)} are the standard model for sequential decision making in stochastic settings. A standard question in the verification of MDPs is: \emph{what is the maximal probability that an error state is reached.} MDP model checking is an active topic in the formal verification community.
\emph{Value iteration (VI)}~\cite{Puterman94} is an iterative and approximate method whose performance in MDP model checking is well-established~\cite{HartmannsK20,BuddeHKKPQTZ20,HartmannsJQW23}. Several extensions with \emph{soundness} have been proposed; they provide, in addition to under-approximations, also over-approximations with a desired precision~\cite{QuatmannK18,HartmannsK20,HaddadM18,Baier0L0W17,PhalakarnTHH20}, so that an approximate answer comes with an error bound. These sound algorithms are implemented in mature model checkers such as Prism~\cite{KwiatkowskaNP11}, Modest~\cite{HartmannsH14}, and Storm~\cite{HenselJKQV22}.

\myparagraph{Compositional Model Checking}
Even with these state-of-the-art algorithms, it is a challenge to model check large MDPs efficiently with high precision.
 Experiments  observe that MDPs with more than $10^8$ states are too large for those  algorithms~\cite{JungesS22,WatanabeEAH23,WatanabeVHRJ24}---they simply do not fit in memory. However, such large MDPs often arise as models of  complicated stochastic systems, e.g.\ in the domains of network and robotics. Furthermore, even small models may be numerically challenging to solve due to their structure~\cite{HaddadM18,HartmannsJQW23,Baier0L0W17}.

\emph{Compositional model checking} is a promising approach to tackle this scalability challenge.
Given a compositional structure of a target system, compositional model checking executes a divide-and-conquer algorithm that avoids loading the entire state space at once, often solving the above memory problem. Moreover, reusing the model checking results for components can lead to speed-up by magnitudes. Although finding a suitable compositional structure for a given ``monolithic'' MDP is still open, many systems come with such an \emph{a priori} compositional structure. For example,  such compositional structures are often assumed in robotics and referred to as \emph{hierarchical models}~\cite{HauskrechtMKDB98,BarryKL11,JungesS22,GopalandLMSTWW17,SaxeER17,NearyVCT22,VienT15}.% Similar structures are found also in which also naturally applies to network protocols~\cite{JungesS22,WatanabeEAH23}.   

\begin{wrapfigure}[6]{r}[0pt]{7.8cm}
\vspace{-2em}
\centering
\scalebox{0.7}{
\begin{tikzpicture}[
innode/.style={draw, rectangle, minimum size=0.5cm},
interface/.style={draw, rectangle, minimum size=0.5cm},]
\fill[orange] (-0.7cm, -1cm)--(-0.7cm, 1.2cm)--(3.8cm, 1.2cm)--(3.8cm, -1cm)--cycle;
\node at (3.6cm, 1cm) {$\mdp{A}$};
\node[interface,fill=white,yshift=0.5cm] (s0) {$\enrarg{1}$};
\node[inner sep=0,right=-1.25cm of s0] (enr1) {};
\node[interface,fill=white,yshift=-0.5cm] (s0o) {$\exlarg{1}$};
\node[inner sep=0,right=-1.25cm of s0o] (exl1) {};
\node[state,right=0.6cm of s0, minimum size=0.5cm,fill=white] (s1) {$s_1$};
\node[state,right=0.6cm of s0o, minimum size=0.5cm,fill=white] (ssink) {$s_2$};

\node[inner sep=2pt, fill=black,right=0.5cm of s1] (a1a) {};
\node[inner sep=2pt, fill=black,right=-1cm of ssink,yshift=0.5cm] (a1b) {};

\node[interface, right=1.3cm of ssink,fill=white] (s2) {$\enlarg{1}$};
\node[interface, right=0.6cm of a1a,fill=white] (s3) {$\exrarg{1}$};
\node[inner sep=0,right= 0.5cm of s2] (exr1) {};
\node[inner sep=0,right= 0.5cm of s3] (exr2) {};
\draw[->] (enr1) -> (s0);
\draw[->] (s0o) -> (exl1);
\draw[->] (s0) -> node [above] {$1$} (s1);
\draw[->] (s1) -> node [above] {} (a1a);
\draw[->] (s1) -> node [above] {} (a1b);
\draw[->] (a1a) -> node [above] {$0.5$} (s3);
\draw[->] (a1a) -> node [right] {$0.5$} (ssink);
\draw[->] (ssink) -> node [above] {$1$} (s0o);
\draw[->] (a1b) -> node [right,yshift=0.2cm] {$1$} (ssink);
\draw[->] (exr1) -> (s2);
\draw[->] (s3) -> (exr2);
\draw[->] (s2) -> node [above] {$0.3$} (ssink);
\draw[->] (s2) -> node [left] {$0.7$} (s3);
\end{tikzpicture}
\hspace{30pt}
\begin{tikzpicture}[
innode/.style={draw, rectangle, minimum size=0.5cm},
interface/.style={draw, rectangle, minimum size=0.5cm},]
\fill[cyan] (-0.7cm, -1cm)--(-0.7cm, 1.2cm)--(3.2cm, 1.2cm)--(3.2cm, -1cm)--cycle;
\node at (3cm, 1cm) {$\mdp{B}$};
\node[interface, yshift=0.5cm,fill=white] (t0) {$\enrarg{1}$};
\node[interface, yshift=-0.5cm,fill=white] (t1) {$\exlarg{1}$};
\node[inner sep=0,right=-1.3cm of t0] (enr1) {};
\node[inner sep=0,right=-1.3cm of t1] (enr2) {};
\node[state,right=0.6cm of t0, yshift=-0.5cm, minimum size=0.5cm,fill=white] (t2) {$t_1$};
\node[interface, right=0.6cm of t2,fill=white] (t3) {$\exrarg{1}$};
\node[inner sep=0,right=0.4cm of t3] (exr1) {};
% \node[state,right=3cm of s1, accepting] (s2) {$s_2$};

% \draw[->] (enr1) -> (s0);
\draw[->] (enr1) -> (t0);
\draw[->] (t1) -> (enr2);
\draw[->] (t0) -> node [below] {$0.3$} (t2);
\draw[->] (t2) -> node [below] {$1$} (t1);
\draw[->] (t0) to [out=10,in=150] node [above] {$0.7$} (t3.west);
\draw[->] (t3) -> (exr1);
\end{tikzpicture}
}
\caption{open MDPs $\mdp{A}$ and $\mdp{B}$. \color{red}}
\label{fig:openMDPs}
\end{wrapfigure}

\begin{figure}[t]
\centering
\vspace{-3mm}
\begin{minipage}[]{0.45\linewidth}
\scalebox{0.9}{
\begin{tikzpicture}[
innode/.style={draw, rectangle, minimum size=0.5cm},
interface/.style={inner sep=0},]
\fill[orange] (-0.5cm, -0.4cm)--(-0.5cm, 0.4cm)--(0.5cm, 0.4cm)--(0.5cm, -0.4cm)--cycle;
\node (mdpA) at (0.4cm, 0.3cm) {\scalebox{0.5}{$\mdp{A}$}};
\node[inner sep=0] at (-0.7cm, 0.2cm) (enrA1d) {};
\node[inner sep=0,right=0.4cm of enrA1d] (enrA1c) {};
\node[inner sep=0] at (-0.7cm, -0.2cm) (exlA1c) {};
\node[inner sep=0,right=0.4cm of exlA1c] (exlA1d) {};
\node[inner sep=0, right=0.15cm of enrA1c, yshift=-0.2cm] (cdotsA1) {\scalebox{0.6}{$\cdots$}};
\node[inner sep=0] at (0.3cm, 0.2cm) (exrA1d) {};
\node[inner sep=0,right=0.4cm of exrA1d] (exrA1c) {};
\node[inner sep=0] at (0.3cm, -0.2cm) (enlA1c) {};
\node[inner sep=0,right=0.4cm of enlA1c] (enlA1d) {};
\draw[->] (enrA1d) -> (enrA1c);
\draw[->] (exlA1d) -> (exlA1c);
\draw[->] (exrA1d) -> (exrA1c);
\draw[->] (enlA1d) -> (enlA1c);
\node[interface, right=0.1cm of exrA1c, yshift=-0.2cm] (seqcomp) {$\seqcomp$};
\fill[cyan] (1.3cm, -0.4cm)--(1.3cm, 0.4cm)--(2.3cm, 0.4cm)--(2.3cm, -0.4cm)--cycle;
\node (mdpA) at (2.2cm, 0.3cm) {\scalebox{0.5}{$\mdp{B}$}};
\node[inner sep=0] at (1.1cm, 0.2cm) (enrB1d) {};
\node[inner sep=0,right=0.4cm of enrB1d] (enrB1c) {};
\node[inner sep=0] at (1.1cm, -0.2cm) (exlB1c) {};
\node[inner sep=0,right=0.4cm of exlB1c] (exlB1d) {};
\node[inner sep=0, right=0.15cm of enrB1c, yshift=-0.2cm] (cdotsB1) {\scalebox{0.6}{$\cdots$}};
\node[inner sep=0] at (2.1cm, 0cm) (exrB1d) {};
\node[inner sep=0,right=0.4cm of exrB1d] (exrB1c) {};
% \node[inner sep=0] at (2.1cm, -0.2cm) (enlB1c) {};
% \node[inner sep=0,right=0.4cm of enlB1c] (enlB1d) {};
\draw[->] (enrB1d) -> (enrB1c);
\draw[->] (exlB1d) -> (exlB1c);
\draw[->] (exrB1d) -> (exrB1c);
% \draw[->] (enlB1d) -> (enlB1c);

\node[inner sep=0,right=0.2cm of exrB1c] (eqseqcomp) {$=$};
\fill[orange] (3.5cm, -0.4cm)--(3.5cm, 0.4cm)--(4.5cm, 0.4cm)--(4.5cm, -0.4cm)--cycle;
\node (mdpA2) at (4.4cm, 0.3cm) {\scalebox{0.5}{$\mdp{A}$}};
\node[inner sep=0] at (3.3cm, 0.2cm) (enrA2d) {};
\node[inner sep=0,right=0.4cm of enrA2d] (enrA2c) {};
\node[inner sep=0] at (3.3cm, -0.2cm) (exlA2c) {};
\node[inner sep=0,right=0.4cm of exlA2c] (exlA2d) {};
\node[inner sep=0, right=0.15cm of enrA2c, yshift=-0.2cm] (cdotsA2) {\scalebox{0.6}{$\cdots$}};
\node[inner sep=0] at (4.3cm, 0.2cm) (exrA2d) {};
\node[inner sep=0,right=0.4cm of exrA2d] (exrA2c) {};
\node[inner sep=0] at (4.3cm, -0.2cm) (enlA2c) {};
\node[inner sep=0,right=0.4cm of enlA2c] (enlA1d) {};
\fill[cyan] (4.7cm, -0.4cm)--(4.7cm, 0.4cm)--(5.7cm, 0.4cm)--(5.7cm, -0.4cm)--cycle;
\node (mdpB2) at (5.6cm, 0.3cm) {\scalebox{0.5}{$\mdp{B}$}};
\node[inner sep=0] at (4.5cm, 0.2cm) (enrB2d) {};
\node[inner sep=0,right=0.4cm of enrB2d] (enrB2c) {};
\node[inner sep=0] at (4.5cm, -0.2cm) (exlB2c) {};
\node[inner sep=0,right=0.4cm of exlB2c] (exlB2d) {};
\node[inner sep=0, right=0.15cm of enrB2c, yshift=-0.2cm] (cdotsB2) {\scalebox{0.6}{$\cdots$}};
\node[inner sep=0] at (5.5cm, 0cm) (exrB2d) {};
\node[inner sep=0,right=0.4cm of exrB2d] (exrB2c) {};
% \node[inner sep=0] at (5.5cm, -0.2cm) (enlB2c) {};
% \node[inner sep=0,right=0.4cm of enlB2c] (enlB2d) {};
\draw[->] (enrA2d) -> (enrA2c);
\draw[->] (exlA2d) -> (exlA2c);
\draw[->] (exrA2d) -> (enrB2c);
\draw[->] (exlB2d) -> (enlA2c);
\draw[->] (exrB2d) -> (exrB2c);
% \draw[->] (enlB2d) -> (enlB2c);
\end{tikzpicture}},
\end{minipage}
\hfill
\begin{minipage}[]{0.45\linewidth}
\scalebox{0.9}{
\begin{tikzpicture}[
innode/.style={draw, rectangle, minimum size=0.5cm},
interface/.style={inner sep=0},]
\fill[orange] (-0.5cm, -0.4cm)--(-0.5cm, 0.4cm)--(0.5cm, 0.4cm)--(0.5cm, -0.4cm)--cycle;
\node (mdpA) at (0.4cm, 0.3cm) {\scalebox{0.5}{$\mdp{A}$}};
\node[inner sep=0] at (-0.7cm, 0.2cm) (enrA1d) {};
\node[inner sep=0,right=0.4cm of enrA1d] (enrA1c) {};
\node[inner sep=0] at (-0.7cm, -0.2cm) (exlA1c) {};
\node[inner sep=0,right=0.4cm of exlA1c] (exlA1d) {};
\node[inner sep=0, right=0.15cm of enrA1c, yshift=-0.2cm] (cdotsA1) {\scalebox{0.6}{$\cdots$}};
\node[inner sep=0] at (0.3cm, 0.2cm) (exrA1d) {};
\node[inner sep=0,right=0.4cm of exrA1d] (exrA1c) {};
\node[inner sep=0] at (0.3cm, -0.2cm) (enlA1c) {};
\node[inner sep=0,right=0.4cm of enlA1c] (enlA1d) {};
\draw[->] (enrA1d) -> (enrA1c);
\draw[->] (exlA1d) -> (exlA1c);
\draw[->] (exrA1d) -> (exrA1c);
\draw[->] (enlA1d) -> (enlA1c);
\node[interface, right=0cm of exrA1c, yshift=-0.2cm] (oplus) {$\oplus$};
\fill[cyan] (1.3cm, -0.4cm)--(1.3cm, 0.4cm)--(2.3cm, 0.4cm)--(2.3cm, -0.4cm)--cycle;
\node (mdpA) at (2.2cm, 0.3cm) {\scalebox{0.5}{$\mdp{B}$}};
\node[inner sep=0] at (1.1cm, 0.2cm) (enrB1d) {};
\node[inner sep=0,right=0.4cm of enrB1d] (enrB1c) {};
\node[inner sep=0] at (1.1cm, -0.2cm) (exlB1c) {};
\node[inner sep=0,right=0.4cm of exlB1c] (exlB1d) {};
\node[inner sep=0, right=0.15cm of enrB1c, yshift=-0.2cm] (cdotsB1) {\scalebox{0.6}{$\cdots$}};
\node[inner sep=0] at (2.1cm, 0cm) (exrB1d) {};
\node[inner sep=0,right=0.4cm of exrB1d] (exrB1c) {};
% \node[inner sep=0] at (2.1cm, -0.2cm) (enlB1c) {};
% \node[inner sep=0,right=0.4cm of enlB1c] (enlB1d) {};
\draw[->] (enrB1d) -> (enrB1c);
\draw[->] (exlB1d) -> (exlB1c);
\draw[->] (exrB1d) -> (exrB1c);

\node[inner sep=0,right=0.2cm of exrB1c] (eqseqcomp) {$=$};
\fill[orange] (3.5cm, 0.1cm)--(3.5cm, 0.9cm)--(4.5cm, 0.9cm)--(4.5cm, 0.1cm)--cycle;
\node (mdpA) at (4.4cm, 0.8cm) {\scalebox{0.5}{$\mdp{A}$}};
\node[inner sep=0] at (3.3cm, 0.7cm) (enrA2d) {};
\node[inner sep=0,right=0.4cm of enrA2d] (enrA2c) {};
\node[inner sep=0] at (3.3cm, 0.3cm) (exlA2c) {};
\node[inner sep=0,right=0.4cm of exlA2c] (exlA2d) {};
\node[inner sep=0, right=0.15cm of enrA2c, yshift=-0.2cm] (cdotsA2) {\scalebox{0.6}{$\cdots$}};
\node[inner sep=0] at (4.3cm, 0.7cm) (exrA2d) {};
\node[inner sep=0,right=0.4cm of exrA2d] (exrA2c) {};
\node[inner sep=0] at (4.3cm, 0.3cm) (enlA2c) {};
\node[inner sep=0,right=0.4cm of enlA2c] (enlA2d) {};
\draw[->] (enrA2d) -> (enrA2c);
\draw[->] (exlA2d) -> (exlA2c);
\draw[->] (exrA2d) -> (exrA2c);
\draw[->] (enlA2d) -> (enlA2c);
\fill[cyan] (3.5cm, -0.1cm)--(3.5cm, -0.9cm)--(4.5cm, -0.9cm)--(4.5cm, -0.1cm)--cycle;
\node (mdpB2) at (4.4cm, -0.2cm) {\scalebox{0.5}{$\mdp{B}$}};
\node[inner sep=0] at (3.3cm, -0.3cm) (enrB2d) {};
\node[inner sep=0,right=0.4cm of enrB2d] (enrB2c) {};
\node[inner sep=0] at (3.3cm, -0.7cm) (exlB2c) {};
\node[inner sep=0,right=0.4cm of exlB2c] (exlB2d) {};
\node[inner sep=0, right=0.15cm of enrB2c, yshift=-0.2cm] (cdotsB2) {\scalebox{0.6}{$\cdots$}};
\node[inner sep=0] at (4.3cm, -0.5cm) (exrB2d) {};
\node[inner sep=0,right=0.4cm of exrB2d] (exrB2c) {};
% \node[inner sep=0] at (2.1cm, -0.2cm) (enlB1c) {};
% \node[inner sep=0,right=0.4cm of enlB1c] (enlB1d) {};
\draw[->] (enrB2d) -> (enrB2c);
\draw[->] (exlB2d) -> (exlB2c);
\draw[->] (exrB2d) -> (exrB2c);
\end{tikzpicture}
}
\end{minipage}
\caption{sequential composition $\mdp{A}\seqcomp\mdp{B}$ and sum $\mdp{A}\oplus\mdp{B}$ of open MDPs. The framework is \emph{bidirectional} (edges can be left- and right-ward); thus loops can arise in  $\mdp{A}\seqcomp\mdp{B}$.}
\label{fig:seqcompAndSum}
\end{figure}
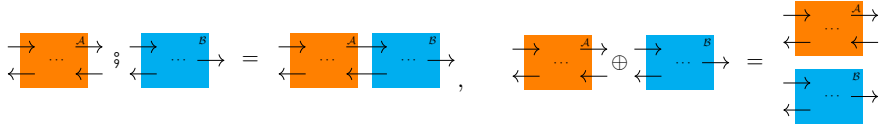

Recently,  \emph{string diagrams of MDPs} are introduced for compositional model checking~\cite{WatanabeEAH23,WatanabeVHRJ24}; the current paper adopts this formalism. There, MDPs are extended with (open) entrances and exits (\cref{fig:openMDPs}), and they get composed by \emph{sequential composition} $\seqcomp$ and \emph{sum} $\oplus$. See~\cref{fig:seqcompAndSum}, where the right-hand sides are simple  juxtapositions of graphs (wires get connected in $\seqcomp$). This makes the formalism focused on sequential (as opposed to parallel) composition. This restriction eases the design of compositional algorithms; yet, the formalism is rich enough to capture the compositional structures of many system models.

\begin{auxproof}
 By exploiting such given compositional structures, compositional probabilistic approaches provide approximations with certain guarantees by assuming certain conditions, and outperform monolithic approaches in such situations. 
 Recently, \emph{String diagrams of MDPs}~\cite{WatanabeEAH23,WatanabeVHRJ24} are proposed as a syntax for such compositional MDPs and naturally capture compositional structures by compositions of \emph{open MDPs (oMDPs)} with algebraic operations (illustrated in~\cref{fig:openMDPs,fig:seqcompAndSum}); we recall a recent work~\cite{WatanabeVHRJ24} on string diagrams of MDPs in~\cref{subsec:sd_MDPs,subsec:TACAS24}. 
\end{auxproof}

\myparagraph{Current Work: Compositional Value Iteration}
In this paper, we present a \emph{compositional value iteration (CVI)} algorithm that solves reachability probabilities of string diagrams of MDPs, operating in a divide-and-conquer manner along compositional structures. Our approximate VI algorithm comes with \emph{soundness}---it produces error bounds---and exploits compositionality for \emph{efficiency}. 

Specifically, for soundness, we lift the recent paradigm of \emph{optimistic value iteration (OVI)}~\cite{HartmannsK20} to the current compositional setting. We use it both for local (component-level) model checking and---in one of the two global VI stopping criteria that we present---for providing a global over-approximation.

For efficiency, firstly, we adopt a \emph{top-down} compositional approach where each component is model-checked repeatedly, each time on a different weight $\convw{w}$, in a \emph{by-need} manner. Secondly, in order to suppress repetitive computation on similar weights, we introduce a novel technique of \emph{Pareto caching} that allows ``approximate reuse'' of model checking results. This closely relates to multi-objective probabilistic model checking~\cite{EtessamiKVY08,ForejtKP12,Quatmann23}, without the explicit goal of building Pareto curves. Our Pareto caching also leads to another (\emph{sound}) global VI stopping criterion that is based on the approximate bottom-up approach~\cite{WatanabeVHRJ24}. 

Our algorithm is approximate (unlike the exact one in~\cite{WatanabeEAH23}), and top-down (unlike the \emph{bottom-up} approximate one in~\cite{WatanabeVHRJ24}).  Experimental evaluation demonstrates its performance thanks to the combination of these two features. 

\myparagraph{Contributions and Organization}~ We start with an overview (\cref{sec:overview})  that presents graphical intuitions. After formalizing the problem setting in \cref{sec:prelinaries},
we move on to describe our technical contributions:
\begin{itemize}
 \item \emph{compositional value iteration} for string diagrams of MDPs where VI is run in a top-down and thus by-need manner (\cref{sec:cvi}),
 \item the \emph{Pareto caching} technique for reusing results for components (\cref{sec:approximation}), 
 \item two \emph{global stopping criteria} that ensure soundness (\cref{sec:pareto_curves}).
\end{itemize}
We evaluate and discuss our approach through experiments (\cref{sec:imp_exp}), show related work (\cref{sec:related_work}), and conclude this paper (\cref{sec:conclusion}).

\myparagraph{Notations}
  For a natural number $m$, we write $[m]$ for  $\{1, \dots, m\}$.
For a set $X$, we write $\mathcal{D}(X)$ for the set of distributions on $X$. For sets $X, Y$,
we write $X\uplus Y$ for their disjoint union  and $f\colon X\rightharpoonup Y$ for a partial function $f$ from $X$ to~$Y$.

\section{Overview}
\label{sec:overview}

This section illustrates our take on CVI with so-called Pareto caches using  graphical intuitions. 
We describe MDPs as string diagrams over so-called \emph{open MDPs}~\cite{WatanabeEAH23}. Open MDPs, such as $\mdp{A}, \mdp{B}$ in \cref{fig:openMDPs}, extend MDPs with \emph{open ends} (entrances and exits).  We use two operations $\seqcomp$ and $\oplus$; see \cref{fig:seqcompAndSum}. 
That figure also illustrates the  \emph{bidirectional} nature of the formalism: arrows can point left and right; thus acyclic MDPs can create cycles when combined.
String diagrams come from category theory (see~\cite{WatanabeEAH23}) and they are used in
many fields of computer science~\cite{BonchiGKSZ22,BonchiHPSZ19,GhaniHWZ18,abs-2112-14058}.

\subsection{Approximate Bottom-Up Model Checking}
\label{subsec:TACAS24}

\begin{wrapfigure}[5]{r}[0pt]{4.5cm}
\centering
\vspace{-2.5em}
\scalebox{0.6}{
    \begin{tikzpicture}[
innode/.style={draw, rectangle, minimum size=0.5cm},
interface/.style={draw, rectangle, minimum size=0.5cm},]
\fill[orange] (0cm, -1cm)--(0cm, 1cm)--(3cm, 1cm)--(3cm, -1cm)--cycle;
\node[inner sep=0] (Aenr1d) at (1.5cm, 0cm) {\scalebox{1}{$\cdots$}};
\node at (2.8cm, 0.9cm) {\scalebox{0.7}{$\mdp{A}$}};
\node[inner sep=0] (Aenr1d) at (-0.3cm, 0cm) {};
\node[interface,fill=white, right = 0.5cm of Aenr1d] (Aenr1c) {\scalebox{0.8}{$i_1$}};
\node[inner sep=0, right = 0.4cm of Aenr1c] (Aenr1out) {};
\node[inner sep=0, right = 1.1cm of Aenr1c, yshift=0.5cm] (Aexr1in) {};
\node[interface,fill=white, right=0.4cm of Aexr1in] (Aexr1d) {\scalebox{0.8}{$o_1$}};
\node[inner sep=0, right = 1.1cm of Aenr1c, yshift=-0.5cm] (Aexr2in) {};
\node[interface,fill=white, right=0.4cm of Aexr2in] (Aexr2d) {\scalebox{0.8}{$o_2$}};

\fill[cyan] (3.4cm, -1cm)--(3.4cm, 1cm)--(6.4cm, 1cm)--(6.4cm, -1cm)--cycle;
\node at (6.2cm, 0.9cm) {\scalebox{0.7}{$\mdp{B}$}};
\node[interface,fill=white,right=0.8cm of Aexr1d] (Benr1c) {\scalebox{0.8}{$i_2$}};
\node[interface,fill=white,right=0.8cm of Aexr2d] (Benr2c) {\scalebox{0.8}{$i_3$}};
\node[inner sep=0, right = 0.4cm of Benr1c] (Benr1out) {};
\node[inner sep=0, right = 0.4cm of Benr2c] (Benr2out) {};
\node[inner sep=0, right = 0.1cm of Benr1out, yshift=-0.5cm] (bcdots) {\scalebox{1}{$\cdots$}};
% \node[state,right=0.4cm of Benr1c, minimum size=0.3cm,fill=white, yshift=-0.5cm] (s1) {};
\node[inner sep=0, right = 1.1cm of Benr1c, yshift=-0.5cm] (Bexr1in) {};
\node[interface,fill=white,right=1.5cm of Benr1c, yshift=-0.5cm] (Bexr1d) {\scalebox{0.8}{$o_3$}};
\node[inner sep=0, right=0.5cm of Bexr1d] (Bexr1c) {};

\draw[->, thick] (Aenr1d) -> (Aenr1c);
\draw[->, thick] (Aenr1c) -> (Aenr1out);
\draw[->, thick] (Aexr1in) -> (Aexr1d);
\draw[->, thick] (Aexr2in) -> (Aexr2d);
\draw[->, thick] (Aexr1d) -> (Benr1c);
\draw[->, thick] (Aexr2d) -> (Benr2c);
\draw[->, thick] (Benr1c) -> (Benr1out);
\draw[->, thick] (Benr2c) -> (Benr2out);
\draw[->, thick] (Bexr1in) -> (Bexr1d);
% \draw[->, thick] (Benr1c) -> node [yshift=-0.1cm,right=-0.5cm] {\scalebox{0.8}{$0.2$}} (s1);
% \draw[->, thick] (Benr2c) -> node [yshift=-0.15cm,right=-0.2cm] {\scalebox{0.8}{$0.7$}} (s1);
% \draw[->, thick] (Benr1c) to [out=10,in=150] node[yshift=0.2cm,right=-0.2cm] {\scalebox{0.8}{$0.8$}}  (Bexr1d);
% \draw[->, thick] (Benr2c) to [out=-10,in=210]
% node[yshift=-0.2cm,right=-0.2cm] {\scalebox{0.8}{$0.3$}} (Bexr1d);
\draw[->, thick] (Bexr1d) -> (Bexr1c);
\end{tikzpicture}
}
\caption{$\mdp{A}\seqcomp\mdp{B}$
}
\label{fig:bbAandBSeqCompOneDir}
\end{wrapfigure}
 The first compositional model checking algorithm for string diagrams of MDPs is in~\cite{WatanabeEAH23}, which is exact. Subsequently, in~\cite{WatanabeVHRJ24}, an \emph{approximate} compositional model checking algorithm is proposed. This is the basis of our algorithm and we shall review it here.
Consider, for illustration, the sequential composition $\mdp{A}\seqcomp\mdp{B}$ in \cref{fig:bbAandBSeqCompOneDir}, where the exit $o_{3}$ is the target. The algorithm from~\cite{WatanabeVHRJ24} proceeds in the following \emph{bottom-up} manner.

\myparagraph{First Step: Model Checking Each Component} Firstly, model checking is conducted for component oMDPs $\mdp{A}$ and $\mdp{B}$ separately, which amounts to identifying an \emph{optimal scheduler} for each.  At this point, however, it is unclear what constitutes an optimal scheduler:

\begin{example} In the MDP $\mdp{A}$ in~\cref{fig:bbAandBSeqCompOneDir},
let's say the reachability probabilities  $\bigl(\,\mathrm{RPr}^{\sigma_{1}}(i_1\to o_1),\,\mathrm{RPr}^{\sigma_{1}}(i_1\to o_2)\bigr)$ are $(0.2,0.7)$ under a scheduler $\sigma_{1}$, and $(0.6,0.2)$ under another  $\sigma_{2}$. One cannot tell which scheduler ($\sigma_{1}$ or $\sigma_{2}$) is better for the global objective (i.e.\ reaching $o_{3}$ in $\mdp{A}\seqcomp\mdp{B}$) since $\mdp{B}$ is a black box.
\end{example}

Concretely, the context $\underline{\;\;}\seqcomp \mdp{B}$ of $\mdp{A}$ is unknown. Therefore we have to compute all candidates of optimal schedulers, instead of one. This set is given by, 
for each component $\mdp{C}$ and its entrance $i$, 
\begin{equation}\label{eq:cav23ParetoEq}
 \bigl\{\,\text{schedulers }\sigma\,\big|\, \text{
$\bigl(\mathrm{RPr}^{\sigma}(i\to o)\bigr)_{o:\text{ $\mdp{C}$'s exit}}$ is Pareto optimal}\,\bigr\}.
\end{equation}
\begin{wrapfigure}[13]{r}[0pt]{5cm}
\centering
%
% \vspace{1.5em}
\vspace{-1em}
 \scalebox{0.6}{
 \begin{tikzpicture}
        \begin{axis}[
            axis lines = left,
            xmin=0, xmax=1,
            ymin=0, ymax=0.9,
            % xlabel = \(\exrarg{1}\),
            % ylabel = \(\exrarg{2}\),
            xtick={0, 0.2, 0.6},
            ytick={0, 0.2, 0.7},
            axis background/.style={fill=white},
            xlabel={\scalebox{1.5}{$\mathrm{RPr}(i_1\to o_1)$}},
            ylabel={\scalebox{1.5}{$\mathrm{RPr}(i_1\to o_2)$}}
        ]
        
        \node[point,label=0:$\sigma_1$] at (axis cs:0.2,0.7) {};
	\node[point,label=220:$\sigma_2$] at (axis cs:0.6,0.2) {};
    \addplot[
            color=blue,
            ultra thick
        ]
        coordinates {
            (0.2,0.7) (0.6, 0.2)
        };
        \addplot[
            color=blue,
            ultra thick,
            dashed
        ]
        coordinates {
            (0.1, 0.73) (0.2,0.7) (0.35, 0.6) (0.5, 0.4) (0.6, 0.2) (0.62, 0.1) 
        };
        \addplot[
            dotted
        ]
        coordinates {
            (0.2,0.7) (0, 0.7)
        };
        \addplot[
            dotted
        ]
        coordinates {
            (0.2,0.7) (0.2, 0)
        };
        \addplot[
            dotted
        ]
        coordinates {
            (0.6,0.2) (0, 0.2)
        };
        \addplot[
            dotted
        ]
        coordinates {
            (0.6,0.2) (0.6, 0)
        };
        \end{axis}
        \end{tikzpicture}
    }
\caption{Pareto-optimal points}
\label{fig:cav23ParetoFig}
\end{wrapfigure}
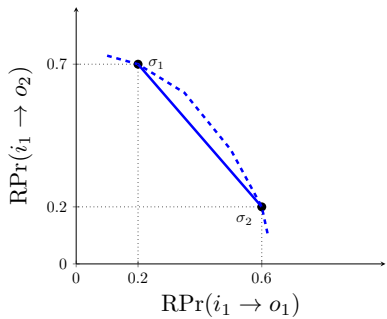
\begin{wrapfigure}[10]{r}[0pt]{5cm}
\centering
\vspace{-2em}
\pgfplotsset{width=4.5cm,compat=1.9}
    \scalebox{1}{
        \begin{tikzpicture}
        \begin{axis}[
	    axis on top=true,
            axis lines = left,
            xmin=0, xmax=0.5,
            ymin=0, ymax=0.5,
            % xlabel = \(\exrarg{1}\),
            % ylabel = \(\exrarg{2}\),
            xticklabel=\empty,
            yticklabel=\empty,
            axis background/.style={fill=white},
            xlabel={$\mathrm{RPr}(i_1\to o_1)$},
            ylabel={$\mathrm{RPr}(i_1\to o_2)$}
        ]
        \addplot[fill=green!50, very thin] coordinates {(0.35,0) (0.35,0.15) (0.32, 0.25) (0.25, 0.32) (0.15,0.35) (0,0.35) (0,0)} -- cycle;
        \addplot[fill=red!50, very thin] coordinates {(0.5, 0.5) (0.5, 0) (0.44,0) (0.42,0.2) (0.38,0.32) (0.32, 0.38) (0.18, 0.42)  (0, 0.44) (0, 0.5)};

        \node[] at (axis cs:0.15,0.2) {$L_{i_1}$};
	\node[] at (axis cs:0.41,0.44) {$\real^{2}\backslash U_{i_1}$};		
        \end{axis}
        \end{tikzpicture}
        }
\caption{approximations $(L_{i_1}, U_{i_1})$.}
\label{fig:tacas24ParetoFig}
\end{wrapfigure}
Here the \emph{Pareto optimality} is a usual notion from \emph{multi-objective model checking} (e.g.~\cite{PapadimitriouY00,EtessamiKVY08}); here, it means that there is no  scheduler $\sigma'$ that \emph{dominates} $\sigma$ in the sense that $\mathrm{RPr}^{\sigma}(i\to o) \le \mathrm{RPr}^{\sigma'}(i\to o)$ holds for each $o$  and $<$ holds for some $o$.  
The two points from the example can be plotted, see~\cref{fig:cav23ParetoFig}. 

The \emph{Pareto curve}---the set of points $\bigl(\mathrm{RPr}^{\sigma}(i\to o)\bigr)_{o}$ for the Pareto optimal schedulers $\sigma$ in~\cref{eq:cav23ParetoEq}---will look like the dashed blue line in~\cref{fig:cav23ParetoFig}.The solid blue line is realizable by a convex combination of the schedulers $\sigma_1$ and $\sigma_2$. It is always below the Pareto curve.

The algorithm in~\cite{WatanabeVHRJ24} computes guaranteed under- and over-approximations $(L, U)$ of Pareto-optimal points~\cref{eq:cav23ParetoEq} for every open MDP. 
See \cref{fig:tacas24ParetoFig}; here the green area indicates the under-approximation, and the red area is the complement of the over-approximation, so that any Pareto-optimal points are guaranteed to be in their gap (white).
These approximations are obtained by repeated application of (optimistic) value iteration on the open MDPs, i.e., a standard approach for verifying MDPs, based on~\cite{ForejtKP12, QuatmannK21}. We formalize these notions in~\cref{subsec:MOOandCompositionalMC}.

\myparagraph{Second Step: Combination along $\seqcomp, \oplus$ }
The second (inductive) step of the bottom-up algorithm in~\cite{WatanabeVHRJ24} is to combine the results of the first step---approximations as in \cref{fig:tacas24ParetoFig}, and the corresponding (near) optimal schedulers~\cref{eq:cav23ParetoEq},
for each component $\mdp{C}$---along the operations $\seqcomp, \oplus$ in a string diagram.

Here we describe this second step through the example 
 in \cref{fig:bbAandBSeqCompOneDir}. It computes reachability probabilities 
\begin{equation}\label{eq:CAV23DecompEq}\small
\begin{array}{ll}
  \mathrm{RPr}^{\sigma,\tau}(i_{1}\to o_{3})
 =& 
 \mathrm{RPr}^{\sigma}(i_{1}\to o_{1})
 \cdot
 \mathrm{RPr}^{\tau}(i_{2}\to o_{3})\\
 &+
 \mathrm{RPr}^{\sigma}(i_{1}\to o_{2})
 \cdot
 \mathrm{RPr}^{\tau}(i_{3}\to o_{3})
\end{array}
\qquad\qquad\qquad\qquad
\end{equation}
for each combination of Pareto-optimal schedulers $\sigma$ (for $\mdp{A}$) and $\tau$ (for $\mdp{B}$) to find which combinations of $\sigma,\tau$ are Pareto optimal for $\mdp{A}\seqcomp\mdp{B}$. 

The equality~\cref{eq:CAV23DecompEq}---called the \emph{decomposition equality} 
in~\cite{WatanabeEAH23}---enables
compositional reasoning on Pareto-optimal points and on their approximations: Pareto-optimal schedulers for $\mdp{A}\seqcomp\mdp{B}$ can be computed from those for $\mdp{A}$ and $\mdp{B}$. This compositional reasoning can be exploited for performance. In particular, when the same component  $\mdp{A}$ occurs multiple times in a string diagram $\sd{D}$, the model checking result of $\mdp{A}$ can be reused multiple times.

\subsection{Key Idea I: from Bottom-Up to Top-Down}\label{subsec:toTopDown}
The bottom-up approaches compute the Pareto curves independent of the context of the open MDP. 
One key idea is to move from bottom-up to \emph{top-down}, a direction followed by other compositional techniques too, see \cref{sec:related_work}. 

\begin{wrapfigure}[6]{r}[0pt]{5cm}
\centering

\vspace{-2.3em}
\scalebox{0.7}{
    \begin{tikzpicture}[
innode/.style={draw, rectangle, minimum size=0.5cm},
interface/.style={draw, rectangle, minimum size=0.5cm},]
\fill[orange] (0cm, -1cm)--(0cm, 1cm)--(3cm, 1cm)--(3cm, -1cm)--cycle;
\node[inner sep=0] (Aenr1d) at (1.5cm, 0cm) {\scalebox{1}{$\cdots$}};
\node at (2.8cm, 0.9cm) {\scalebox{0.7}{$\mdp{A}$}};
\node[inner sep=0] (Aenr1d) at (-0.3cm, 0cm) {};
\node[interface,fill=white, right = 0.5cm of Aenr1d] (Aenr1c) {\scalebox{0.8}{$i_1$}};
\node[inner sep=0, right = 0.4cm of Aenr1c] (Aenr1out) {};
\node[inner sep=0, right = 1.1cm of Aenr1c, yshift=0.5cm] (Aexr1in) {};
\node[interface,fill=white, right=0.4cm of Aexr1in] (Aexr1d) {\scalebox{0.8}{$o_1$}};
\node[inner sep=0, right = 1.1cm of Aenr1c, yshift=-0.5cm] (Aexr2in) {};
\node[interface,fill=white, right=0.4cm of Aexr2in] (Aexr2d) {\scalebox{0.8}{$o_2$}};

\fill[cyan] (3.4cm, -1cm)--(3.4cm, 1cm)--(5.9cm, 1cm)--(5.9cm, -1cm)--cycle;
\node at (5.7cm, 0.9cm) {\scalebox{0.7}{$\mdp{B}$}};
\node[interface,fill=white,right=0.8cm of Aexr1d] (Benr1c) {\scalebox{0.8}{$i_2$}};
\node[interface,fill=white,right=0.8cm of Aexr2d] (Benr2c) {\scalebox{0.8}{$i_3$}};
\node[state,right=0.4cm of Benr1c, minimum size=0.3cm,fill=white, yshift=-0.5cm] (s1) {};
\node[interface,fill=white,right=0.3cm of s1] (Bexr1d) {\scalebox{0.8}{$o_3$}};
\node[inner sep=0, right=0.5cm of Bexr1d] (Bexr1c) {};

\draw[->, thick] (Aenr1d) -> (Aenr1c);
\draw[->, thick] (Aenr1c) -> (Aenr1out);
\draw[->, thick] (Aexr1in) -> (Aexr1d);
\draw[->, thick] (Aexr2in) -> (Aexr2d);
\draw[->, thick] (Aexr1d) -> (Benr1c);
\draw[->, thick] (Aexr2d) -> (Benr2c);
\draw[->, thick] (Benr1c) -> node [yshift=-0.1cm,right=-0.5cm] {\scalebox{0.8}{$0.2$}} (s1);
\draw[->, thick] (Benr2c) -> node [yshift=-0.15cm,right=-0.2cm] {\scalebox{0.8}{$0.7$}} (s1);
\draw[->, thick] (Benr1c) to [out=10,in=150] node[yshift=0.2cm,right=-0.2cm] {\scalebox{0.8}{$0.8$}}  (Bexr1d);
\draw[->, thick] (Benr2c) to [out=-10,in=210]
node[yshift=-0.2cm,right=-0.2cm] {\scalebox{0.8}{$0.3$}} (Bexr1d);
\draw[->, thick] (Bexr1d) -> (Bexr1c);
\end{tikzpicture}
}
\caption{$\mdp{A}\seqcomp\mdp{B}$}
\label{fig:motivateTopDown}
\end{wrapfigure}
For illustration, consider the sequential composition $\mdp{A}\seqcomp\mdp{B}$ in \cref{fig:motivateTopDown};  we have concretized $\mdp{B}$ in \cref{fig:bbAandBSeqCompOneDir}. For this $\mdp{B}$, it follows that
 $\mathrm{RPr}(i_{2}\to o_{3})=0.8$ 
and
 $\mathrm{RPr}(i_{3}\to o_{3})=0.3$.
Therefore the  equality~\cref{eq:CAV23DecompEq} boils down to
\begin{equation}\label{eq:CAV23DecompEqWeighted}
  \mathrm{RPr}^{\sigma}(i_{1}\to o_{3})
= 
0.8\cdot \mathrm{RPr}^{\sigma}(i_{1}\to o_{1})
+
0.3\cdot \mathrm{RPr}^{\sigma}(i_{1}\to o_{2}).
\end{equation}
The equation~\cref{eq:CAV23DecompEqWeighted} is a significant simplification compared to~\cref{eq:CAV23DecompEq}: 
\begin{itemize}
 \item 
 in~\cref{eq:CAV23DecompEq}, since the \emph{weight} $\bigl(\mathrm{RPr}^{\tau}(i_{2}\to o_{3}),\mathrm{RPr}^{\tau}(i_{3}\to o_{3})\bigr)$ is unknown,  we must compute multidimensional Pareto curves as in \cref{fig:cav23ParetoFig,fig:tacas24ParetoFig}; 
 \item 
in~\cref{eq:CAV23DecompEqWeighted}, since the weight is known to be $(0.8,0.3)$, we can solve the equation using standard single-objective model checking.
\end{itemize}

Exploiting this simplification is our first key idea. 
We introduce a systematic procedure for deriving weights (such as $(0.8,0.3)$  above) that uses the context of an oMDP, i.e., it goes \emph{top-down} along the string diagram.
The procedure works for bi-directional sequential composition (thus for loops, cf.\ \cref{fig:seqcompAndSum}), not only for uni-directional as in \cref{fig:motivateTopDown}.
In the procedure, we first examine the context
of a component $\mdp{C}$, approximate a weight $\convw{w}$ for $\mdp{C}$, and then compute maximum weighted reachability probabilities in $\mdp{C}$.  We formalize the approach in \cref{sec:cvi}.

Potential performance advantages compared to the bottom-up algorithm in~\cite{WatanabeVHRJ24} 
should be obvious from \cref{fig:motivateTopDown}. Specifically,  the bottom-up algorithm draws a complete picture for Pareto-optimal points  (such as \cref{fig:tacas24ParetoFig}) \emph{once for all}, but a large part of this complete picture may not be used. 
In contrast, the top-down one draws the picture in a \emph{by-need} manner, for a weight $\convw{w}$ only when the weight $\convw{w}$  is suggested by the context. 

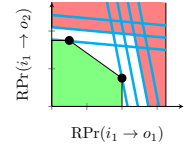
\begin{wrapfigure}[6]{r}[0pt]{5cm}
    \centering
    
    % \vspace{-2em}
    \pgfplotsset{width=4.5cm,compat=1.9}
        \scalebox{0.6}{
            \begin{tikzpicture}
            \begin{axis}[
                axis lines = left,
                axis on top=true,
                xmin=0, xmax=0.75,
                ymin=0, ymax=0.55,
                xticklabel=\empty,
                yticklabel=\empty,
                axis background/.style={fill=white},
                xlabel={$\mathrm{RPr}(i_1\to o_1)$},
                ylabel={$\mathrm{RPr}(i_1\to o_2)$}
            ]
            \addplot[fill=red!50, very thin] coordinates {(0.65, 0.55) (0.65, 0) (0.55,0) (0.47, 0.38) (0,0.42) (0, 0.65)};
    
     % \node[point,label=220:$\point^l_3$] at (axis cs:0.33,0.33) {};
      % \node[point,label=150:$\point^u_2$] at (axis cs:0.32,0.42) {};
            % \node[] at (axis cs:0.15,0.2) {$L_i$};		
        \addplot[
                domain = 0:0.55,
                color=cyan,
                ultra thick,
            ]
            {-4.75*x+2.6125};
        \addplot[
                domain = 0:0.55,
                color=cyan,
                ultra thick,
            ]
            {-4.75*x+2.05};
        \addplot[
                domain = 0:0.55,
                color=cyan,
                ultra thick,
            ]
            {-4.75*x+2.33};
            \addplot[
                domain = 0:0.65,
                color=cyan,
                ultra thick,
            ]
            {-4.75*x+2.9};
        \addplot[
                domain = 0:0.65,
                color=cyan,
                ultra thick,
            ]
            {-0.08510638297*x+0.42};
        % \addplot[
        %         domain = 0:0.65,
        %         color=cyan,
        %         ultra thick,
        %     ]
        %     {-0.08510638297*x+0.39};
        \addplot[
                domain = 0:0.65,
                color=cyan,
                ultra thick,
            ]
            {-0.08510638297*x+0.355};
        \addplot[
                domain = 0:0.65,
                color=cyan,
                ultra thick,
            ]
            {-0.08510638297*x+0.485};
        \addplot[fill=green!50, very thin] coordinates {(0.4,0) (0.4,0.15)  (0.1,0.35) (0,0.35) (0, 0)} -- cycle;
        % \addplot[
        %         domain = 0.43:0.55,
        %         color=black,
        %         thick,
        %         -latex,
        %     ]
        %     {0.1*x+0.15};
        % \node[] at (axis cs:0.55,0.27) {$h|_{\ex^{\mdp{A}}}$};
        % \node[] at (axis cs:0.2,0.48) {$\real^{2}\backslash U_i$};
        \node[point] at (axis cs:0.4,0.15) {};
        \node[point] at (axis cs:0.1,0.35) {};
        \end{axis}
        \end{tikzpicture}
        }
    \caption{top-down approximation. }
    \label{fig:topDownFig}
    \end{wrapfigure}

The top-down approximation of Pareto-optimal points is illustrated in \cref{fig:topDownFig}. Here a weight $\convw{w}$ is the normal vector of the blue lines; the figure shows a situation after considering two weights.

\subsection{Key Idea II: Pareto Caching}\label{subsec:overviewParetoCaching}

\begin{figure}[t]
    \centering
    \scalebox{0.8}{
        \begin{tikzpicture}[
innode/.style={draw, rectangle, minimum size=0.5cm},
interface/.style={draw, rectangle, minimum size=0.5cm},]
\fill[orange] (0cm, -1cm)--(0cm, 1cm)--(1.5cm, 1cm)--(1.5cm, -1cm)--cycle;
\node[inner sep=0] (Aenr1d) at (0.75cm, 0cm) {\scalebox{1}{$\cdots$}};
\node at (1.3cm, 0.8cm) {\scalebox{1}{$\mdp{A}$}};
\node[inner sep=0] (Aenr1d) at (-0.3cm, 0cm) {};
\node[inner sep=0, right = 0.6cm of Aenr1d] (Aenr1c) {};

\node[inner sep=0] (Aexr1d) at (1.2cm, 0.5cm) {};
\node[inner sep=0] (Aexr2d) at (1.2cm, -0.5cm) {};

\fill[cyan] (1.8cm, -1cm)--(1.8cm, 1cm)--(4.5cm, 1cm)--(4.5cm, -1cm)--cycle;
\node at (4.2cm, 0.8cm) {\scalebox{1}{$\mdp{B}$}};
\node[interface,fill=white,right=1cm of Aexr1d] (Benr1c) {\scalebox{0.8}{$i_1$}};
\node[interface,fill=white,right=1cm of Aexr2d] (Benr2c) {\scalebox{0.8}{$i_2$}};
\node[state,right=0.4cm of Benr1c, minimum size=0.3cm,fill=white, yshift=-0.5cm] (s1) {};
\node[interface,fill=white,right=0.3cm of s1] (Bexr1d) {\scalebox{0.8}{$o_1$}};
\node[inner sep=0, right=0.9cm of Bexr1d] (Bexr1c) {};

\fill[orange] (4.8cm, -1cm)--(4.8cm, 1cm)--(6.3cm, 1cm)--(6.3cm, -1cm)--cycle;
\node[inner sep=0] (cdots2) at (5.55cm, 0cm) {\scalebox{1}{$\cdots$}};
\node at (6.1cm, 0.8cm) {\scalebox{1}{$\mdp{A}$}};
\node[inner sep=0] (A2exr1d) at (6cm, 0.5cm) {};
\node[inner sep=0] (A2exr2d) at (6cm, -0.5cm) {};

\fill[lime] (6.6cm, -1cm)--(6.6cm, 1cm)--(9.3cm, 1cm)--(9.3cm, -1cm)--cycle;
\node at (9.1cm, 0.8cm) {\scalebox{1}{$\mdp{D}$}};
\node[interface,fill=white,right=1cm of A2exr1d] (Denr1c) {\scalebox{0.8}{$i_3$}};
\node[interface,fill=white,right=1cm of A2exr2d] (Denr2c) {\scalebox{0.8}{$i_4$}};
\node[state,right=0.4cm of Denr1c, minimum size=0.3cm,fill=white, yshift=-0.5cm] (s2) {};
\node[interface,fill=white,right=0.3cm of s2] (Dexr1d) {\scalebox{0.8}{$o_2$}};
\node[inner sep=0, right=0.9cm of Dexr1d] (Dexr1c) {};

\fill[orange] (9.6cm, -1cm)--(9.6cm, 1cm)--(11.1cm, 1cm)--(11.1cm, -1cm)--cycle;
\node[inner sep=0] (cdots3) at (10.35cm, 0cm) {\scalebox{1}{$\cdots$}};
\node at (10.9cm, 0.8cm) {\scalebox{1}{$\mdp{A}$}};
\node[inner sep=0] (A3exr1d) at (10.8cm, 0.5cm) {};
\node[inner sep=0] (A3exr2d) at (10.8cm, -0.5cm) {};

\fill[teal] (11.4cm, -1cm)--(11.4cm, 1cm)--(14.4cm, 1cm)--(14.4cm, -1cm)--cycle;
\node at (14.1cm, 0.8cm) {\scalebox{1}{$\mdp{E}$}};
\node[interface,fill=white,right=1cm of A3exr1d] (Eenr1c) {\scalebox{0.8}{$i_5$}};
\node[interface,fill=white,right=1cm of A3exr2d] (Eenr2c) {\scalebox{0.8}{$i_6$}};
\node[state,right=0.4cm of Eenr1c, minimum size=0.3cm,fill=white, yshift=-0.5cm] (s3) {};
\node[interface,fill=white,right=0.3cm of s3] (Eexr1d) {\scalebox{0.8}{$o_3$}};
\node[inner sep=0, right=0.9cm of Eexr1d] (Eexr1c) {};

\draw[->, thick] (Aenr1d) -> (Aenr1c);
\draw[->, thick] (Aexr1d) -> (Benr1c);
\draw[->, thick] (Aexr2d) -> (Benr2c);
\draw[->, thick] (Benr1c) -> node [yshift=-0.1cm,right=-0.6cm] {\scalebox{0.8}{$0.25$}} (s1);
\draw[->, thick] (Benr2c) -> node [yshift=-0.15cm,right=-0.2cm] {\scalebox{0.8}{$0.7$}} (s1);
\draw[->, thick] (Benr1c) to [out=10,in=150] node[yshift=0.2cm,right=-0.2cm] {\scalebox{0.8}{$0.75$}}  (Bexr1d);
\draw[->, thick] (Benr2c) to [out=-10,in=210]
node[yshift=-0.2cm,right=-0.2cm] {\scalebox{0.8}{$0.3$}} (Bexr1d);
\draw[->, thick] (Bexr1d) -> (Bexr1c);
\draw[->, thick] (A2exr1d) -> (Denr1c);
\draw[->, thick] (A2exr2d) -> (Denr2c);
\draw[->, thick] (Denr1c) -> node [yshift=-0.1cm,right=-0.5cm] {\scalebox{0.8}{$0.8$}} (s2);
\draw[->, thick] (Denr2c) -> node [yshift=-0.15cm,right=-0.2cm] {\scalebox{0.8}{$0.3$}} (s2);
\draw[->, thick] (Denr1c) to [out=10,in=150] node[yshift=0.2cm,right=-0.2cm] {\scalebox{0.8}{$0.2$}}  (Dexr1d);
\draw[->, thick] (Denr2c) to [out=-10,in=210]
node[yshift=-0.2cm,right=-0.2cm] {\scalebox{0.8}{$0.7$}} (Dexr1d);
\draw[->, thick] (Dexr1d) -> (Dexr1c);
\draw[->, thick] (A3exr1d) -> (Eenr1c);
\draw[->, thick] (A3exr2d) -> (Eenr2c);
\draw[->, thick] (Eenr1c) -> node [yshift=-0.1cm,right=-0.5cm] {\scalebox{0.8}{$0.2$}} (s3);
\draw[->, thick] (Eenr2c) -> node [yshift=-0.15cm,right=-0.2cm] {\scalebox{0.8}{$0.7$}} (s3);
\draw[->, thick] (Eenr1c) to [out=10,in=150] node[yshift=0.2cm,right=-0.2cm] {\scalebox{0.8}{$0.8$}}  (Eexr1d);
\draw[->, thick] (Eenr2c) to [out=-10,in=210]
node[yshift=-0.2cm,right=-0.2cm] {\scalebox{0.8}{$0.3$}} (Eexr1d);
\draw[->, thick] (Eexr1d) -> (Eexr1c);
\end{tikzpicture}
}
    \caption{$\mdp{A}\seqcomp \mdp{B}\seqcomp \mdp{A}\seqcomp \mdp{D}\seqcomp \mdp{A}\seqcomp \mdp{E}$, an example}
    \label{fig:contribution_2}
\end{figure}
Our second key idea (\emph{Pareto caching}) arises when we try to combine the last idea (top-down compositionality) with the key advantage of the bottom-up approach~\cite{WatanabeVHRJ24}, namely \emph{exploiting duplicates}. 
Consider the string diagram  $\mdp{A}\seqcomp\mdp{B}\seqcomp\mdp{A}\seqcomp\mdp{D}\seqcomp\mdp{A}\seqcomp\mdp{E}$ in \cref{fig:contribution_2}, for motivation, where we designate multiple occurrences of $\mdp{A}$ by $\mdp{A}_{1}, \mdp{A}_{2}, \mdp{A}_{3}$ for distinction, from left to right.

Let us run the top-down algorithm. The component $\mdp{E}$ suggests the weight $(0.8,0.3)$ for the two exits of $\mdp{A}_{3}$, and $\mdp{D}$ suggest the weight $(0.2,0.7)$ for the exits of $\mdp{A}_{2}$.  Recalling that $\mdp{A}_{2}$ and $\mdp{A}_{3}$ are identical,  the weighted optimization results for these two weights can be combined, leading to a picture like \cref{fig:topDownFig}. 

Now, in \cref{fig:contribution_2}, we go on to the component $\mdp{B}$. It suggests the weight $(0.75, 0.3)$. 
\begin{itemize}
 \item In the bottom-up approach~\cite{WatanabeVHRJ24}, performance advantages are brought by \emph{exploiting duplicates}, that is, by reusing the model checking result of a component $\mdp{C}$ for its multiple occurrences. 
 \item Therefore, also here, we wish to use the previous analysis results for $\mdp{A}$---for the weights $(0.8,0.3)$ and $(0.2,0.7)$---for the weight $(0.75,0.3)$.
 \item Intuitively, $(0.75,0.3)$ seems close enough to $(0.8,0.3)$, suggesting that we can use the previously obtained result for $(0.8,0.3)$. 
\end{itemize}
But this casts the following questions: what is it for two weights to be ``close enough''? Is  $(0.75,0.3)$ really closer to $(0.8,0.3)$ than to $(0.2,0.7)$?
Can we bound errors---much like in \cref{subsec:TACAS24}---that arise from this ``approximate reuse''? 

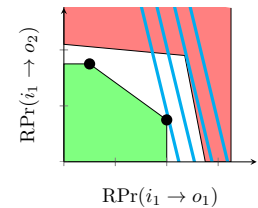
\begin{wrapfigure}[9]{r}[0pt]{4cm}
\centering

\vspace{-2em}
\pgfplotsset{width=5cm,compat=1.9}
    \scalebox{0.75}{
        \begin{tikzpicture}
        \begin{axis}[
            axis lines = left,
            axis on top = true,
            xmin=0, xmax=0.75,
            ymin=0, ymax=0.55,
            xticklabel=\empty,
            yticklabel=\empty,
            axis background/.style={fill=white},
            xlabel={$\mathrm{RPr}(i_1\to o_1)$},
            ylabel={$\mathrm{RPr}(i_1\to o_2)$}
        ]
        \addplot[fill=green!50, very thin] coordinates {(0.4,0) (0.4,0.15)  (0.1,0.35) (0,0.35) (0, 0)} -- cycle;
        \addplot[fill=red!50, very thin] coordinates {(0.65, 0.55) (0.65, 0) (0.55,0) (0.47, 0.38) (0,0.42) (0, 0.65)};		
    \addplot[
            domain = 0:0.65,
            color=cyan,
            ultra thick,
        ]
        {-3.8*x+2.19};
    \addplot[
            domain = 0:0.55,
            color=cyan,
            ultra thick,
        ]
        {-3.8*x+1.7};
    \addplot[
            domain = 0:0.55,
            color=cyan,
            ultra thick,
        ]
        {-3.8*x+1.93};
        \addplot[
            domain = 0:0.65,
            color=cyan,
            ultra thick,
        ]
        {-3.8*x+2.42};
    \node[point] at (axis cs:0.4,0.15) {};
    \node[point] at (axis cs:0.1,0.35) {};
    \end{axis}
    \end{tikzpicture}
    }
\caption{Pareto caching}
\label{fig:paretoCachingFig}
\end{wrapfigure}
 In \cref{sec:approximation}, we use the existing theory on Pareto curves in multi-objective model checking from~\cite{EtessamiKVY08,ForejtKP12,Quatmann23} to answer these questions. Intuitively,  the previous analysis result (red and green regions) gets \emph{queried} on a new weight $\convw{w}$ (the normal vector of the blue lines), as illustrated in~\cref{fig:paretoCachingFig}. We call answering weighted reachability based on the Pareto curve  \emph{Pareto caching}. The technique can prevent many invocations of using VI to compute the weighted reachability for $\convw{w}$. 
The distance between the under- and over-approximations computed this way can be big; 
if so (``cache miss''), we run VI again for the weight $\convw{w}$.

\subsection{Global Stopping Criteria (GSCs)}
On top of two key ideas, we provide two global stopping criteria (GSCs) in~\cref{sec:pareto_curves}: one is based on the ideas from OVI~\cite{HartmannsK20} and the other is a symbiosis of the Pareto caches with the bottom-up approach. 
Although ensuring the termination of our algorithm in finite steps with our GSCs remains future work, we show that our GSCs are \emph{sound}, that is, its output satisfies a given precision upon termination.

\section{Formal Problem Statement}
\label{sec:prelinaries}
We recall (weighted) reachability in Markov decision processes (MDPs) and formalize string diagrams as their compositional representation. Together, this is the formal basis for our problem statement as already introduced above.  
\subsection{Markov Decision Process (MDP)}
\label{subsec:prelimMDPs}
\begin{mydefinition}[MDP] \label{def:MDP}
    An \emph{MDP} $\mdp{M} = (S, A, P)$ is a tuple with a finite set $S$ of \emph{states}, a finite set $A$ of \emph{actions}, and a \emph{probabilistic transition function} $P\colon S\times A \rightharpoonup \dist{S}$ (which is a partial function, cf.\ notations in \cref{sec:intro}). 
\end{mydefinition}%

A \emph{(finite) path}
(on $\mdp{M}$) is a finite sequence of states $\pi\defeq (\pi_i)_{i\in \nset{m}}$.  We write $\FinPaths{\mdp{M}}$ for the set of finite paths on $\mdp{M}$. A \emph{memoryless scheduler} $\sigma$ is a function $\sigma\colon S\rightarrow \dist{A}$; in this paper, memoryless schedulers suffice~\cite{Puterman94,ForejtKP12}. We say $\sigma$ is \emph{deterministic memoryless (DM)} if for each $s\in S$, $\sigma(s)$ is  Dirac. We also write $\sigma\colon S\rightarrow A$ for a DM scheduler $\sigma$. The set of all memoryless schedulers on $\mdp{M}$ is  $\Possched{\mdp{M}}$, and the set of all DM schedulers on $\mdp{M}$ is  $\Dmsched{\mdp{M}}$.

For a memoryless scheduler $\sigma$ and a \emph{target state}  $t\in S$, the \emph{reachability probability} $\Reacha{\mdp{M}, \sigma}{s}{t}$ from a state $s$ is given by  $\Reacha{\mdp{M}, \sigma}{s}{t}\defeq \sum_{\pi \in \TFinPaths{\mdp{M}}{t}} \prmeas{\mdp{M}}{\sigma,s}(\pi)$, where (i) the set $\TFinPaths{\mdp{M}}{t}\subseteq \FinPaths{\mdp{M}}$ is defined by $\TFinPaths{\mdp{M}}{t} \defeq \{(\pi_i)_{i\in \nset{m}} \in \FinPaths{\mdp{M}}\mid \last{\pi} =  t,\text{ and } \pi_i \not = t \text{ for }i\in \nset{m-1}\}$, and (ii) the probability $\prmeas{\mdp{M}}{\sigma,s}(\pi)$ is defined by $\prmeas{\mdp{M}}{\sigma,s}(\pi)\defeq \prod_{i\in \nset{m-1}} \sum_{a\in A} P(\pi_{i}, a, \pi_{i+1})\cdot \sigma(\pi_{i-1})(a)$ if $\pi_1 = s$ and $\prmeas{\mdp{M}}{\sigma,s}(\pi)\defeq 0$ otherwise.

Towards our compositional approach for a reachability objective, we must generalize the objective to a \emph{weighted reachability probability} objective: we want to compute the \emph{weighted sum}---with respect to a certain weight vector $\convw{w}$---over reachability probabilities to multiple target states. The standard reachability probability problem is a special case of this weighted reachability problem using a suitable unit vector 
$\convw{e}$ as the weight $\convw{w}$.

\begin{mydefinition}[weighted reachability probability]\label{def:weightedReachProb}
Let $\mdp{M}$ be an MDP, and $T$ be a set of target states.
A \emph{weight} $\convw{w}$ on $T$ is a vector $\convw{w} \defeq (w_t)_{t\in T}\in \probinterval^{T}$. 

Let $s$ be a state, and $\sigma$ be a scheduler.
The \emph{weighted reachability probability} $\WReacha{\mdp{M},\sigma, T}{\convw{w}}{s}\in \probinterval$ from $s$ to $T$ over $\sigma$ with respect to a weight  $\convw{w}$ is defined naturally by a weighted sum, that is, 
 $\WReacha{\mdp{M},\sigma, T}{\convw{w}}{s} \defeq \sum_{t\in T} w_t\cdot \Reacha{\mdp{M},\sigma}{s}{t}$. We write $\MaxWReacha{\mdp{M}, T}{\convw{w}}{s}$ for the \emph{maximum weighted reachability probability} 
\begin{math}
 \sup_{\sigma} \WReacha{\mdp{M},\sigma, T}{\convw{w}}{s}
\end{math}. (The supremum is realizable; see e.g.~\cite{HartmannsJKQ20}.)

\end{mydefinition}

\subsection{String Diagram of MDPs}
\begin{mydefinition}[oMDP]
\label{def:openMDPs}
An \emph{open MDP} (oMDP) $\mdp{A} = (M, \interface)$ is a pair consisting of an MDP $M$ with \emph{open ends} $\interface =(\enr, \enl, \exr, \exl)$, where % given by the following data:
  $\enr, \enl, \exr, \exl\subseteq S$ are pairwise disjoint and each of them is totally ordered. The states in $\en \defeq \enr\cup \enl$ are the \emph{entrances}, and the states in  $\ex \defeq \exr\cup\exl$ are the \emph{exits}, respectively. We often use superscripts to designate the oMDP $\mdp{A}$ in question, such as $\en^{\mdp{A}}$ and $\ex^{\mdp{A}}$.
\end{mydefinition}
We write $\typemdp{\mdp{A}}\colon (\arr{m}, \arl{m})\rightarrow  (\arr{n}, \arl{n})$ for the \emph{arities} of $\mdp{A}$, where $\arr{m} \defeq |\enr|$, $\arl{m} \defeq |\exl|$, $\arr{n} \defeq |\exr|$, and $\arl{n} \defeq |\enl|$.
We assume that every exit $s$ is a sink state, that is, $P(s, a)$ is undefined for any $a\in A$. We can naturally lift the definitions of schedulers and weighted reachability probabilities from MDPs to oMDPs: we will be particularly interested in the following instances; 
1) 
the \emph{weighted reachability probability} $\WReacha{\mdp{A},\sigma}{\convw{w}}{i}\defeq\WReacha{\mdp{A},\sigma, \ex^{\mdp{A}}}{\convw{w}}{i}$ from a chosen entrance $i$ to the set $\ex^{\mdp{A}}$ of all exits; and
2)
the \emph{maximum weighted reachability probability} $\MaxWReacha{\mdp{A}}{\convw{w}}{i}\defeq\sup_{\sigma}\WReacha{\mdp{A}, \sigma}{\convw{w}}{i}$ from $i$ to $\ex^{\mdp{A}}$ weighted by $\convw{w}$.

We define \emph{string diagrams of MDPs}~\cite{WatanabeEAH23} syntactically, as  syntactic trees
 whose leaves are oMDPs and non-leaf nodes are algebraic operations. The latter are syntactic operations and they are yet to be interpreted.

\begin{mydefinition}[string diagram of MDPs]
\label{def:sd_mdps}
A \emph{string diagram $\sd{D}$ of MDPs} is a term adhering to the  grammar $\sd{D}\; ::= \;\mathsf{c}_\mdp{A} \mid \sd{D} \seqcomp \sd{D} \mid \sd{D} \oplus \sd{D}$, where $\constMDP{\mdp{A}}$ is a constant designating an oMDP $\mdp{A}$. 
\end{mydefinition}
The above syntactic operations $\seqcomp, \oplus$ are interpreted by the \emph{semantic} operations below. 
The following definitions explicate the graphical intuition in \cref{fig:seqcompAndSum}.

\begin{mydefinition}[sequential composition $\seqcomp$]\label{def:seqOMDP}
    Let $\mdp{A}$, $\mdp{B}$ be oMDPs, $\typemdp{\mdp{A}} = (\arr{m}, \arl{m}) \rightarrow (\arr{l}, \arl{l})$, and $\typemdp{\mdp{B}} = (\arr{l}, \arl{l}) \rightarrow (\arr{n}, \arl{n})$. Their \emph{sequential composition} $\mdp{A} \seqcomp \mdp{B}$  is the oMDP $(M, \interface')$ where  $\interface' = (\enr^{\mdp{A}}, \enl^{\mdp{B}}, \exr^{\mdp{B}}, \exl^{\mdp{A}})$,  $M \defeq \big((S^{\mdp{A}} \uplus S^{\mdp{B}})\setminus (\exr^{\mdp{A}}\uplus \exl^{\mdp{B}}), A^{\mdp{A}} \uplus A^{\mdp{B}}, P\big)$ and $P$  is
    \begin{align*}\small
\begin{array}{ll}
         P(s, a, s') &\defeq \begin{cases}
            P^{\mdp{D}}(s, a, s') &\text{if $\mdp{D} \in \{\mdp{A}, \mdp{B}\}$, $s\in S^{\mdp{D}}$, $a\in A^{\mdp{D}}$, and $s'\in S^{\mdp{D}}$, }\\
            P^{\mdp{A}}(s, a, \exrarg{i}^{\mdp{A}}) &\text{if $s\in S^{\mdp{A}}$, $a\in A^{\mdp{A}}$, $s' = \enrarg{i}^{\mdp{B}}$ for some $1 \leq i \leq \arr{l}$},\\
            P^{\mdp{B}}(s, a, \exlarg{i}^{\mdp{B}}) &\text{if $s\in S^{\mdp{B}}$, $a\in A^{\mdp{B}}$, $s' = \enlarg{i}^{\mdp{A}}$ for some $1 \leq i \leq \arl{l}$},\\
            0 &\text{otherwise. }
        \end{cases}
\end{array}    
\end{align*}
\end{mydefinition}

\begin{mydefinition}[sum $\oplus$]
\label{def:sumOMDP}
    Let $\mdp{A}, \mdp{B}$ be oMDPs.
    Their \emph{sum} $\mdp{A}\oplus\mdp{B}$ is the oMDP $(M, \interface')$ where  $\interface' = (\enr^{\mdp{A}}\uplus \enr^{\mdp{B}}, \enl^{\mdp{A}}\uplus \enl^{\mdp{B}}, \exr^{\mdp{A}}\uplus \exr^{\mdp{B}}, \exl^{\mdp{A}}\uplus \exl^{\mdp{B}})$,  $M = (S^{\mdp{A}} \uplus S^{\mdp{B}}, A^{\mdp{A}} \uplus A^{\mdp{B}}, P)$, and $P$ is given by $P(s, a, s') \defeq P^{\mdp{D}}(s, a, s')$ if $\mdp{D} \in \{\mdp{A}, \mdp{B}\}$, $s\in S^{\mdp{D}}$, $a\in A^{\mdp{D}}$, and $s'\in S^{\mdp{D}}$, and otherwise $P(s, a, s')\defeq 0$.
\end{mydefinition}

\begin{mydefinition}[operational semantics $\semantics{\sd{D}}$]
\label{def:opsem}
 Let $\sd{D}$ be a string diagram of MDPs. 
 The \emph{operational semantics} $\semantics{\sd{D}}$ is the oMDP which is inductively defined by Defs~\ref{def:seqOMDP} and~\ref{def:sumOMDP}, with the base case $\semantics{\constMDP{\mdp{A}}}=\mdp{A}$. Here we assume that every string diagram $\sd{D}$ has matching arities so that compositions are well-defined.
 We call $\en^{\semantics{\sd{D}}}$ and $\ex^{\semantics{\sd{D}}}$ \emph{global entrances} and \emph{global exits} of $\sd{D}$, respectively. 
\end{mydefinition}
\begin{wrapfigure}[6]{r}[0pt]{7.5cm}
\centering
\vspace{-2em}
\scalebox{0.7}{
    \begin{tikzpicture}[
innode/.style={draw, rectangle, minimum size=0.5cm},
interface/.style={draw, rectangle, minimum size=0.5cm},]
\fill[orange] (0cm, -1cm)--(0cm, 1cm)--(3.1cm, 1cm)--(3.1cm, -1cm)--cycle;
\node[inner sep=0] (Aenr1d) at (1.6cm, 0cm) {\scalebox{1}{$\cdots$}};
\node at (2.8cm, 0.9cm) {\scalebox{0.7}{$\mdp{A}_1$}};
\node[inner sep=0] (Aenr1d) at (-0.3cm, 0.5cm) {};
\node[interface,fill=white, right = 0.5cm of Aenr1d] (Aenr1c) {\scalebox{0.6}{$i^{\mdp{A}_1}_1$}};
\node[inner sep=0] (Aexl1c) at (-0.3cm, -0.5cm) {};
\node[interface,fill=white, right = 0.5cm of Aexl1c] (Aexl1d) {\scalebox{0.6}{$o^{\mdp{A}_1}_1$}};
\node[inner sep=0, right = 0.4cm of Aenr1c] (Aenr1out) {};
\node[inner sep=0, right = 0.4cm of Aexl1d] (Aexl1in) {};
\node[inner sep=0, right = 1.1cm of Aenr1c] (Aexr1in) {};
\node[interface,fill=white, right=0.4cm of Aexr1in] (Aexr1d) {\scalebox{0.6}{$o^{\mdp{A}_1}_2$}};
\node[inner sep=0, right = 1.1cm of Aexl1d] (Aenl1out) {};
\node[interface,fill=white, right=0.4cm of Aenl1out] (Aenl1c) {\scalebox{0.6}{$i^{\mdp{A}_1}_2$}};

\fill[orange] (3.4cm, -1cm)--(3.4cm, 1cm)--(6.5cm, 1cm)--(6.5cm, -1cm)--cycle;
\node[inner sep=0] (A2enr1d) at (5cm, 0cm) {\scalebox{1}{$\cdots$}};
\node at (6.2cm, 0.9cm) {\scalebox{0.7}{$\mdp{A}_2$}};
\node[inner sep=0] (A2enr1d) at (3.1cm, 0.5cm) {};
\node[interface,fill=white, right = 0.5cm of A2enr1d] (A2enr1c) {\scalebox{0.6}{$i^{\mdp{A}_2}_1$}};
\node[inner sep=0] (A2exl1c) at (3.1cm, -0.5cm) {};
\node[interface,fill=white, right = 0.5cm of A2exl1c] (A2exl1d) {\scalebox{0.6}{$o^{\mdp{A}_2}_1$}};
\node[inner sep=0, right = 0.4cm of A2enr1c] (A2enr1out) {};
\node[inner sep=0, right = 0.4cm of A2exl1d] (A2exl1in) {};
\node[inner sep=0, right = 1.1cm of A2enr1c] (A2exr1in) {};
\node[interface,fill=white, right=0.4cm of A2exr1in] (A2exr1d) {\scalebox{0.6}{$o^{\mdp{A}_2}_2$}};
\node[inner sep=0, right = 1.1cm of A2exl1d] (A2enl1out) {};
\node[interface,fill=white, right=0.4cm of A2enl1out] (A2enl1c) {\scalebox{0.6}{$i^{\mdp{A}_2}_2$}};

\fill[cyan] (6.8cm, -1cm)--(6.8cm, 1cm)--(9.8cm, 1cm)--(9.8cm, -1cm)--cycle;
\node at (9.6cm, 0.9cm) {\scalebox{0.7}{$\mdp{B}$}};
\node[interface,fill=white,right=0.8cm of A2exr1d] (Benr1c) {\scalebox{0.6}{$i^{\mdp{B}}_1$}};
\node[interface,fill=white,right=0.8cm of A2enl1c] (Bexl1d) {\scalebox{0.6}{$o^{\mdp{B}}_1$}};
\node[inner sep=0, right = 0.4cm of Benr1c] (Benr1out) {};
\node[inner sep=0, right = 0.4cm of Bexl1d] (Bexl1in) {};
\node[inner sep=0, right = 0.1cm of Benr1out, yshift=-0.5cm] (bcdots) {\scalebox{1}{$\cdots$}};
% \node[state,right=0.4cm of Benr1c, minimum size=0.3cm,fill=white, yshift=-0.5cm] (s1) {};
\node[inner sep=0, right = 1.1cm of Benr1c, yshift=-0.5cm] (Bexr1in) {};
\node[interface,fill=white,right=1.5cm of Benr1c, yshift=-0.5cm] (Bexr1d) {\scalebox{0.6}{$o^{\mdp{B}}_2$}};
\node[inner sep=0, right=0.5cm of Bexr1d] (Bexr1c) {};

\draw[->, thick] (Aenr1d) -> (Aenr1c);
\draw[->, thick] (Aenr1c) -> (Aenr1out);
\draw[->, thick] (Aexl1d) -> (Aexl1c);
\draw[->, thick] (Aexl1in) -> (Aexl1d);
\draw[->, thick] (Aexr1in) -> (Aexr1d);
\draw[->, thick] (Aenl1c) -> (Aenl1out);
\draw[->, thick] (Aexr1d) -> (A2enr1c);
\draw[->, thick] (A2exl1d) -> (Aenl1c);
\draw[->, thick] (A2enr1c) -> (A2enr1out);
\draw[->, thick] (A2exl1in) -> (A2exl1d);
\draw[->, thick] (A2exr1in) -> (A2exr1d);
\draw[->, thick] (A2enl1c) -> (A2enl1out);
\draw[->, thick] (A2exr1d) -> (Benr1c);
\draw[->, thick] (Bexl1d) -> (A2enl1c);
\draw[->, thick] (Benr1c) -> (Benr1out);
\draw[->, thick] (Bexl1in) -> (Bexl1d);
\draw[->, thick] (Bexr1in) -> (Bexr1d);
% \draw[->, thick] (Benr1c) -> node [yshift=-0.1cm,right=-0.5cm] {\scalebox{0.8}{$0.2$}} (s1);
% \draw[->, thick] (Benr2c) -> node [yshift=-0.15cm,right=-0.2cm] {\scalebox{0.8}{$0.7$}} (s1);
% \draw[->, thick] (Benr1c) to [out=10,in=150] node[yshift=0.2cm,right=-0.2cm] {\scalebox{0.8}{$0.8$}}  (Bexr1d);
% \draw[->, thick] (Benr2c) to [out=-10,in=210]
% node[yshift=-0.2cm,right=-0.2cm] {\scalebox{0.8}{$0.3$}} (Bexr1d);
\draw[->, thick] (Bexr1d) -> (Bexr1c);
\end{tikzpicture}
}
\caption{$\semantics{\sd{D}}$ in Ex.~\ref{ex:auxiliaryOnStrDiag}}
\label{fig:entExitNames}
\end{wrapfigure}

For describing the occurrence of oMDPs and their duplicates in a string diagram $\sd{D}$, we formally define \emph{nominal components} $\leaf{\sd{D}}$ and \emph{components} $\CP{\sd{D}}$. The latter for graph-theoretic operations in our compositional VI (CVI) (\cref{alg:CVIprototype}), while the former is for Pareto caching (\cref{sec:approximation}). 
Examples are provided later in Ex.~\ref{ex:auxiliaryOnStrDiag}. 
\begin{mydefinition}[$\leaf{\sd{D}}$, $\CP{\sd{D}}$]\label{def:compCompDesig}
The set $\leaf{\sd{D}}$ of \emph{nominal components} is the set of constants occurring in $\sd{D}$ (as a term). 
The set $\CP{\sd{D}}$ of \emph{components} is inductively defined by the following: 
$\CP{\constMDP{\mdp{A}}}\defeq \{\mdp{A}\}$,  and
$
 \CP{\sd{E}\ast\sd{F}}\defeq \CP{\sd{E}}\uplus\CP{\sd{F}}
$ for $\ast\in \{\seqcomp, \oplus\}$; here we count multiplicities, unlike $\leaf{\sd{D}}$. 
\end{mydefinition}
 We introduce \emph{local open ends} of string diagrams, in contrast to global open ends defined in Def.~\ref{def:opsem}. 
\begin{mydefinition}[$\enLocal{\sd{D}}$, $\exLocal{\sd{D}}$ (local)]\label{def:entExitLocalGlobal}
The sets $\enLocal{\sd{D}}$ and $\exLocal{\sd{D}}$ of \emph{local entrances} and \emph{exits} of $\sd{D}$ are given by $\enLocal{\sd{D}}\defeq\biguplus_{\mdp{A}\in \CP{\sd{D}}}\en^{\mdp{A}}$, and  $\exLocal{\sd{D}}\defeq\biguplus_{\mdp{A}\in \CP{\sd{D}}}\ex^{\mdp{A}}$, respectively. 
Clearly we have $\enGlobal{\sd{D}}\subseteq \enLocal{\sd{D}}$, 
 $\exGlobal{\sd{D}}\subseteq \exLocal{\sd{D}}$.
\end{mydefinition}
\begin{myexample}\label{ex:auxiliaryOnStrDiag}
Let $\sd{D}=\constMDP{\mdp{A}}\seqcomp\constMDP{\mdp{A}}\seqcomp\constMDP{\mdp{B}}$, where $\mdp{A}$ and $\mdp{B}$ are from \cref{fig:openMDPs}. The oMDP $\semantics{\sd{D}}$ is shown in \cref{fig:entExitNames}. 
Then $\leaf{\sd{D}}=\{\constMDP{\mdp{A}},\constMDP{\mdp{B}}
\}$, 
while $\CP{\sd{D}}=\{\mdp{A}_{1}, \mdp{A}_{2}, \mdp{B}\}$ with subscripts added for distinction. We have 
 $\enGlobal{\sd{D}}=\{i^{\mdp{A}_{1}}_{1}\}$ and $\exGlobal{\sd{D}}=\{o^{\mdp{A}_{1}}_{1}, o^{\mdp{B}}_{2}\}$, and $\enLocal{\sd{D}}=
\{
i^{\mdp{A}_{1}}_{1},
i^{\mdp{A}_{1}}_{2},
i^{\mdp{A}_{2}}_{1},
i^{\mdp{A}_{2}}_{2},
i^{\mdp{B}}_{1}
\}
$ and $\exLocal{\sd{D}}=
\{
o^{\mdp{A}_{1}}_{1},
o^{\mdp{A}_{1}}_{2},
o^{\mdp{A}_{2}}_{1},
o^{\mdp{A}_{2}}_{2},
o^{\mdp{B}}_{1},
o^{\mdp{B}}_{2}
\}
$. Note also that $\exLocal{\sd{D}}$ does \emph{not} suppress exits removed in sequential composition, such as
$\{
o^{\mdp{A}_{1}}_{2},
o^{\mdp{A}_{2}}_{1},
o^{\mdp{A}_{2}}_{2},
o^{\mdp{B}}_{1}
\}$.
\end{myexample}

\medskip
\begin{mdframed}
{\bf Problem: Near-Optimal Weighted Reachability Probability}
\quad

\noindent Given a string diagram $\sd{D}$, an entrance $i\in \enGlobal{\sd{D}}$, 
a weight $\convw{w}
\in \probinterval^{\exGlobal{\sd{D}}}$ over exits, 
and an error bound $\epsilon \in \probinterval$,
compute an under-approximation $l\in \probinterval$ such that 
    $l\leq \MaxWReacha{\semantics{\sd{D}}}{\convw{w}}{i}\leq l+\epsilon$. 
\end{mdframed}
We remark that as a straightforward extension, we can also extract a scheduler that achieves the under-approximation.

\section{VI
in a Compositional Setting}
\label{sec:VIandMOO}
We recap \emph{value iteration (VI)}~\cite{Puterman94, BaierK08} and its extension to \emph{optimistic value iteration (OVI)}~\cite{HartmannsK20} before presenting our \emph{compositional VI (CVI)}. 
\subsection{Value Iteration (VI) and Optimistic Value Iteration (OVI)}
\label{subsec:VIOVI}

VI relies on the characterization of maximum reachability probabilities as a \emph{least fixed point (lfp)}, specifically the lfp $\mu \bellman{\mdp{M}}{T}$ of the \emph{Bellman operator} $\bellman{\mdp{M}}{T}$: the Bellman operator $\bellman{\mdp{M}}{T}$ is an operator on the set $\probinterval^S$ that intuitively returns the $t+1$-step reachability probabilities given the $t$-step reachability probabilities. 
\iffull
Def.~\ref{def:bellman_operator} contains a formal treatment.
\else 
A formal treatment can be found in~\cite[Appendix B]{WVJH2024accepted}.   
\fi
Then the \emph{Kleene sequence} $\bot \le \bellman{\mdp{M}}{T}(\bot)\le\bellman{\mdp{M}}{T}^{2}(\bot)\le\cdots$ gives a monotonically increasing sequence that converges to the lfp  $\mu \bellman{\mdp{M}}{T}$,
where $\bot$ is the least element. This also applies to weighted reachability probabilities. %(\cref{sec:details_VI}). 

While VI gives guaranteed under-approximations, it does not say \emph{how close} the current approximation is to the solution  $\mu \bellman{\mdp{M}}{T}$\footnote{The challenge applies to VI in (our) undiscounted setting,  where the Bellman operator is not a contraction operator. With discounting, one can easily approximate the gap.}. The capability of providing guaranteed over-approximations as well is called \emph{soundness} in  VI, and many techniques come with soundness~\cite{QuatmannK18,HartmannsK20,HaddadM18,PhalakarnTHH20}. Soundness is useful for \emph{stopping criteria}: one can fix an \emph{error bound} $\eta \in \probinterval$; VI can terminate when the distance between under- and over-approximations is at most $\eta$.

Among sound VI techniques, in this paper we focus on optimistic VI (OVI) due to its proven performance record~\cite{BuddeHKKPQTZ20,HartmannsJQW23}. We use OVI in many places, specifically for 1) stopping criteria for local VIs in~\cref{sec:cvi}, 2) caching heuristics in~\cref{sec:approximation}, and 3) a stopping criterion for global (compositional) VI
 in~\cref{sec:pareto_curves}.

The main steps of OVI proceed as follows: 1) a VI iteration produces an under-approximation $l$ for every state; 2) we heuristically pick an over-approximation candidate $u$, for example by $u\defeq l+\epsilon$; and 3) we verify the candidate $u$ by checking if $\bellman{\mdp{M}}{T}(u)\le u$. If the last holds, then by the \emph{Park induction principle}~\cite{park1969fixpoint},
 $u$ is guaranteed to over-approximate the lfp  $\mu \bellman{\mdp{M}}{T}$. If it does not, then we refine $l,u$ and try again. See~\cite{HartmannsK20} for details.

\subsection{Going Top-Down in Compositional Value Iteration}
\label{sec:cvi}

\begin{algorithm}[t]
\caption{A prototype of compositional value iteration (CVI)
}
\begin{algorithmic}[1]
\Statex \textbf{Input:} a string diagram $\sd{D}$ of MDPs and a weight $\convw{w}\in\probinterval^{\exGlobal{\sd{D}}}$, as in the target problem.
\Statex \textbf{Output:} a function $f\colon \enGlobal{\sd{D}}\rightarrow \probinterval$. 
\State \label{line:CVI_init1}initialize $g\colon \enLocal{\sd{D}}\rightarrow \probinterval$ as the least element $\bot$ (i.e.\ everywhere $0$)
\State\label{line:CVI_init2}initialize $h\colon \exLocal{\sd{D}}\rightarrow \probinterval$ as everywhere $0$, except for global exits $o$ where $h(o)\defeq w_o$ (depending on the weight $\convw{w}=(w_o)_{o}$)
 \setcounter{ALG@line}{3}
\While{$\mathbf{not}\ \mathtt{GlobalStoppingCriterion}(g)$}
\label{line:CVI_beginwhile}
 \For{each $\mdp{A}\in \CP{\sd{D}}$}  \Comment{for each component  $\mdp{A}$, counting multiplicities}
  \State \label{line:prototypeLocalVI} $g_{\mdp{A}}\gets\mathtt{LocalVI}(\mdp{A}, h|_{\ex^{\mdp{A}}})$
       \Comment{run VI locally in $\mdp{A}$ and obtain $g_{\mdp{A}}\colon \en^{\mdp{A}}\to \probinterval$}
 \EndFor
 \setcounter{ALG@line}{10}
 \State $g\gets \coprod_{\mdp{A}\in \CP{\sd{D}}}g_{\mdp{A}}$ 
 \label{line:CVI_patching}\Comment{$g\colon \enLocal{\sd{D}}\rightarrow\probinterval$ is obtained by patching $(g_{\mdp{A}})_{\mdp{A}}$}

 \State $h\gets\texttt{PropagateSeqComp}(g, \convw{w})$  
 \label{line:prototypePropag}			

 \Statex\Comment{$g$'s update is propagated to $h$ along sequential composition, see \cref{subfig:propagation}}

\EndWhile
\State \Return $ g|_{\enGlobal{\sd{D}}}$   \Comment{restrict $g\colon \enLocal{\sd{D}}\rightarrow \probinterval$ along $\enGlobal{\sd{D}}\subseteq\enLocal{\sd{D}}$}
\end{algorithmic}
\label{alg:CVIprototype}
\end{algorithm}

 We move on to formalize the story of \cref{subsec:toTopDown}. \cref{alg:CVIprototype} is a prototype of our proposed algorithm, where compositional VI is run in a top-down manner. It will be combined with Pareto caching (\cref{sec:approximation}) and the stopping criteria introduced in~\cref{sec:pareto_curves}. 
A high-level view of~\cref{alg:CVIprototype} is the iteration of the following operations: 1) running local VI in each component oMDP, and 2) propagating its result along sequential composition, from an entrance of a succeeding component, to the corresponding exit of a preceding component.
See \cref{fig:prototypeOverview} for illustration.
\begin{auxproof}
\begin{myremark}
  We note that $g$ and $h$ can be merged in implementation. Here the domain of the merged function will be $\enGlobal{\sd{D}}\cup \exGlobal{\sd{D}}$ minus the exits that get removed in sequential composition. This way, we no longer need the propagation step. 
\end{myremark}
\end{auxproof}
\begin{figure}[t]
    \centering
    \begin{minipage}[t]{0.3\hsize}
    % \vspace{-1.85cm}
    \scalebox{0.7}{
    \begin{tikzpicture}[
innode/.style={draw, rectangle, minimum size=0.5cm},
interface/.style={draw, rectangle, minimum size=0.5cm},]
        
        \fill[lightgray] (-0.5,0) -- (-0.5,2.2) -- (1.7,2.2) -- (1.7,0) -- cycle;
        \node at (1.5cm, 2cm) {\scalebox{1}{$\mdp{A}$}};
        \fill[lightgray] (2,0) -- (2,2.2) -- (4.1,2.2) -- (4.1,0) -- cycle;
        \node at (3.9cm, 2cm) {\scalebox{1}{$\mdp{B}$}};
        \node[interface,fill=white,yshift=1cm] (s0) {\scalebox{0.7}{$i^{\mdp{A}}_1$}};
        \node[inner sep=0,right=-0.5cm of s0,yshift=0.4cm] (venr1) {{\color{blue} $0$}};
        \node[inner sep=0, right = 0.2cm of s0] (acdots) {$\cdots$};
        \node[inner sep=0,right=-1cm of s0] (enr1) {};
        \node[interface,right=0.75cm of s0,fill=white,yshift=0.5cm] (exr1) {\scalebox{0.7}{$o^{\mdp{A}}_1$}};
        \node[inner sep=0,right=-0.5cm of exr1,yshift=0.4cm] (vexr1) {{\color{red} $0$}};
        \node[interface,right=0.75cm of s0,fill=white,yshift=-0.5cm] (exr2) {\scalebox{0.7}{$o^{\mdp{A}}_2$}};
        \node[inner sep=0,right=-0.5cm of exr2,yshift=0.4cm] (vexr2) {{\color{red} $0$}};
        \node[interface,fill=white,right=0.7cm of exr1] (s3) {\scalebox{0.7}{$i^{\mdp{B}}_1$}};
        \node[inner sep=0,right=-0.5cm of s3,yshift=0.4cm] (vs3) {{\color{blue} $0$}};
        \node[interface,fill=white,right=0.7cm of exr2] (s4) {\scalebox{0.7}{$i^{\mdp{B}}_2$}};
        \node[inner sep=0,right=-0.5cm of s4,yshift=0.4cm] (vs4) {{\color{blue} $0$}};
        \node[interface,right=0.75cm of s0,fill=white,yshift=0.5cm,right=0.75cm of s4] (exr3) {\scalebox{0.7}{$o^{\mdp{B}}_1$}};
        \node[inner sep=0, right = -1.1cm of exr3] (bcdots) {$\cdots$};
        \node[inner sep=0,right=-0.5cm of exr3,yshift=0.4cm] (vexr3) {{\color{red} $1$}};
        \node[inner sep=0,right=0.4cm of exr3] (s5) {};
        \draw[->] (enr1) -> (s0);
        % \draw[->,dotted] (s0) -> (exr1);
        % \draw[->,dotted] (s0) -> (exr2);
        \draw[->] (exr1) -> (s3);
        \draw[->] (exr2) -> (s4);
        % \draw[->,dotted] (s3) -> (exr3);
        % \draw[->,dotted] (s4) -> (exr3);
        \draw[->] (exr3) -> (s5);
        \draw [white, arrows = {Stealth-}, ultra thick] (1.1,-0.3) -- (2.7,-0.3);
    \end{tikzpicture}
    }
    \centering
    \subcaption{Initialization,  lines~\ref{line:CVI_init1}--\ref{line:CVI_init2}.}\label{subfig:initialization}
    \end{minipage}\hfill
    \begin{minipage}[t]{0.3\hsize}
    \centering
    \scalebox{0.7}{
    \begin{tikzpicture}[
innode/.style={draw, rectangle, minimum size=0.5cm},
interface/.style={draw, rectangle, minimum size=0.5cm},]
        
        \fill[lightgray] (-0.5,0) -- (-0.5,2.2) -- (1.7,2.2) -- (1.7,0) -- cycle;
        \node at (1.5cm, 2cm) {\scalebox{1}{$\mdp{A}$}};
        
        \fill[lightgray] (2,0) -- (2,2.2) -- (4.1,2.2) -- (4.1,0) -- cycle;
        \node at (3.9cm, 2cm) {\scalebox{1}{$\mdp{B}$}};
        \node[interface,fill=white,yshift=1cm] (s0) {\scalebox{0.7}{$i^{\mdp{A}}_1$}};
        \node[inner sep=0, right = 0.2cm of s0] (acdots) {$\cdots$};
        \node[inner sep=0,right=-0.5cm of s0,yshift=0.4cm] (venr1) {{\color{blue} $0$}};
        \node[inner sep=0,right=-1cm of s0] (enr1) {};
        \node[interface,right=0.75cm of s0,fill=white,yshift=0.5cm] (exr1) {\scalebox{0.7}{$o^{\mdp{A}}_1$}};
        \node[inner sep=0,right=-0.5cm of exr1,yshift=0.4cm] (vexr1) {{\color{red} $0$}};
        \node[interface,right=0.75cm of s0,fill=white,yshift=-0.5cm] (exr2) {\scalebox{0.7}{$o^{\mdp{A}}_2$}};
        \node[inner sep=0,right=-0.5cm of exr2,yshift=0.4cm] (vexr2) {{\color{red} $0$}};
        \node[interface,fill=white,right=0.7cm of exr1] (s3) {\scalebox{0.7}{$i^{\mdp{B}}_1$}};
        \node[inner sep=0,right=-0.5cm of s3,yshift=0.4cm] (vs3) {{\color{blue} $0\to 0.5$}};
        \node[interface,fill=white,right=0.7cm of exr2] (s4) {\scalebox{0.7}{$i^{\mdp{B}}_2$}};
        \node[inner sep=0,right=-0.5cm of s4,yshift=0.4cm] (vs4) {{\color{blue} $0\to 0.2$}};
        \node[interface,right=0.75cm of s0,fill=white,yshift=0.5cm,right=0.75cm of s4] (exr3) {\scalebox{0.7}{$o^{\mdp{B}}_1$}};
        \node[inner sep=0, right = -1.1cm of exr3, yshift=0.2cm] (bcdots) {$\cdots$};
        \node[inner sep=0,right=-0.5cm of exr3,yshift=0.4cm] (vexr3) {{\color{red} $1$}};
        \node[inner sep=0,right=0.4cm of exr3] (s5) {};
        \draw[->] (enr1) -> (s0);
        % \draw[->,dotted] (s0) -> (exr1);
        % \draw[->,dotted] (s0) -> (exr2);
        \draw[->] (exr1) -> (s3);
        \draw[->] (exr2) -> (s4);
        % \draw[->,dotted] (s3) -> (exr3);
        % \draw[->,dotted] (s4) -> (exr3);
        \draw[->] (exr3) -> (s5);
        \draw [teal, arrows = {Stealth-}, ultra thick] (2.2,-0.3) -- (3.8,-0.3);
    \end{tikzpicture}
    }
    \subcaption{Local VI, line~\ref{line:prototypeLocalVI}.}\label{subfig:localVI}
    \end{minipage}\hfill
    \begin{minipage}[t]{0.3\hsize}
    \centering
    \pgfplotsset{width=4.5cm,compat=1.9}
    \scalebox{0.7}{
    \begin{tikzpicture}[
innode/.style={draw, rectangle, minimum size=0.5cm},
interface/.style={draw, rectangle, minimum size=0.5cm},]
        
        \fill[lightgray] (-0.5,0) -- (-0.5,2.2) -- (1.7,2.2) -- (1.7,0) -- cycle;
        \node at (-0.3cm, 2cm) {\scalebox{1}{$\mdp{A}$}};
        \fill[lightgray] (2,0) -- (2,2.2) -- (4.1,2.2) -- (4.1,0) -- cycle;
        \node at (3.9cm, 2cm) {\scalebox{1}{$\mdp{B}$}};
        \node[interface,fill=white,yshift=1cm] (s0) {\scalebox{0.7}{$i^{\mdp{A}}_1$}};
        \node[inner sep=0, right = 0.2cm of s0, yshift=0.2cm] (acdots) {$\cdots$};
        \node[inner sep=0,right=-0.5cm of s0,yshift=0.4cm] (venr1) {{\color{blue} $0$}};
        \node[inner sep=0,right=-1cm of s0] (enr1) {};
        \node[interface,right=0.75cm of s0,fill=white,yshift=0.5cm] (exr1) {\scalebox{0.7}{$o^{\mdp{A}}_1$}};
        \node[inner sep=0,right=-1cm of exr1,yshift=0.4cm] (vexr1) {{\color{red} $0\to 0.5$}};
        \node[interface,right=0.75cm of s0,fill=white,yshift=-0.5cm] (exr2) {\scalebox{0.7}{$o^{\mdp{A}}_2$}};
        \node[inner sep=0,right=-1cm of exr2,yshift=0.4cm] (vexr2) {{\color{red} $0\to 0.2$}};
        \node[interface,fill=white,right=0.7cm of exr1] (s3) {\scalebox{0.7}{$i^{\mdp{B}}_1$}};
        \node[inner sep=0,right=-0.5cm of s3,yshift=0.4cm] (vs3) {{\color{blue} $0.5$}};
        \node[interface,fill=white,right=0.7cm of exr2] (s4) {\scalebox{0.7}{$i^{\mdp{B}}_2$}};
        \node[inner sep=0,right=-0.5cm of s4,yshift=0.4cm] (vs4) {{\color{blue} $0.2$}};
        \node[interface,right=0.75cm of s0,fill=white,yshift=0.5cm,right=0.75cm of s4] (exr3) {\scalebox{0.7}{$o^{\mdp{B}}_1$}};
        \node[inner sep=0, right = -1.1cm of exr3] (bcdots) {$\cdots$};
        \node[inner sep=0,right=-0.5cm of exr3,yshift=0.4cm] (vexr3) {{\color{red} $1$}};
        \node[inner sep=0,right=0.4cm of exr3] (s5) {};
        \draw[->] (enr1) -> (s0);
        % \draw[->,dotted] (s0) -> (exr1);
        % \draw[->,dotted] (s0) -> (exr2);
        \draw[->] (exr1) -> (s3);
        \draw[->] (exr2) -> (s4);
        % \draw[->,dotted] (s3) -> (exr3);
        % \draw[->,dotted] (s4) -> (exr3);
        \draw[->] (exr3) -> (s5);
        \draw [teal, arrows = {Stealth-}, ultra thick] (1.1,-0.3) -- (2.7,-0.3);
    \end{tikzpicture}
    }
    \subcaption{Propag.\ along $\seqcomp$, line~\ref{line:prototypePropag}.}\label{subfig:propagation}
    \end{minipage}
    \caption{an overview of  \cref{alg:CVIprototype}. In the MDP $\semantics{\sd{D}}$, the exit $o^{\mdp{A}}_{1}$ and the entrance $i^{\mdp{B}}_{1}$ get merged in $\mdp{A}\seqcomp\mdp{B}$ (Def.~\ref{def:seqOMDP}); here they are distinguished, much like in Def.~\ref{def:entExitLocalGlobal}. Numbers in red are the values of $h$; those in blue are the values of $g$. }
    \label{fig:prototypeOverview}
\end{figure} 
\begin{auxproof}
 A weight $\convw{w}\in\probinterval^{\exGlobal{\sd{D}}}$ is an input; it is from
 Def.~\ref{def:weightedReachProb}.
 The output is a function that assigns, to each global entrance $i$ of $\sd{D}$, a (guaranteed) under approximation of the optimal reachability probability $\MaxWReacha{\semantics{\sd{D}}}{\convw{w}}{i}$. 
\end{auxproof}
The algorithm maintains two main constructs: functions $g\colon \enLocal{\sd{D}}\rightarrow \probinterval$ and  $h\colon \exLocal{\sd{D}}\rightarrow \probinterval$ that assign values
to local entrances and exits, respectively. They are analogues of the value function $f\colon S\to \probinterval$ in (standard) VI (\cref{subsec:VIOVI});  $g$ and $h$ get iteratively increased as the algorithm proceeds. 
\begin{auxproof}
 This intuition should also explain the initialization (lines~\ref{line:CVI_init1}-\ref{line:CVI_init2}), where we assign $0$ to almost everywhere except for target states.
 We do not care about other states since they are addressed in $\texttt{LocalVI}$ in line~\ref{line:prototypeLocalVI}. 
\end{auxproof}

Lines~\ref{line:CVI_beginwhile}--\ref{line:prototypePropag} are the main VI loop, where we combine local VI (over each component $\mdp{A}$) and propagation along sequential composition. The algorithm $\texttt{LocalVI}$ takes the target oMDP $\mdp{A}$ and its ``local weight'' as arguments; the latter is  the restriction $h|_{\ex^{\mdp{A}}}\colon \ex^{\mdp{A}}\rightarrow\probinterval $ of the function $h\colon\exLocal{\sd{D}}\rightarrow \probinterval$.
Any VI algorithm will do for $\texttt{LocalVI}$; we use OVI as announced in \cref{subsec:VIOVI}. 
\begin{auxproof}
  in this work since the guaranteed over-approximation given by OVI is useful as both local VI (in line~\ref{line:prototypeLocalVI}) and global stopping criteria (in line~\ref{line:CVI_beginwhile}). Its local use is already described in~\cref{subsec:VIOVI}; its global use is discussed in \cref{subsec:over_approximation}. 
\end{auxproof}
The result of local VI is a function $g_{\mdp{A}}\colon \en^{\mdp{A}}\to \probinterval$ for values over entrances of $\mdp{A}$. These get patched up to form $g\colon \enLocal{\sd{D}}\rightarrow \probinterval$ in line~\ref{line:CVI_patching}. The function $\coprod_{\mdp{A}\in \CP{\sd{D}}}g_{\mdp{A}}$  is defined by obvious case-distinction: it returns $g_{\mdp{A}}(i)$ for a local entrance $i\in \en^{\mdp{A}}$. Recall from Def.~\ref{def:entExitLocalGlobal} that $\enLocal{\sd{D}}=\biguplus_{\mdp{A}\in \CP{\sd{D}}}I^{\mdp{A}}$.
In line~\ref{line:prototypePropag}, the values at entrances are propagated to the connected exits.

On $\texttt{PropagateSeqComp}$ in line~\ref{line:prototypePropag}, 
its graphical intuition is in \cref{subfig:propagation}; here are some details.
We first note that the set $\exLocal{\mathbb{D}}$ of local exits is partitioned into 1) global exits (i.e.\ those in $\exGlobal{\mathbb{D}}$) and 2) those local exits that get removed by sequential composition. Indeed, by examining Defs~\ref{def:seqOMDP} and \ref{def:sumOMDP}, we see that sequential composition $\seqcomp$ is the only operation that removes local exits, and the local exits that are not removed eventually become global exits.
It is also obvious (Def.~\ref{def:seqOMDP}) that each local exit $o$ removed in sequential composition has a corresponding local entrance $i_o$. Using these, we define the function $h\defeq\texttt{PropagateSeqComp}(g,\convw{w})$, of the type  $\exLocal{\mathbb{D}}\rightarrow \probinterval$, as follows: $h(o)=w_{o}$ if $o$ is a global exit (much like line~2); $h(o)=g(i_o)$ otherwise.

\begin{mytheorem}
\label{thm:cvi_correct}
\cref{alg:CVIprototype} satisfies the following properties: 
\begin{enumerate}
 \item (Guaranteed under-approximation) For the output $f$ of \cref{alg:CVIprototype}, we have $f(i)\leq \MaxWReacha{\semantics{\sd{D}}}{\convw{w}}{i}$ for each $i\in \enGlobal{\sd{D}}$.
 \item\label{item:prototypeConvergence} (Convergence) Assume  that $\texttt{GlobalStoppingCriterion}$ is $\textbf{false}$. \cref{alg:CVIprototype} converges to the optimal value, that is, $f$ converges to $ \MaxWReacha{\semantics{\sd{D}}}{\convw{w}}{\enGlobal{\sd{D}}}$. \qed
\end{enumerate}
\end{mytheorem}
The correctness of the under-approximation
of \cref{alg:CVIprototype} follows easily from those of (non-compositional, asynchronous) VI.
The convergence depends on the fact that line~\ref{line:prototypeLocalVI} of~\cref{alg:CVIprototype} iterates over \emph{all} components.

\section{Pareto Caching in Compositional VI}
\label{sec:generalizingCVI}
In our formulation of \cref{alg:CVIprototype}, there is no explicit notion of Pareto curves. However, in line~\ref{line:prototypeLocalVI}, we do (implicitly) compute under-approximations on points on the Pareto curves.  Here we recap approximate Pareto curves. We then show how we conduct \emph{Pareto caching}, the key idea sketched in \cref{subsec:overviewParetoCaching}.

\subsection{Approximating Pareto Curves}
\label{subsec:MOOandCompositionalMC}
We formalize the Pareto curves illustrated in \cref{sec:overview}. For details, see~\cite{PapadimitriouY00,ForejtKP12,Quatmann23,EtessamiKVY08}.  Model checking oMDPs is a multi-objective problem, that determines different trade-offs between reachability probabilities for the individual exits. 
  
\begin{mydefinition}[Pareto curve for an oMDP~\cite{WatanabeVHRJ24}]
Let $\mdp{A}$ be an oMDP, and $i$ be a (chosen) entrance. 
Let $\point,\point' \in \probinterval^{\ex^{\mdp{A}}}$.  The relation $\preceq$ between them is defined by
$\point\preceq \point'$ if $\point(o)\le\point'(o)$ for each $o\in \ex^{\mdp{A}}$. When $\point\prec \point'$ (i.e.\ $\point\preceq \point'$ and $\point\neq \point'$), we say $\point'$ \emph{dominates} $\point$. Let $\sigma$ be a scheduler for $\mdp{A}$. We define the point \emph{realized by} $\sigma$, denoted by $\point^{\sigma}_{i}$, by $\point^{\sigma}_{i}(o)\defeq \Reacha{\mdp{A}, \sigma}{i}{o}$,  the reachability probability from $i$ to $o$ under $\sigma$.

The set $\achievablesched{\mdp{A}}{i}{\sigma}$ of points \emph{achievable} by $\sigma$ is  $\achievablesched{\mdp{A}}{i}{\sigma}\defeq\{\point\mid\point\preceq\point^{\sigma}_{i}\}$.  The set $\achievablesched{\mdp{A}}{i}{}$ of \emph{achievable points} is given by $\achievablesched{\mdp{A}}{i}{}\defeq \bigcup_{\sigma\in \Possched{\mdp{A}}}  \achievablesched{\mdp{A}}{i}{\sigma}$. 
The \emph{Pareto curve}
$\mathsf{Pareto}_{i}\subseteq \probinterval^{\ex^{\mdp{A}}}$ 
 is the set of maximal elements in $\achievable{\mdp{A}}{i}$ wrt.\ $\preceq$.
We say a scheduler $\sigma$ is \emph{Pareto-optimal}  if 
$
 \point^{\sigma}_{i}
\in \mathsf{Pareto}_{i}
$.

\end{mydefinition}
The set  $\achievablesched{\mdp{A}}{i}{}\subseteq \probinterval^{\ex^{\mdp{A}}}$ is convex, downward closed, and finitely generated by DM schedulers; 
it follows that,
for our target problem,  Pareto-optimal DM schedulers suffice. This is illustrated in \cref{fig:paretoCachingFig}, where a weight $\convw{w}$ is the normal vector of blue lines, and the maximum is achieved by a generating point for $\achievable{\mdp{A}}{i}$.

The last observations are formally stated as follows.
\begin{myproposition}[\!\!\cite{EtessamiKVY08,ForejtKP12,Quatmann23}]
    \label{prop:finGen}
    For any entrance $i\in \en$, the set $\achievable{\mdp{A}}{i}$ of achievable points is finitely generated by DM schedulers, that is, 
    \begin{math}
        \achievable{\mdp{A}}{i} = \dwconvcl{\achievablesched{\mdp{A}}{i}{\Dmsched{\mdp{A}}}}
    \end{math}.
    Here, $\dwconvcl{X}$ denotes the downward
    and convex closed set  generated by $X\subseteq \real^n$, and $\achievablesched{\mdp{A}}{i}{\Dmsched{\mdp{A}}}$ is given by $\achievablesched{\mdp{A}}{i}{\Dmsched{\mdp{A}}}\defeq \bigcup_{\sigma\in \Dmsched{\mdp{A}}}  \achievablesched{\mdp{A}}{i}{\sigma}$, where $\Dmsched{\mdp{A}}$ is the set of DM schedulers. 
    \end{myproposition}
    
    \begin{myproposition}[\!\!\cite{EtessamiKVY08,ForejtKP12,Quatmann23}]
    \label{prop:paretosuffice}
    Given a weight $\convw{w}\in \probinterval^{\ex^{\mdp{A}}}$ and an entrance $i$, there is a scheduler $\sigma$ such that $ \WReacha{\semantics{\mathbb{D}},\sigma}{\convw{w}}{i} = \MaxWReacha{\semantics{\mathbb{D}}}{\convw{w}}{i}$. 
    Moreover, this $\sigma$ can be chosen to be DM and Pareto-optimal. 
    \end{myproposition} 
\begin{auxproof}
 \begin{figure}[h]
    \centering
    \begin{minipage}[t]{0.35\hsize}
    \scalebox{0.8}{
    \begin{tikzpicture}[
 innode/.style={draw, rectangle, minimum size=0.5cm},
 interface/.style={draw, rectangle, minimum size=0.5cm},]
        \node[interface,fill=white,yshift=1cm] (s0) {\scalebox{0.7}{$\enrarg{1}$}};
        \node[inner sep=0,right=-1cm of s0] (enr1) {};
        \node[inner sep=2pt, fill=black,right=0.5cm of s0, yshift=0.5cm] (as0) {};
        \node[inner sep=2pt, fill=black,right=0.5cm of s0, yshift=-0.5cm] (bs0) {};
        \node[state,right=0.5cm of as0, minimum size=0.5cm,fill=white] (s1) {\scalebox{0.7}{$s_1$}};
        \node[inner sep=2pt, fill=black,right=0.5cm of s1, yshift=0.6cm] (as1) {};
        \node[inner sep=2pt, fill=black,right=0.5cm of s1, yshift=-0.6cm] (bs1) {};
        \node[interface,right=1.5cm of s1,fill=white] (s3) {\scalebox{0.7}{$\exrarg{1}$}};
        \node[interface,right=-0.65cm of s3,fill=white, yshift=-1cm] (s4) {\scalebox{0.7}{$\exrarg{2}$}};
        \node[inner sep=0,right=0.4cm of s3] (exr1) {};
        \node[inner sep=0,right=0.4cm of s4] (exr2) {};
        \draw[->] (enr1) -> (s0);
        \draw[->] (s0) -> node [above] {$\scalebox{0.7}{a}$} (as0);
        \draw[->] (s0) -> node [above] {$\scalebox{0.7}{b}$} (bs0);
        \draw[->] (as0) -> node [above] {$\scalebox{0.7}{1}$} (s1);
        \draw[->] (s1) -> node [above] {$\scalebox{0.7}{a}$} (as1);
        \draw[->] (s1) -> node [above] {$\scalebox{0.7}{b}$} (bs1);
        \draw[->] (bs0) -> node [left, xshift=-0.5cm] {$\scalebox{0.7}{0.4}$} (s3);
        \draw[->] (bs0) -> node [below] {$\scalebox{0.7}{0.2}$} (s4);
        \draw[->] (as1) -> node [above] {$\scalebox{0.7}{0.2}$} (s3); 
        % \draw[->] (as1) -> node [left] {$\scalebox{0.7}{.4}$} (s2); 
        \draw[->] (as1) -> node [left, yshift=0.2cm, xshift=-0.2cm] {$\scalebox{0.7}{0.4}$} (s4); 
        \draw[->] (bs1) -> node [below] {$\scalebox{0.7}{0.35}$} (s3);
        % \draw[->] (bs1) -> node [right] {$\scalebox{0.7}{.6}$} (s2);
        \draw[->] (bs1) -> node [below,xshift=-0.1cm, yshift=0.05cm] {$\scalebox{0.7}{0.35}$} (s4);
        \draw[->] (s3) -> (exr1);
        \draw[->] (s4) -> (exr2);
    \end{tikzpicture}
    }
    \centering
    \subcaption{oMDP $\mdp{A}$, $\typemdp{\mdp{A}} = (1, 0) \rightarrow (2, 0)$}\label{subfig:ExOpenMDP}
    \end{minipage}
    \begin{minipage}[t]{0.3\hsize}
    \centering
    \pgfplotsset{width=4.5cm,compat=1.9}
 \scalebox{0.9}{
 \begin{tikzpicture}
        \begin{axis}[
            axis lines = left,
            xmin=0, xmax=0.5,
            ymin=0, ymax=0.5,
            % xlabel = \(\exrarg{1}\),
            % ylabel = \(\exrarg{2}\),
            xtick={0, 0.2, 0.4, 0.6},
            ytick={0, 0.2, 0.4, 0.6},
            axis background/.style={fill=white},
        ]
        \addplot[fill=green!50, very thin] coordinates {(0.4,0) (0.4,0.2) (0.35, 0.35) (0.2,0.4) (0,0.4) (0, 0)} -- cycle;
        \addplot[
            color=blue,
            ultra thick
        ]
        coordinates {
            (0.2,0.4) (0.35,0.35)
        };
        \addplot[
            color=blue,
            ultra thick
        ]
        coordinates {
            (0.35,0.35) (0.4, 0.2)
        };
        \node[point,label=0:$\point_1$] at (axis cs:0.4,0.2) {};
	\node[point,label=90:$\point_2$] at (axis cs:0.2,0.4) {};
 \node[point,label=230:$\point_3$] at (axis cs:0.35,0.35) {};	
        \node[] at (axis cs:0.15,0.2) {$\achievable{\mdp{A}}{\enrarg{1}}$};
        \end{axis}
        \end{tikzpicture}
    }
    \subcaption{$\achievable{\mdp{A}}{\enrarg{1}}$.}\label{subfig:ExPareto}
    \end{minipage}
    \begin{minipage}[t]{0.3\hsize}
    \centering
    \pgfplotsset{width=4.5cm,compat=1.9}
    \scalebox{0.9}{
        \begin{tikzpicture}
        \begin{axis}[
            axis lines = left,
            xmin=0, xmax=0.5,
            ymin=0, ymax=0.5,
            % xlabel = \(\exrarg{1}\),
            % ylabel = \(\exrarg{2}\),
            xtick={0, 0.2, 0.4, 0.5},
            ytick={0, 0.2, 0.4, 0.5},
            axis background/.style={fill=white},
        ]
        \addplot[fill=green!50, very thin] coordinates {(0.38,0) (0.38,0.15) (0.15,0.38) (0,0.38) (0, 0)} -- cycle;
        \addplot[fill=red!50, very thin] coordinates {(0.5, 0.5) (0.5, 0) (0.42,0) (0.42,0.42) (0,0.42) (0, 0.5)};

        \node[point,label=180:$\point^l_1$] at (axis cs:0.38,0.15) {};
	\node[point,label=200:$\point^l_2$] at (axis cs:0.15,0.38) {};
 \node[point,label=240:$\point^u_1$] at (axis cs:0.42,0.42) {};
        \node[] at (axis cs:0.15,0.2) {$L_i$};
	\node[] at (axis cs:0.42,0.45) {$\real^{2}\backslash U_i$};		
        \end{axis}
        \end{tikzpicture}
        }
    \subcaption{Sound approx.\ $(L_i,U_i)$ }\label{subfig:sound_approx}
    \end{minipage}
    \caption{consider the oMDP $\mdp{A}$ shown in~\cref{subfig:ExOpenMDP}. The blue curve is the Pareto curve.  Its under approximation is colored by green, and  its over approximation is colored by white and green. }
    \label{fig:exParetoOpenMDPs}
 \end{figure} 
\end{auxproof}
We now formulate \emph{sound approximations} of Pareto curves, which is a foundation of our Pareto caching (and a global stopping criterion in~\cref{sec:pareto_curves}). 

\begin{mydefinition}[sound approximation~\cite{WatanabeVHRJ24}]
\label{def:sound_approximation}
    Let
 $i$ be an entrance. An \emph{under-approximation $L_i$} of the Pareto curve $\mathsf{Pareto}_{i}$ is a downward closed subset $L_i\subseteq \achievable{\mdp{A}}{i}$; an \emph{over-approximation} is a downward closed superset $U_i\supseteq \achievable{\mdp{A}}{i}$.     A pair $(L_i,U_i)$ is called a \emph{sound approximation} of the Pareto curve $\mathsf{Pareto}_{i}$. In this paper, we focus on $L_i$ and $U_i$ that are \emph{finitely generated}, i.e.\  the convex and downward closures of some finite generators $L_i^{\mathrm{g}}, U_i^{\mathrm{g}}\subseteq \probinterval^{\ex^{\mdp{A}}}$, respectively. A \emph{sound approximation} of an oMDP $\mathcal{A}$ is a pair $(L,U)$, where $L=(L_{i})_{i\in\en^{\mdp{A}}}$, $U=(U_{i})_{i\in\en^{\mdp{A}}}$, and $(L_i, U_i)$ is a sound approximation for each entrance $i$. 
\end{mydefinition}

\begin{auxproof}
\begin{myremark}
The approximate bottom-up model checking algorithm~\cite{WatanabeVHRJ24}, outlined in \cref{subsec:TACAS24}, first computes a sound approximation for each component oMDP $\mdp{A}$ (one can rely on existing algorithms~\cite{ForejtKP12,QuatmannK21,Quatmann23}), and then combines them along $\seqcomp$ and $\oplus$. An example shows it is hard to compose local (component-level) error bounds to obtain a global error bound~\cite{WatanabeVHRJ24}. Therefore the algorithm backtracks and refines sound approximations for components when a global error is excessive. 
\end{myremark}
\end{auxproof}

\subsection{Pareto Caching}
\label{sec:approximation}

We go on to formalize our second key idea, \emph{Pareto caching}, outlined in \cref{subsec:overviewParetoCaching}.

In Def.~\ref{def:ParetoCache},  an index $\constMDP{\mdp{A}}\in \leaf{\sd{D}}$ is a \emph{nominal} component that ignores multiplicities, since we want to 
  reuse results for different occurrences of $\mdp{A}$.

\begin{mydefinition}[Pareto cache]\label{def:ParetoCache}
Let $\sd{D}$ be a string diagram of MDPs. A \emph{Pareto cache} $\paretoAssign$ is an indexed family $\paretoAssign\defeq \big((L^{\mdp{A}}, U^{\mdp{A}})\big)_{\constMDP{\mdp{A}}\in \leaf{\sd{D}}}$, where  $ (L^{\mdp{A}}, U^{\mdp{A}})$ is a   sound approximation for each nominal component $\constMDP{\mdp{A}}$, defined in Def.~\ref{def:sound_approximation}. 
\end{mydefinition}
As announced in~\cref{subsec:overviewParetoCaching}, a Pareto cache $\paretoAssign$---its component $(L^{\mdp{A}}, U^{\mdp{A}})$, to be precise---gets \emph{queried} on a weight $\convw{w}\in\probinterval^{\ex^{\mdp{A}}}$. It is not trivial what to return, however, since the specific weight $\convw{w}$ may not have been used before to construct $\paretoAssign$. The query is answered in the way depicted in \cref{fig:paretoCachingFig},  finding an extremal point where $L^{\mdp{A}}$ intersects with a plane with its normal vector $\convw{w}$.

\begin{mydefinition}[cache read]
\label{def:interpolation}
Assume the above setting, and let $i$ be an entrance of interest. The \emph{cache read} $(L^{\mdp{A}}_{i}(\convw{w}), U^{\mdp{A}}_{i}(\convw{w}))\in\probinterval^{2}$ on $\convw{w}$ at $i$ is defined by $ L^{\mdp{A}}_i(\convw{w})\defeq \sup_{\point\in L_i}\convw{w}\cdot \point$ and 
$U^{\mdp{A}}_i(\convw{w})\defeq \sup_{\point\in U_i}\convw{w}\cdot \point$.
\end{mydefinition}
Recall from \cref{subsec:MOOandCompositionalMC} that we can assume $L_{i}$ and $U_{i}$ are finitely generated as convex and downward closures.
It follows~\cite{ForejtKP12,Quatmann23} that each supremum above is realized by some generating point, 
much like in Prop.~\ref{prop:paretosuffice}, easing computation.

\begin{algorithm}[t]
\caption{Updating $g_{\mdp{A}}$ with a Pareto cache $\paretoAssign$ and a bound $\eta\in \probinterval$
}
\label{alg:UpdateWithParetoCache}
\begin{algorithmic}[1]

\setcounter{ALG@line}{2}
\State\label{line:CVIPC_init3}initialize a Pareto cache $\paretoAssign$ by $((\emptyset,\probinterval^{\ex^{\mdp{A}}}))_{\constMDP{\mdp{A}}}$
\algrule
\setcounter{ALG@line}{5}
\State\hspace{\algorithmicindent}\hspace{\algorithmicindent} \textbf{if} $\max_{i\in \en^{\mdp{A}}} \big(U^{\mdp{A}}_i(h|_{\ex^{\mdp{A}}}) - L^{\mdp{A}}_i(h|_{\ex^{\mdp{A}}})\big)\leq \eta$ \textbf{then}
\label{line:PC_checkError}
\Statex\Comment{
 computing the error of cache $\paretoAssign$ for  weight $h|_{\ex^{\mdp{A}}}$ (cf.\ Def.~\ref{def:interpolation})
}
\State\hspace{\algorithmicindent}\hspace{\algorithmicindent} \hspace{\algorithmicindent} $g_{\mdp{A}}\gets L^{\mdp{A}}(h|_{\ex^{\mdp{A}}})$ 
 \label{line:ParetoInterpolation}\Comment{``cache hit''; use the cache read $L^{\mdp{A}}(h|_{\ex^{\mdp{A}}})$}
\State\hspace{\algorithmicindent}\hspace{\algorithmicindent} \textbf{else}
\State\hspace{\algorithmicindent}\hspace{\algorithmicindent} \hspace{\algorithmicindent}\label{line:ParetoCacheLocalVI}$(l, \sigma)\gets\mathtt{LocalOVI}(\mdp{A}, h|_{\ex^{\mdp{A}}}, \eta)$

  \Statex       \Comment{``cache miss''; run local VI as in \cref{alg:CVIprototype}
}
  \State\hspace{\algorithmicindent}\hspace{\algorithmicindent} \hspace{\algorithmicindent} $g_{\mdp{A}}\gets l$ and $\paretoAssign\gets\mathrm{Update}(\mdp{A}, \paretoAssign, l, \sigma, h|_{\ex^{\mdp{A}}}, \eta)$ 
  \label{line:PC_update}
  \Comment{see the end of~\cref{sec:approximation}} 

\end{algorithmic}
\end{algorithm}

We complement~\cref{alg:CVIprototype} by~\cref{alg:UpdateWithParetoCache} that introduces our Pareto caching. 
Specifically, for the weight $h|_{\ex^{\mdp{A}}}$ in question, we first compute the  \emph{error} $\max_{i\in \en^{\mdp{A}}}U^{\mdp{A}}_i(h|_{\ex^{\mdp{A}}}) - L^{\mdp{A}}_i(h|_{\ex^{\mdp{A}}})$ of the Pareto cache $\paretoAssign= \big((L^{\mdp{A}}, U^{\mdp{A}})\big)_{\constMDP{\mdp{A}}}$ with respect to this weight. The error can be greater than a prescribed bound $\eta$---we call this \emph{cache miss}---in which case we run OVI locally for $\mdp{A}$ (line~\ref{line:ParetoCacheLocalVI}). When the error is no greater than $\eta$---we call this \emph{cache hit}---we use the cache read (Def.~\ref{def:interpolation}), sparing OVI on a component $\mdp{A}\in \CP{\sd{D}}$. In the case of a cache miss, the result $(l,\sigma)$ of local OVI (line~\ref{line:ParetoCacheLocalVI}) is used also to update the Pareto cache $\paretoAssign$ (line~\ref{line:PC_update}); see below. 

Using a Pareto cache may prevent the execution of local VI on every component, which can be critical for the convergence of~\cref{alg:CVIprototype}; see Thm.~\ref{thm:cvi_correct}. A simple solution is to disregard Pareto caches eventually.  

\myparagraph{Updating the cache}
Pareto caches get incrementally updated using the results for weighted reachabilities with different weights $\convw{w}$. We build upon data structures in~\cite{ForejtKP12,Quatmann23}. Notable is the asymmetry between under- and over-approximations $(L_{i},U_{i})$: we obtain 1)~a \emph{point in}   $L_{i}$ and 2)~a  \emph{plane that bounds}~$U_{i}$. 

We update the cache after running OVI 
on a weight $\convw{w}\in\probinterval^{\ex^{\mdp{A}}}$, which  approximately computes the optimal weighted  reachability to exits $o\in\ex^{\mdp{A}}$. That is, it returns $l,u\in\probinterval$ such that 
\begin{equation}\label{eq:VIAndMOO}
l\;\le\;
\textstyle\sup_{\sigma}
 \bigl(\,\convw{w}\cdot \bigl(\,
\mathrm{RPr}^{\sigma}(i\to o)
\,\bigr)_{o\in \ex^{\mdp{A}}}\,\bigr)
\;\le\;
u.
\end{equation}
Here   $i$ is any entrance and $\mathrm{RPr}^{\sigma}(i\to o)$ is the  probability $\mathrm{RPr}^{\mdp{A},\sigma,\{o\}}(i)$ in \cref{subsec:prelimMDPs}.

What are the ``graphical''
 roles of $l,u$ in the Pareto curve? The role of $u$ is easier: it follows from~\cref{eq:VIAndMOO}  that any achievable reachability vector
\begin{math}
 \bigl(
\mathrm{RPr}^{\sigma}(i\to o)
\bigr)_{o}
\end{math}
resides under the plane
 $\{\point\mid \convw{w}\cdot\point= u\}$. 
 This plane thus bounds an over-approximation $U_{i}$.
The use of $l$ takes some computation. By~\cref{eq:VIAndMOO}, the existence of a good scheduler $\sigma$ is guaranteed; but this alone does not carry any graphical information e.g.\ in \cref{fig:paretoCachingFig}. We have to go constructive, by extracting a near-optimal DM scheduler $\sigma_{0}$ (we can do so in VI) and using this fixed $\sigma_{0}$ to compute
\begin{math}
 \bigl(
\mathrm{RPr}^{\sigma_{0}}(i\to o)
\bigr)_{o}
\end{math}. This way we can plot an achievable point---a corner point in \cref{fig:paretoCachingFig}---in $L_i$.

\section{Global Stopping Criteria (GSC)}
\label{sec:pareto_curves}
We present the last missing piece, namely global stopping criteria (\emph{GSC} in short, in line~\ref{line:CVI_beginwhile} of~\cref{alg:CVIprototype}).
It has to ensure that the computed underapproximation $f$ is  $\epsilon$ close to the exact reachability probability.
We provide two criteria, called \emph{optimistic} and \emph{bottom-up}. 

\myparagraph{Optimistic GSC (Opt-GSC)} The challenge in adapting the idea of OVI (see~\cref{subsec:VIOVI}) to CVI is to define a suitable Bellman operator for CVI.
Once we define such a Bellman operator for CVI, we can immediately apply the idea of OVI.
For simplicity, we assume that CVI solves exactly in each local component (line~\ref{line:prototypeLocalVI} in~\cref{alg:CVIprototype}) without Pareto caching; this can be done, for example,  by policy iteration~\cite{HartmannsJQW23}.
Then, CVI (without Pareto caching and a global stopping criterion) on $\sd{D}$ is exactly the same as the (non-compositional) VI on a suitable \emph{shortcut MDP}~\cite{WatanabeVHRJ24} of $\sd{D}$.
Intuitively, a shortcut MDP summarizes a Pareto-optimal scheduler by a single action from a local entrance to exit, 
\iffull 
see Def.~\ref{def:shortcut_MDP}.
\else 
see~\cite[Appendix C]{WVJH2024accepted} for the definition.
\fi 
Thus, we can regard the standard Bellman operator on the shortcut MDP as the Bellman operator for CVI, and define Opt-GSC as the standard OVI based on this characterisation.
CVI with Opt-GSC (and Pareto caching) actually uses local under-approximations (not exact solutions) for obtaining a global under-approximation (line~\ref{line:ParetoInterpolation} in~\cref{alg:UpdateWithParetoCache} and line~\ref{line:ParetoCacheLocalVI} in~\cref{alg:UpdateWithParetoCache}), where the desired soundness property still holds. 
\iffull 
See~\cref{sec:details_pareto_curves} for more details. 
\else 
See~\cite[Appendix C]{WVJH2024accepted} for more details. 
\fi

\myparagraph{Bottom-up GSC (BU-GSC)}
We obtain another global stopping criterion by composing  Pareto caches---computed in \cref{alg:UpdateWithParetoCache} for each component $\mdp{A}$---in the bottom-up manner in~\cite{WatanabeVHRJ24} (outlined in~\cref{subsec:TACAS24}). Specifically, 1) \cref{alg:UpdateWithParetoCache} produces an over-approximation $U^{\mdp{A}}$ for the Pareto curve of each component $\mdp{A}$; 2) we combine $(U^{\mdp{A}})_{\mdp{A}}$ along $\seqcomp$ and $\oplus$ to derive an over-approximation $U$ of the global Pareto curve; and 3) this $U$ is queried on the weight $\convw{w}$ in question (i.e.\ the input of CVI), in order to obtain an over-approximation $u$ of the weighted reachability probabilities.
BU-GSC checks if this over-approximation $u$ is close enough to the under-approximation $l$ derived from $g$ in \cref{alg:CVIprototype}.

\myparagraph{Correctness} CVI (\cref{alg:CVIprototype} with Pareto caching under either GSC) is sound. 
\iffull 
The proof is in~\cref{sec:details_pareto_curves}.
\else 
The proof is in~\cite[Appendix C]{WVJH2024accepted}.
\fi

\begin{mytheorem}[$\epsilon$-soundness of CVI]
\label{thm:epsilon_soundness}
    Given a string diagram $\sd{D}$, a weight $\convw{w}$, and $\epsilon\in \probinterval$, if CVI terminates,
then the output $f$ satisfies 
\begin{align*}
    f(i)\leq\MaxWReacha{\semantics{\sd{D}}}{\convw{w}}{\enGlobal{\sd{D}}} \leq f(i)+\epsilon,
\end{align*}
for each $i\in \enGlobal{\sd{D}}$. 
\end{mytheorem}
Our algorithm currently comes with no termination guarantee; this is future work.  Termination of VI (with soundness) is a tricky problem: most known termination proofs exploit the uniqueness of a fixed point of the Bellman operator, which must be algorithmically enforced e.g.\ by eliminating end components~\cite{HaddadM18,BrazdilCCFKKPU14}. In the current compositional setting, end components can arise by composing components, so detecting them is much more challenging.

\section{Empirical evaluation}
\label{sec:imp_exp}

\newcommand{\approachfont}[1]{\textsc{#1}}
\newcommand{\approachmonolithic}{\approachfont{Mono}}
\newcommand{\approachbottomup}{\approachfont{BU}}
\newcommand{\approachocvi}{\approachfont{OCVI}}
\newcommand{\approachocviexact}{\approachocvi{}$^e$}
\newcommand{\approachocvipareto}{\approachocvi{}$^p$}
\newcommand{\approachsymbiosis}{\approachfont{Symb}}
\newcommand{\approachbaseline}{\approachfont{Baseline}}
\newcommand{\approachnovel}{\approachfont{Novel}}

In this section, we compare the scalability of our approaches both among each other and in comparison with some existing baselines. We discuss the setup, give the results, and then give our interpretation of them.

\myparagraph{Approaches}
We examine our three main algorithms. Opt-GSC with either exact caching  (\approachocviexact{}) or Pareto-caching (\approachocvipareto{}), and BU-GSC with Pareto-caching (\approachsymbiosis{}). BU-GSC needs a Pareto cache, so we cannot run BU-GSC with an exact cache.
We compare our approaches against two baselines: a \emph{monolithic} (\approachmonolithic) algorithm building the complete MDP $\semantics{\sd{D}}$ and the \emph{bottom-up} (\approachbottomup) as explained in~\cite{WatanabeVHRJ24}.
We use two \emph{virtual} approaches that use a perfect oracle to select the fastest out of the specified algorithms: \emph{baselines} is the best-of-the-baselines, while \emph{novel} is the best of the three new algorithms.  All algorithms are built on top of the probabilistic model checker Storm~\cite{HenselJKQV22}, which is primarily used for model building and (O)VI on component MDPs as well as operating on Pareto curves.

\myparagraph{Setup}
We run all benchmarks on a single core of an AMD Ryzen TRP~5965WX, with a 900s time-out and a 16GB memory limit. 
We use \emph{all} (scalable) benchmark instances from~\cite{WatanabeVHRJ24}.
While these benchmarks are synthetic, they reflect typical structures found in network protocols and high-level planning domains. 
We require an overall precision of $10^{-4}$, we run the Pareto cache with an acceptance precision of $10^{-5}$, and solve the LPs in the upper-bound queries for the Pareto cache with an exact LP solver and a tolerance of $10^{-4}$. 
The components are reverse topologically ordered, i.e., we always first analyse component MDPs towards the end of a given MDP $\semantics{\sd{D}}$.
To solve the component MDPs inside the VI, we use OVI for the lower bounds and precise policy iteration for the upper bounds.
We use algorithms and data structures already present in Storm for maintaining Pareto curves~\cite{Quatmann23}, which use exact rational arithmetic for numerical stability.
Although our implementation supports exact arithmetic throughout the code, in practice this leads to a significant performance penalty, performing up to 100 times slower.
For algorithms not related to maintaining the Pareto cache, we opted for using 64-bit floating point arithmetic, which is standard in probabilistic model checking~\cite{BuddeHKKPQTZ20}.
Using floating point arithmetic can produce unsound results~\cite{Hartmanns22}; we attempt to prevent unsound results in our benchmark.
First, we check with our setup that our results are very close (error $<10^{-5}$) to the exact solutions (when they could be computed).
Second, we check that all results, obtained with different methods, are close. 
We evaluate the stopping criteria after ten iterations.
These choices can be adapted using our prototypical implementation, we discuss some of these choices at the end of the discussion below.

\begin{figure}[t]
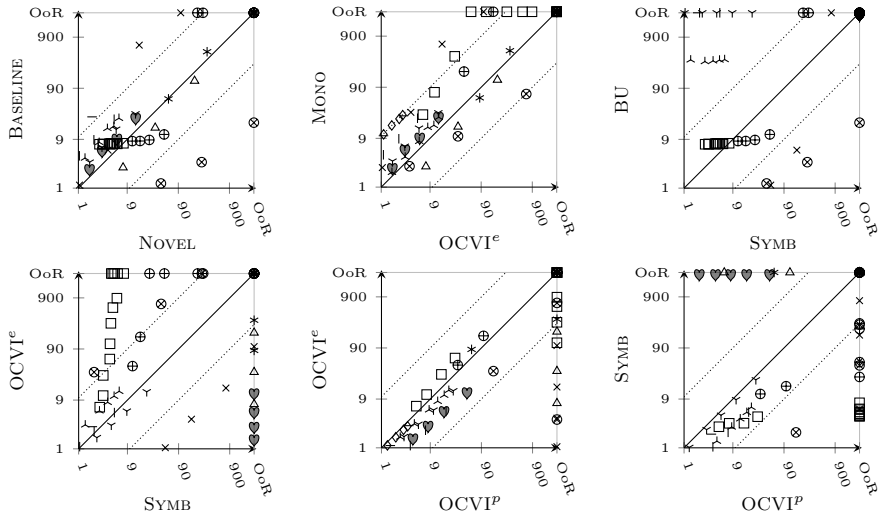

    \centering
    \begin{minipage}[t]{0.33\textwidth}%
    \centering
    \scatterplot{plots/results_plot}{Total time Novel}{\approachnovel}{Total time Baseline}{\approachbaseline}
    %\subcaption{Novel vs Baseline}%
    \end{minipage}%
    \begin{minipage}[t]{0.33\textwidth}%
    \centering
    \scatterplot{plots/results_plot}{Total time ALGORITHM_A1}{\approachocviexact}{Total time Monolithic}{\approachmonolithic}
    %\subcaption{OCVI vs monolithic}%
    \end{minipage}%
    \begin{minipage}[t]{0.33\textwidth}%
    \centering
    \scatterplot{plots/results_plot}{Total time ALGORITHM_B}{\approachsymbiosis}{Total time Pareto4}{\approachbottomup}
    %\subcaption{Symbiosis vs Bottom-up}%
    \end{minipage}%
    
    \begin{minipage}[t]{0.33\textwidth}%
    \centering
    \scatterplot{plots/results_plot}{Total time ALGORITHM_B}{\approachsymbiosis}{Total time ALGORITHM_A1}{\approachocviexact}
    %\subcaption{Symbiosis vs Bottom-up}%
    \end{minipage}%
    \begin{minipage}[t]{0.33\textwidth}%
    \centering
    \scatterplot{plots/results_plot}{Total time ALGORITHM_A2}{\approachocvipareto}{Total time ALGORITHM_A1}{\approachocviexact}
    %\subcaption{Symbiosis vs Bottom-up}%
    \end{minipage}%
   \begin{minipage}[t]{0.33\textwidth}%
    \centering
    \scatterplot{plots/results_plot}{Total time ALGORITHM_A2}{\approachocvipareto}{Total time ALGORITHM_B}{\approachsymbiosis}
    %\subcaption{Symbiosis vs Bottom-up}%
    \end{minipage}%
    \caption{Benchmark scatter plots, time in seconds, OoR=Out of Resources}
    \label{fig:scatterplots}
\end{figure}

\medskip\noindent\textbf{Results}
We provide pairwise comparisons of the runtimes on all benchmarks using the scatter plots in Fig.~\ref{fig:scatterplots}\footnote{A point $(x,y)$ means that the approach on the x-axis took $x$ seconds and the tool on the $y$-axis took $y$ seconds. Different shapes refer to different benchmark sets.}.
Notice the {log-log} scale.
For some of the benchmark instances, we provide detailed information in Tables~\ref{tab:runtimesetc} and \ref{tab:caches}, respectively.
In Table~\ref{tab:runtimesetc}, we give the identifier for the string diagram and the component MDPs, as well as the number of states in $\semantics{\sd{D}}$.
Then, for each of the five algorithms, we provide the timings in $t$, for each algorithm maintaining Pareto points, we give the number of Pareto points stored $|P|$, and for the three novel VI-based algorithms, we give the amount of time spent in an attempt to prove convergence ($t_s$).
In Table~\ref{tab:caches}, we focus on our three novel algorithms and the performance of the caches.
We again provide identifiers for the models, and then for each algorithm, the total time spent by the algorithm, the time spent on inserting and retrieving items from the cache, as well as the fraction of cache hits $H$ and the number of total queries $Q$.
Thus, the number of cache hits is given by $H\cdot Q$.
The full tables and more figures are given in \iffull Appendix~\ref{app:benchmarkresults}.
\else ~\cite[Appendix A]{WVJH2024accepted}.
\fi

\newcommand{\btime}{$t$}
\newcommand{\btimestop}{$t_s$}
\newcommand{\bpareto}{$|P|$}
\newcommand{\bhit}{$H$}
\newcommand{\bquery}{$Q$}
\newcommand{\bcache}{$t_c$}
\newcommand{\bcacheinsert}{$t_i$}
\newcommand{\bcacheretrieve}{$t_r$}
\newcommand{\cacheexact}{\textsc{Exact}}
\newcommand{\cachepareto}{\textsc{Pareto}}
\newcommand{\benchmark}[1]{\texttt{#1}}

\begin{table}[t]
    \centering
     \caption{Performance for different algorithms. See \textbf{Results} for explanations.}
{    \scriptsize
    \begin{NiceTabular}{rrrr  r  rr  rr rrr  rrr}
\toprule
& & & &
\multicolumn{3}{c}{\approachbaseline} & 
\multicolumn{8}{c}{\approachnovel}\\
\cmidrule(lr){5-7}\cmidrule(lr){8-15}

& & & &
\multicolumn{1}{c}{\approachmonolithic{}} &
\multicolumn{2}{c}{\approachbottomup{}} &
\multicolumn{2}{c}{\approachocviexact{}} &
\multicolumn{3}{c}{\approachocvipareto{}} &
\multicolumn{3}{c}{\approachsymbiosis{}} \\
\cmidrule(lr){5-5}\cmidrule(lr){6-7}\cmidrule(lr){8-9}\cmidrule(lr){10-12}\cmidrule(lr){13-15}

\multicolumn{1}{c}{$\sd{D}$} &
\multicolumn{1}{c}{$\mdp{M}$} &
\multicolumn{1}{c}{$|S|$} &
\multicolumn{1}{c}{$|L|$} &
\multicolumn{1}{c}{\btime} &
\multicolumn{1}{c}{\btime} &
\multicolumn{1}{c}{\bpareto} &
\multicolumn{1}{c}{\btime} &
\multicolumn{1}{c}{\btimestop} &
\multicolumn{1}{c}{\btime} &
\multicolumn{1}{c}{\btimestop} &
\multicolumn{1}{c}{\bpareto} &
\multicolumn{1}{c}{\btime} &
\multicolumn{1}{c}{\btimestop} &
\multicolumn{1}{c}{\bpareto}\\
\midrule

\begin{filecontents*}{plots/table1.csv}
sd,leaf,statecount,uniqueleaves,monolithictotaltime,pareto4totaltime,pareto4lowerparetopoints,algorithma1totaltime,algorithma1terminationtime,algorithma2totaltime,algorithma2terminationtime,algorithma2lowerparetopoints,algorithmbtotaltime,algorithmbterminationtime,algorithmblowerparetopoints
\texttt{Birooms10},\texttt{RmB},1.1e+06,16,56,TO,TO,84,25,58,25,1155,TO,TO,TO
\texttt{Birooms100},\texttt{RmS},8.5e+05,16,15,TO,TO,32,8,TO,TO,TO,TO,TO,TO
\texttt{Birooms200},\texttt{RmS},3.4e+06,16,126,TO,TO,187,58,TO,TO,TO,TO,TO,TO
\texttt{Chains100},\texttt{RmB},1.1e+06,6,TO,7,46,57,26,27,24,40,4,0,44
\texttt{Chains3500},\texttt{RmB},3.7e+07,6,OOM,8,46,TO,TO,TO,TO,TO,8,1,40
\texttt{ChainsLoop500},\texttt{Dice4},4.9e+06,4,28,TO,TO,13,13,25,12,54,21,16,50
\texttt{ChainsLoop500},\texttt{Dice5},4.9e+06,4,25,TO,TO,13,13,47,20,66,TO,TO,TO
\texttt{Chains500},\texttt{RmS},4.2e+04,6,1,0,67,0,0,TO,TO,TO,0,0,38
\texttt{Rooms10},\texttt{RmB},1.1e+06,14,183,8,84,42,21,31,20,108,11,0,108
\texttt{Rooms250},\texttt{RmB},6.6e+08,14,OOM,OOM,OOM,TO,TO,TO,TO,TO,273,226,134
\texttt{Rooms500},\texttt{RmS},2.1e+07,13,TO,OOM,OOM,100,57,TO,TO,TO,TO,TO,TO
\end{filecontents*}
\csvreader[late after line = \\, head to column names]{plots/table1.csv}{}{\csvlinetotablerow}
\bottomrule
    \end{NiceTabular}
    }
   \label{tab:runtimesetc}
\end{table}
\begin{table}[t]
    \centering
     \caption{Cache access times for CVI algorithms. See \textbf{Results} for explanations. }
{\scriptsize
    \begin{NiceTabular}{rr rrrrr rrrrrr rrrr}
\toprule

& &
\multicolumn{5}{c}{\approachocviexact{}} &
\multicolumn{5}{c}{\approachocvipareto{}} &
\multicolumn{5}{c}{\approachsymbiosis{}} \\
\cmidrule(lr){3-7}\cmidrule(lr){8-12}\cmidrule(lr){13-17}

\multicolumn{1}{c}{$\mathbb{D}$} &
\multicolumn{1}{c}{$\mdp{M}$} &
\multicolumn{1}{c}{\btime} &
\multicolumn{1}{c}{\bcacheinsert} &
\multicolumn{1}{c}{\bcacheretrieve} &
\multicolumn{1}{c}{\bhit} &
\multicolumn{1}{c}{\bquery} &

\multicolumn{1}{c}{\btime} &
\multicolumn{1}{c}{\bcacheinsert} &
\multicolumn{1}{c}{\bcacheretrieve} &
\multicolumn{1}{c}{\bhit} &
\multicolumn{1}{c}{\bquery} &

\multicolumn{1}{c}{\btime} &
\multicolumn{1}{c}{\bcacheinsert} &
\multicolumn{1}{c}{\bcacheretrieve} &
\multicolumn{1}{c}{\bhit} &
\multicolumn{1}{c}{\bquery}
\\
\midrule
\begin{filecontents*}{plots/table2.csv}
sd,leaf,algorithma1totaltime,algorithma1cacheinsertiontime,algorithma1cacheretrievaltime,algorithma1cachehitratio,algorithma1weightedreachabilityqueries,algorithma2totaltime,algorithma2cacheinsertiontime,algorithma2cacheretrievaltime,algorithma2cachehitratio,algorithma2weightedreachabilityqueries,algorithmbtotaltime,algorithmbcacheinsertiontime,algorithmbcacheretrievaltime,algorithmbcachehitratio,algorithmbweightedreachabilityqueries
\texttt{Birooms10},\texttt{RmB},84,0,0,0.03,206,58,14,4,0.66,1500,TO,TO,TO,TO,TO
\texttt{Birooms100},\texttt{RmS},32,0,0,0.24,65545,TO,TO,TO,TO,TO,TO,TO,TO,TO,TO
\texttt{Birooms200},\texttt{RmS},187,1,3,0.25,477157,TO,TO,TO,TO,TO,TO,TO,TO,TO,TO
\texttt{Chains100},\texttt{RmB},57,0,0,0.50,204,27,0,0,0.62,306,4,0,0,0.84,102
\texttt{Chains3500},\texttt{RmB},TO,TO,TO,TO,TO,TO,TO,TO,TO,TO,8,0,1,1.00,3502
\texttt{ChainsLoop500},\texttt{Dice4},13,0,0,0.84,15508,25,0,13,0.84,15592,21,0,5,1.00,5532
\texttt{ChainsLoop500},\texttt{Dice5},13,0,0,0.84,15511,47,0,26,0.84,15592,TO,TO,TO,TO,TO
\texttt{Chains500},\texttt{RmS},0,0,0,0.50,1004,TO,TO,TO,TO,TO,0,0,0,0.97,502
\texttt{Rooms10},\texttt{RmB},42,0,0,0.50,200,31,1,0,0.50,300,11,1,0,0.50,100
\texttt{Rooms250},\texttt{RmB},TO,TO,TO,TO,TO,TO,TO,TO,TO,TO,273,2,30,1.00,62499
\texttt{Rooms500},\texttt{RmS},100,0,0,0.50,500000,TO,TO,TO,TO,TO,TO,TO,TO,TO,TO
\end{filecontents*}
\csvreader[late after line = \\, head to column names]{plots/table2.csv}{}{\csvlinetotablerow}
  
\bottomrule
    \end{NiceTabular}
}
    \label{tab:caches}
\end{table}

\myparagraph{Discussion}
We make some observations. \emph{We notice that the CVI algorithms collectively solve more benchmarks within the time out and speed up most benchmarks, see Fig.~\ref{fig:scatterplots}(top-l).}\footnote{We highlight that we use the benchmark suite that accompanied the bottom-up approach.} We refer to benchmark results in Tab.~\ref{tab:runtimesetc}.

\myparagraph{\approachocviexact{} Mostly Outperforms \approachmonolithic{}, Fig.~\ref{fig:scatterplots}(top-c).}
The monolithic VI as typical in Storm requires a complete model, which can be prohibitively large.
However, even for medium-sized models such as \benchmark{Chains100-RmB}, the VI can run into time outs due to slow convergence.
CVI with the exact cache (and even with no cache) quickly converges -- highlighting that the grouping of states helps VI to converge.
On the other hand, a model such as \benchmark{Birooms100-RmS} highlights that the harder convergence check can yield a significant overhead. 

\myparagraph{\approachsymbiosis{} Mostly Outperforms \approachbottomup{}, Fig.~\ref{fig:scatterplots}(top-r).}
For many models, the top-down approach as motivated in \cref{sec:cvi} indeed ensures that we avoid the undirected exploration of the Pareto curves.
However, if the VI repeatedly asks for weights that are not relevant for the optimal scheduler, the termination checks fail and this yields a significant overhead.

\myparagraph{\approachocviexact{} and \approachsymbiosis{} Both Provide Clear Added Value, Fig.~\ref{fig:scatterplots}(bot-l).} 
Both approaches can solve benchmarks within ten seconds that the other approach does not solve within the time-out.
Both approaches are able to save significantly upon the number of iterations necessary.
\approachsymbiosis{} suffers from the overhead of the Pareto cache, see below, whereas \approachocviexact{} requires somewhat optimal values in all leaves, regardless of whether these leaves are important for reaching the global target.
Therefore, \approachsymbiosis{} may profit from ideas from asynchronous VI and \approachocviexact{} from adaptive schemes to decide when to run the termination check. 

\myparagraph{Pareto Cache Has a Significant Overhead, Fig.~\ref{fig:scatterplots}(bot-c/r) and Tab.~\ref{tab:caches}.}
We observe that the Pareto cache consistently yields an overhead: In particular, \approachocviexact{} often outperforms \approachocvipareto{}.
The Pareto cache is essential for \approachsymbiosis.
The overhead has three different sources. 
(1)~\emph{More iterations}: \texttt{Birooms10-RmB} illustrates how  \approachocviexact{} requires only 14\% of the iterations of \approachocvipareto{}.
Even with a 66\% cache hit rate in \approachocvipareto{}, this means an overhead in the number of component MDPs analysed.
The main reason is that reusing approximation can delay convergence\footnote{Towards global convergence, we may eventually deactivate the cache.}.  
(2)~\emph{Cache retrieval}: To obtain an upper bound, we must optimize over Pareto curves that contain tens of halfspaces, which are numerically not very stable.
Therefore, Pareto curves in Storm are represented exactly.
The linear program that must be solved is often equally slow\footnote{We use the Soplex LP solver~\cite{GamrathAndersonBestuzhevaetal.2020} for exact LP solving, which is significantly faster than using, e.g., Z3. Soplex may return unknown, which we interpret as a cache miss.}
 as actually solving the LP, especially for small MDPs. 
(3)~\emph{Cache insertion}: Cache insertion of lower bounds requires model checking Markov chains, as many as there are exits in the open MDPs.
These times are pure overhead if this lower bound is never retrieved and can be substantial for large open MDPs. 

\myparagraph{Opportunities for Heuristics and Hyperparameters.}
We extensively studied variations of the presented algorithms. For example, a much higher tolerance in the Pareto cache can significantly speed up \approachocvipareto{} on the cost of not terminating on many benchmark cases and one can investigate a per-query strategy for retrieving and/or inserting cache results. 

\myparagraph{Interpretation of Results.}
\approachmonolithic{} works well on models that fit into memory and exhibit little sharing of open MDPs.
\approachbottomup{} works well when the Pareto curves of the open MDPs can be accurately be approximated with few Pareto points, which, in practice, excludes open MDPs with more than 3 exits.
CVI without caching and termination criteria resembles a basic kind of topological VI\footnote{Topological VI orders strongly connected components, whereas CVI uses the hierarchical structure. This can also lead to advantages.} on the monolithic MDP. 
CVI can thus improve upon topological VI either via the cache or via the alternative stopping criteria. 
Based on the experiments, we conjecture that
\begin{itemize}
 \item the cache is efficient when the cost of performing a single reachability query is expensive --- such as in the \texttt{Room10} model --- while the cache hit rate is high. 
 \item the symbiotic termination criterion (\approachsymbiosis{}) works well when some exits are not relevant for the global target, such as the \texttt{Chains3500} model, in which going backwards is not productive.
 \item the compositional OVI stopping criterion (\approachocviexact{}/\approachocvipareto{}) works well when the likelihood of reaching all individual open MDPs is high, such as can be seen in the \texttt{ChainsLoop500-Dice4} model.
\end{itemize}

\section{Related Work}
\label{sec:related_work}
We group our related work into variations of value iteration, compositional verification of MDPs, and multi-objective verification.

\myparagraph{Value Iteration}
Value iteration as standard analysis of MDPs~\cite{HartmannsJQW23} is widely studied. In the undiscounted, indefinite horizon case we study, value iteration requires an exponential number of iterations in theory, but in practice converges earlier. This motivates the search for sound termination criteria.
Optimistic value iteration~\cite{HartmannsK20} is now widely adopted as the default approach~\cite{BuddeHKKPQTZ20,HartmannsJQW23}.
To accelerate VI, various asynchronous variations have been suggested that prevent operating on the complete state space. In particular \emph{topological VI}~\cite{AzeemEKSW22,DaiMWG11} and \emph{(uni-directional) sequential VI}~\cite{0001BCDDKM016,HahnH16,HartmannsJKQ20} aim to exploit an acyclic structure similar to what exists in uni-directional MDPs. %Various asynchronous alternatives aim to exploit that not all states are equally important. 

\myparagraph{Sequentially Composed MDPs}
The exploitation of a compositional structure in MDPs is widely studied. 
In particular, the sequential composition in our paper is closely related to hierarchical compositions that capture how tasks are often composed of repetitive subtasks~\cite{HauskrechtMKDB98,BarryKL11,JungesS22,GopalandLMSTWW17,SaxeER17,VienT15,BartoM03,SuttonPS99}.
While we study a fully model-based approach, Jothimurugan et al.~\cite{JothimuruganBBA21} provide a compositional reinforcement learning method whose sub-goals are induced by specifications. Neary et al.~\cite{NearyVCT22} update the learning goals based on the analysis of the component MDPs, but do not consider the possibility of reaching multiple exits. 
The widespread \emph{option-framework} and variations such as \emph{abstract VI}~\cite{JothimuruganBA21}, aggregate policies~\cite{SilverC12,CiosekS15} into additional actions to speed up convergence of value iterations and is often applied in model-free approaches. In the context of OVI, we must converge everywhere and the bottom-up stopping criterion is not easily lifted to a model-free setting.

\myparagraph{Further Related Work}
As a different type of compositional reasoning, \emph{assume-guarantee reasoning}~\cite{ChatterjeeH07,BloemCJK15,MajumdarMSZ20,FinkbeinerP22,DewesD23,KwiatkowskaNPQ13,ForejtKNPQ11} is a central topic, and a compositional probabilistic framework~\cite{KwiatkowskaNPQ13} with the parallel composition $\parallel$  is also based on Pareto curves: extending string diagrams of MDPs for the parallel composition $\parallel$ is challenging, but an interesting future work. 
We mention that there are VIs on Pareto curves solving multi-objective simple stochastic games~\cite{ChenFKSW13,AshokCKWW20}. Due to the multi-objectivity, they maintain a \emph{set of points} for each state during iterations; CVI solves single-objective oMDPs determined by weights, thus we maintain a single value for each state during iterations.

\section{Conclusion}
\label{sec:conclusion}
This paper investigates the verification of compositional MDPs, with a particular focus on approximating the behavior of the component MDPs via a Pareto cache and sound stopping criteria for value iteration. The empirical evaluation does not only demonstrate the efficacy of the novel algorithms, but also demonstrates the potential for further improvements, using asynchronous value iteration, efficient Pareto caches manipulations, and powerful compositional stopping criteria.   

\clearpage

\bibliographystyle{splncs04}
\bibliography{CAV2024}
\iffull 
\clearpage
\appendix

\section{Benchmark Details}

\subsection{Approaches and Hyperparameters}
Our implementation has the following hyperparameters:
\begin{itemize}
    \item Global OCVI $\epsilon$,
    \item Local OCVI $\eta$,
    \item Cache tolerance $\tau$,
    \item $N_{\approachocvi{}}, N_{\approachbottomup{}}$, number of CVI steps before performing OVI/Bottom-up termination check, respectively.
\end{itemize}
All benchmark cases are formed by some combination of a string diagram $\sd{D}$ and leaf models $\mdp{M}$. 

\subsection{String Diagrams}
\paragraph{Unidirectional Grid (\texttt{Rooms})}
The unidirectional grid string diagram contains a $N \times N$ grid of connected rooms. In the initial room at $(1,1)$, there is an exit to the north and the east, leading to $(1,2)$ and $(2,1)$ respectively. Once an edge of the grid has been reached, i.e., $(x, N)$ or $(N, y)$, only the east and north exit are available, respectively. The goal of the string diagram is to reach the unique exit located in $(N,N)$.

\paragraph{Bidirectional Grid (\texttt{Birooms})}
The bidirectional grid is defined in the same way as the unidirectional grid, except that it is also possible to traverse between the rooms in the opposite directions, south and west.

\paragraph{Unidirectional Chain with a loop (\texttt{Chains}, \texttt{ChainsLoop})}
The unidirectional chain string diagram (\texttt{Chains}) can be seen as a 1D version of the unidirectional grid. However, instead of having a unique exit at the end of the chain, in the \texttt{ChainsLoop} string diagram, there is another exit that brings us back to the start of the chain, which makes the chain not rightward.

\subsection{Leaf Models}
\paragraph{Small Room (\texttt{RmS})}
There are four different variants of the small room model, indicated by a pair out of the set $\{\mathit{Safe}, \mathit{Unsafe}\} \times \{ \mathit{Windy}, \mathit{Calm} \}$. This pair determines the dynamics of the room.

The small room model consists of a $7 \times 7$ grid world with imprecise movement. After each movement action (north, east, south or west), there is some probability that the agent does not end up where it intended to move. This behaviour is more likely if the room is windy instead of calm. Furthermore, there are some holes in the grid, which cannot be exited once entered. The rooms that are unsafe contain more holes. The exits of the room are at the four center positions of each edge of the grid.

\paragraph{Big Room (\texttt{RmB})}
The big room model is defined in the same way as the small room, except that the dimensions of the grid are 101 instead of 7.

\paragraph{Dice Game (\texttt{Dice})}
The dice models contain a small dice game. The game is played in rounds, and the goal of the game is to score as many points as possible. In each round, there is a choice of three biased dice. After each round, the controller picks a die and throws it, and adds the (potentially negative) number on the die to their score. There are 100 rounds, and the score is clamped between 0 and 100. For the four-exit variant of the dice game, obtaining a score between 0 and 24 means that the first exit will be taken. Similarly, a score between 25 and 49 means that the second exit will be taken, and so forth.

%
%\clearpage
\subsection{Plot Data}\label{app:benchmarkresults}
The plot marks used in the scatter plots of~\cref{fig:scatterplots} can be found in~\cref{table:plotmarks}. The raw data used for these plots can be found in~\cref{tab:runtimesetclong,tab:cacheslong}.
\newcommand{\showpgfmark}[1]{\tikz[baseline=-0.4ex]\node[mark size=0.7ex]{\pgfuseplotmark{#1}};}
\begin{table}[t]
        \centering
        \caption{Plot marks}\label{table:plotmarks}
        \begin{NiceTabular}{ c c c }%
        \toprule
        $\mathbb{D}$ & $\mdp{M}$ & Mark \\\midrule
        \begin{filecontents*}{plots/plot_marks.csv}
stringdiagram,model,mark
\texttt{Birooms},\texttt{RmS},\showpgfmark{triangle}
\texttt{Birooms},\texttt{RmB},\showpgfmark{asterisk}
\texttt{Chains},\texttt{RmS},\showpgfmark{pentagon}
\texttt{Chains},\texttt{RmB},\showpgfmark{square}
\texttt{Chains},\texttt{Dice3},\showpgfmark{diamond}
\texttt{Chains},\texttt{Dice4},\showpgfmark{-}
\texttt{Chains},\texttt{Dice5},\showpgfmark{|}
\texttt{ChainsLoop},\texttt{Dice3},\showpgfmark{Mercedes star}
\texttt{ChainsLoop},\texttt{Dice4},\showpgfmark{Mercedes star flipped}
\texttt{ChainsLoop},\texttt{Dice5},\showpgfmark{heart}
\texttt{Rooms},\texttt{RmS},\showpgfmark{x}
\texttt{Rooms},\texttt{RmB},\showpgfmark{oplus}
\texttt{Rooms},\texttt{Dice2},\showpgfmark{otimes}
\end{filecontents*}
\csvreader[late after line = \\, head to column names]{plots/plot_marks.csv}{}{\csvlinetotablerow}
             \bottomrule
        \end{NiceTabular}
\end{table}
\clearpage
\begin{table}[H]
    \centering
     \caption{Performance for different algorithms. See \textbf{Results} for explanations.}
\scalebox{0.40}{
    \begin{NiceTabular}{rrrr  r  rr  rr rrr  rrr}
\toprule
%\Block[c]{3-1}{$\sd{D}$} &
%\Block[c]{3-1}{$\mdp{M}$} &
%\Block[c]{3-1}{$|S|$} &
& & & &
\multicolumn{3}{c}{\approachbaseline} & 
\multicolumn{8}{c}{\approachnovel}\\
\cmidrule(lr){5-7}\cmidrule(lr){8-15}

& & & &
\multicolumn{1}{c}{\approachmonolithic{}} &
\multicolumn{2}{c}{\approachbottomup{}} &
\multicolumn{2}{c}{\approachocviexact{}} &
\multicolumn{3}{c}{\approachocvipareto{}} &
\multicolumn{3}{c}{\approachsymbiosis{}} \\
\cmidrule(lr){5-5}\cmidrule(lr){6-7}\cmidrule(lr){8-9}\cmidrule(lr){10-12}\cmidrule(lr){13-15}

\multicolumn{1}{c}{$\mathbb{D}$} &
\multicolumn{1}{c}{$\mdp{M}$} &
\multicolumn{1}{c}{$|S|$} &
\multicolumn{1}{c}{$|L|$} &
\multicolumn{1}{c}{\btime} &
\multicolumn{1}{c}{\btime} &
\multicolumn{1}{c}{\bpareto} &
\multicolumn{1}{c}{\btime} &
\multicolumn{1}{c}{\btimestop} &
\multicolumn{1}{c}{\btime} &
\multicolumn{1}{c}{\btimestop} &
\multicolumn{1}{c}{\bpareto} &
\multicolumn{1}{c}{\btime} &
\multicolumn{1}{c}{\btimestop} &
\multicolumn{1}{c}{\bpareto}\\
\midrule

\begin{filecontents*}{plots/table1full.csv}
sd,leaf,statecount,uniqueleaves,monolithictotaltime,pareto4totaltime,pareto4lowerparetopoints,algorithma1totaltime,algorithma1terminationtime,algorithma2totaltime,algorithma2terminationtime,algorithma2lowerparetopoints,algorithmbtotaltime,algorithmbterminationtime,algorithmblowerparetopoints
\texttt{Birooms10},\texttt{RmB},1.1e+06,16,56,TO,TO,58,25,58,25,1155,TO,TO,TO
\texttt{Birooms20},\texttt{RmB},4.2e+06,16,467,TO,TO,329,102,TO,TO,TO,TO,TO,TO
\texttt{Birooms10},\texttt{RmS},8.5e+03,16,0,TO,TO,0,0,6,0,2023,TO,TO,TO
\texttt{Birooms20},\texttt{RmS},3.4e+04,16,0,TO,TO,1,0,118,0,6193,TO,TO,TO
\texttt{Birooms50},\texttt{RmS},2.1e+05,16,3,TO,TO,7,3,TO,TO,TO,TO,TO,TO
\texttt{Birooms100},\texttt{RmS},8.5e+05,16,15,TO,TO,32,8,TO,TO,TO,TO,TO,TO
\texttt{Birooms200},\texttt{RmS},3.4e+06,16,126,TO,TO,187,58,TO,TO,TO,TO,TO,TO
\texttt{Birooms500},\texttt{RmS},2.1e+07,15,OOM,TO,TO,TO,TO,TO,TO,TO,TO,TO,TO
\texttt{Chains50},\texttt{RmB},5.3e+05,6,7,7,46,3,12,15,12,34,3,0,34
\texttt{Chains100},\texttt{RmB},1.1e+06,6,TO,7,46,4,26,27,24,40,4,0,44
\texttt{Chains200},\texttt{RmB},2.1e+06,6,TO,7,46,4,52,TO,TO,TO,4,0,34
\texttt{Chains500},\texttt{RmB},5.3e+06,6,TO,7,46,4,130,TO,TO,TO,4,0,42
\texttt{Chains1000},\texttt{RmB},1.1e+07,6,OOM,7,46,5,263,TO,TO,TO,5,0,40
\texttt{Chains1500},\texttt{RmB},1.6e+07,6,OOM,7,46,6,399,TO,TO,TO,6,1,42
\texttt{Chains2000},\texttt{RmB},2.1e+07,6,OOM,7,46,TO,TO,TO,TO,TO,5,0,34
\texttt{Chains2500},\texttt{RmB},2.7e+07,6,OOM,7,46,TO,TO,TO,TO,TO,6,1,38
\texttt{Chains3000},\texttt{RmB},3.2e+07,6,OOM,7,46,TO,TO,TO,TO,TO,4,1,32
\texttt{Chains3500},\texttt{RmB},3.7e+07,6,OOM,8,46,TO,TO,TO,TO,TO,8,1,40
\texttt{Chains4000},\texttt{RmB},4.2e+07,6,OOM,7,46,TO,TO,TO,TO,TO,5,1,34
\texttt{Chains3},\texttt{Dice3},3.0e+04,3,0,110,1242,0,0,0,0,19,0,0,19
\texttt{Chains10},\texttt{Dice3},9.9e+04,3,1,110,1242,0,0,0,0,19,0,0,19
\texttt{Chains20},\texttt{Dice3},2.0e+05,3,1,109,1240,0,0,0,0,19,0,0,19
\texttt{Chains50},\texttt{Dice3},4.9e+05,3,3,109,1242,0,0,0,0,19,0,0,19
\texttt{Chains100},\texttt{Dice3},9.9e+05,3,5,109,1240,0,1,1,1,19,0,0,19
\texttt{Chains200},\texttt{Dice3},2.0e+06,3,11,110,1242,0,1,1,1,19,0,0,19
\texttt{Chains300},\texttt{Dice3},3.0e+06,3,16,112,1242,0,2,2,2,19,0,0,19
\texttt{Chains400},\texttt{Dice3},3.9e+06,3,22,111,1240,1,2,3,2,19,1,0,19
\texttt{Chains500},\texttt{Dice3},4.9e+06,3,26,112,1242,1,3,3,3,19,1,0,19
\texttt{Chains3},\texttt{Dice4},3.0e+04,3,0,TO,TO,0,0,0,0,24,0,0,24
\texttt{Chains10},\texttt{Dice4},9.9e+04,3,0,TO,TO,0,0,0,0,24,0,0,24
\texttt{Chains20},\texttt{Dice4},2.0e+05,3,1,TO,TO,0,0,0,0,24,0,0,24
\texttt{Chains50},\texttt{Dice4},4.9e+05,3,2,TO,TO,0,0,0,0,24,0,0,24
\texttt{Chains100},\texttt{Dice4},9.9e+05,3,5,TO,TO,0,1,1,1,24,0,0,24
\texttt{Chains200},\texttt{Dice4},2.0e+06,3,10,TO,TO,1,1,1,1,24,1,1,24
\texttt{Chains500},\texttt{Dice4},4.9e+06,3,25,TO,TO,2,3,3,3,24,2,1,24
\texttt{Chains3},\texttt{Dice5},3.0e+04,3,0,TO,TO,0,0,0,0,29,0,0,29
\texttt{Chains10},\texttt{Dice5},9.9e+04,3,0,TO,TO,0,0,0,0,29,0,0,29
\texttt{Chains20},\texttt{Dice5},2.0e+05,3,1,TO,TO,0,0,1,1,29,0,0,29
\texttt{Chains50},\texttt{Dice5},4.9e+05,3,2,TO,TO,1,1,2,2,29,1,0,29
\texttt{Chains100},\texttt{Dice5},9.9e+05,3,4,TO,TO,1,1,4,3,29,1,1,29
\texttt{Chains200},\texttt{Dice5},2.0e+06,3,9,TO,TO,2,2,7,6,29,2,2,29
\texttt{Chains500},\texttt{Dice5},4.9e+06,3,22,TO,TO,5,5,18,16,29,5,4,29
\texttt{ChainsLoop3},\texttt{Dice3},3.0e+04,4,1,307,3515,0,0,0,0,39,0,0,36
\texttt{ChainsLoop10},\texttt{Dice3},9.9e+04,4,0,304,2822,0,0,1,0,39,0,0,36
\texttt{ChainsLoop20},\texttt{Dice3},2.0e+05,4,1,296,2593,0,1,1,1,39,0,0,36
\texttt{ChainsLoop50},\texttt{Dice3},4.9e+05,4,2,311,2535,1,1,2,1,39,1,0,36
\texttt{ChainsLoop100},\texttt{Dice3},9.9e+05,4,4,323,2164,1,3,4,3,39,1,1,36
\texttt{ChainsLoop200},\texttt{Dice3},2.0e+06,4,8,300,2254,3,5,9,5,39,3,1,36
\texttt{ChainsLoop300},\texttt{Dice3},3.0e+06,4,15,309,2207,4,8,13,8,39,4,2,36
\texttt{ChainsLoop400},\texttt{Dice3},3.9e+06,4,16,321,2204,5,10,17,10,39,5,2,36
\texttt{ChainsLoop500},\texttt{Dice3},4.9e+06,4,23,329,2187,6,13,21,13,39,6,3,36
\texttt{ChainsLoop3},\texttt{Dice4},3.0e+04,4,0,TO,TO,0,0,0,0,54,0,0,50
\texttt{ChainsLoop10},\texttt{Dice4},9.9e+04,4,1,TO,TO,1,0,1,0,54,1,0,50
\texttt{ChainsLoop20},\texttt{Dice4},2.0e+05,4,2,TO,TO,1,1,1,1,54,1,1,50
\texttt{ChainsLoop50},\texttt{Dice4},4.9e+05,4,3,TO,TO,2,1,3,1,54,2,2,50
\texttt{ChainsLoop100},\texttt{Dice4},9.9e+05,4,6,TO,TO,3,3,5,2,54,4,3,50
\texttt{ChainsLoop200},\texttt{Dice4},2.0e+06,4,14,TO,TO,5,5,10,5,54,9,7,50
\texttt{ChainsLoop500},\texttt{Dice4},4.9e+06,4,28,TO,TO,13,13,25,12,54,21,16,50
\texttt{ChainsLoop3},\texttt{Dice5},3.0e+04,4,0,TO,TO,0,0,1,0,66,1,0,61
\texttt{ChainsLoop10},\texttt{Dice5},9.9e+04,4,1,TO,TO,1,0,1,0,66,TO,TO,TO
\texttt{ChainsLoop20},\texttt{Dice5},2.0e+05,4,1,TO,TO,1,1,2,1,66,TO,TO,TO
\texttt{ChainsLoop50},\texttt{Dice5},4.9e+05,4,3,TO,TO,2,1,4,2,66,TO,TO,TO
\texttt{ChainsLoop100},\texttt{Dice5},9.9e+05,4,6,TO,TO,3,3,8,4,66,TO,TO,TO
\texttt{ChainsLoop200},\texttt{Dice5},2.0e+06,4,10,TO,TO,6,5,17,8,66,TO,TO,TO
\texttt{ChainsLoop500},\texttt{Dice5},4.9e+06,4,25,TO,TO,13,13,47,20,66,TO,TO,TO
\texttt{Chains10},\texttt{RmS},8.6e+02,6,0,0,67,0,0,0,0,32,0,0,32
\texttt{Chains20},\texttt{RmS},1.7e+03,6,0,0,67,0,0,0,0,36,0,0,36
\texttt{Chains50},\texttt{RmS},4.2e+03,6,0,0,67,0,0,0,0,36,0,0,36
\texttt{Chains100},\texttt{RmS},8.5e+03,6,0,0,67,0,0,0,0,46,0,0,46
\texttt{Chains200},\texttt{RmS},1.7e+04,6,0,0,67,0,0,TO,TO,TO,0,0,36
\texttt{Chains500},\texttt{RmS},4.2e+04,6,0,0,67,0,0,TO,TO,TO,0,0,38
\texttt{Rooms10},\texttt{RmB},1.1e+06,14,8,8,84,11,21,31,20,108,11,0,108
\texttt{Rooms20},\texttt{RmB},4.2e+06,14,TO,8,84,16,84,99,84,133,16,1,133
\texttt{Rooms50},\texttt{RmB},2.7e+07,14,OOM,9,84,TO,TO,TO,TO,TO,24,8,138
\texttt{Rooms100},\texttt{RmB},1.1e+08,14,OOM,11,84,TO,TO,TO,TO,TO,48,29,120
\texttt{Rooms200},\texttt{RmB},4.2e+08,14,OOM,OOM,OOM,TO,TO,TO,TO,TO,213,178,138
\texttt{Rooms250},\texttt{RmB},6.6e+08,14,OOM,OOM,OOM,TO,TO,TO,TO,TO,273,226,134
\texttt{Rooms500},\texttt{RmB},2.7e+09,13,OOM,OOM,OOM,TO,TO,TO,TO,TO,TO,TO,TO
\texttt{Rooms10},\texttt{Dice2},5.3e+05,14,1,1,48,1,3,TO,TO,TO,1,0,93
\texttt{Rooms20},\texttt{Dice2},2.3e+06,14,1,1,48,2,30,155,148,142,2,1,142
\texttt{Rooms50},\texttt{Dice2},1.4e+07,14,1,1,48,41,679,TO,TO,TO,41,39,219
\texttt{Rooms100},\texttt{Dice2},5.4e+07,14,OOM,3,48,TO,TO,TO,TO,TO,257,248,262
\texttt{Rooms200},\texttt{Dice2},2.1e+08,14,OOM,19,48,TO,TO,TO,TO,TO,TO,TO,TO
\texttt{Rooms250},\texttt{Dice2},3.3e+08,14,OOM,OOM,OOM,TO,TO,TO,TO,TO,TO,TO,TO
\texttt{Rooms300},\texttt{Dice2},4.8e+08,14,OOM,OOM,OOM,TO,TO,TO,TO,TO,TO,TO,TO
\texttt{Rooms350},\texttt{Dice2},6.5e+08,14,OOM,OOM,OOM,TO,TO,TO,TO,TO,TO,TO,TO
\texttt{Rooms400},\texttt{Dice2},8.5e+08,14,OOM,OOM,OOM,TO,TO,TO,TO,TO,TO,TO,TO
\texttt{Rooms450},\texttt{Dice2},1.1e+09,14,OOM,OOM,OOM,TO,TO,TO,TO,TO,TO,TO,TO
\texttt{Rooms500},\texttt{Dice2},1.3e+09,13,OOM,OOM,OOM,TO,TO,TO,TO,TO,TO,TO,TO
\texttt{Rooms10},\texttt{RmS},8.4e+03,14,0,0,136,0,0,TO,TO,TO,0,0,143
\texttt{Rooms20},\texttt{RmS},3.4e+04,14,0,1,136,0,0,TO,TO,TO,6,5,240
\texttt{Rooms50},\texttt{RmS},2.1e+05,14,1,1,136,1,1,TO,TO,TO,50,47,283
\texttt{Rooms100},\texttt{RmS},8.5e+05,14,6,6,136,4,2,TO,TO,TO,161,152,242
\texttt{Rooms200},\texttt{RmS},3.4e+06,14,629,OOM,OOM,15,9,TO,TO,TO,758,725,252
\texttt{Rooms500},\texttt{RmS},2.1e+07,13,TO,OOM,OOM,100,57,TO,TO,TO,TO,TO,TO
\end{filecontents*}
\csvreader[late after line = \\, head to column names]{plots/table1full.csv}{}{\csvlinetotablerow}
\bottomrule
    \end{NiceTabular}
}
   \label{tab:runtimesetclong}
\end{table}
\begin{table}[H]
    \centering
     \caption{Cache access times for CVI algorithms. See \textbf{Results} for explanations. }
%\resizebox{\textwidth}{!}{
\scalebox{0.4}{
    \begin{NiceTabular}{rr rrrrr rrrrrr rrrr}
\toprule
%& &
%\multicolumn{10}{c}{\approachocvi{}} &
%\multicolumn{5}{c}{\approachsymbiosis{}}\\
%\cmidrule(lr){2-11}\cmidrule(lr){12-16}

& &
\multicolumn{5}{c}{\approachocviexact{}} &
\multicolumn{5}{c}{\approachocvipareto{}} &
\multicolumn{5}{c}{\approachsymbiosis{}} \\
\cmidrule(lr){3-7}\cmidrule(lr){8-12}\cmidrule(lr){13-17}

\multicolumn{1}{c}{$\mathbb{D}$} &
\multicolumn{1}{c}{$\mdp{M}$} &
\multicolumn{1}{c}{\btime} &
\multicolumn{1}{c}{\bcacheinsert} &
\multicolumn{1}{c}{\bcacheretrieve} &
\multicolumn{1}{c}{\bhit} &
\multicolumn{1}{c}{\bquery} &

\multicolumn{1}{c}{\btime} &
\multicolumn{1}{c}{\bcacheinsert} &
\multicolumn{1}{c}{\bcacheretrieve} &
\multicolumn{1}{c}{\bhit} &
\multicolumn{1}{c}{\bquery} &

\multicolumn{1}{c}{\btime} &
\multicolumn{1}{c}{\bcacheinsert} &
\multicolumn{1}{c}{\bcacheretrieve} &
\multicolumn{1}{c}{\bhit} &
\multicolumn{1}{c}{\bquery}
\\
\midrule
\begin{filecontents*}{plots/table2full.csv}
sd,leaf,algorithma1totaltime,algorithma1cacheinsertiontime,algorithma1cacheretrievaltime,algorithma1cachehitratio,algorithma1weightedreachabilityqueries,algorithma2totaltime,algorithma2cacheinsertiontime,algorithma2cacheretrievaltime,algorithma2cachehitratio,algorithma2weightedreachabilityqueries,algorithmbtotaltime,algorithmbcacheinsertiontime,algorithmbcacheretrievaltime,algorithmbcachehitratio,algorithmbweightedreachabilityqueries
\texttt{Birooms10},\texttt{RmB},58,0,0,0.03,206,58,14,4,0.66,1500,TO,TO,TO,TO,TO
\texttt{Birooms20},\texttt{RmB},329,0,0,0.01,806,TO,TO,TO,TO,TO,TO,TO,TO,TO,TO
\texttt{Birooms10},\texttt{RmS},0,0,0,0.00,200,6,0,6,0.50,1500,TO,TO,TO,TO,TO
\texttt{Birooms20},\texttt{RmS},1,0,0,0.00,800,118,1,116,0.59,6000,TO,TO,TO,TO,TO
\texttt{Birooms50},\texttt{RmS},7,0,0,0.18,21441,TO,TO,TO,TO,TO,TO,TO,TO,TO,TO
\texttt{Birooms100},\texttt{RmS},32,0,0,0.24,65545,TO,TO,TO,TO,TO,TO,TO,TO,TO,TO
\texttt{Birooms200},\texttt{RmS},187,1,3,0.25,477157,TO,TO,TO,TO,TO,TO,TO,TO,TO,TO
\texttt{Birooms500},\texttt{RmS},TO,TO,TO,TO,TO,TO,TO,TO,TO,TO,TO,TO,TO,TO,TO
\texttt{Chains50},\texttt{RmB},3,0,0,0.50,104,15,0,0,0.60,156,3,0,0,0.79,52
\texttt{Chains100},\texttt{RmB},4,0,0,0.50,204,27,0,0,0.62,306,4,0,0,0.84,102
\texttt{Chains200},\texttt{RmB},4,0,0,0.50,404,TO,TO,TO,TO,TO,4,0,0,0.95,202
\texttt{Chains500},\texttt{RmB},4,0,0,0.50,1004,TO,TO,TO,TO,TO,4,0,0,0.97,502
\texttt{Chains1000},\texttt{RmB},5,0,0,0.50,2004,TO,TO,TO,TO,TO,5,0,0,0.99,1002
\texttt{Chains1500},\texttt{RmB},6,0,0,0.50,3004,TO,TO,TO,TO,TO,6,0,0,0.99,1502
\texttt{Chains2000},\texttt{RmB},TO,TO,TO,TO,TO,TO,TO,TO,TO,TO,5,0,1,0.99,2002
\texttt{Chains2500},\texttt{RmB},TO,TO,TO,TO,TO,TO,TO,TO,TO,TO,6,0,1,0.99,2502
\texttt{Chains3000},\texttt{RmB},TO,TO,TO,TO,TO,TO,TO,TO,TO,TO,4,0,1,1.00,3002
\texttt{Chains3500},\texttt{RmB},TO,TO,TO,TO,TO,TO,TO,TO,TO,TO,8,0,1,1.00,3502
\texttt{Chains4000},\texttt{RmB},TO,TO,TO,TO,TO,TO,TO,TO,TO,TO,5,0,1,1.00,4002
\texttt{Chains3},\texttt{Dice3},0,0,0,0.50,10,0,0,0,0.40,15,0,0,0,0.20,5
\texttt{Chains10},\texttt{Dice3},0,0,0,0.61,31,0,0,0,0.56,36,0,0,0,0.67,12
\texttt{Chains20},\texttt{Dice3},0,0,0,0.64,61,0,0,0,0.61,66,0,0,0,0.82,22
\texttt{Chains50},\texttt{Dice3},0,0,0,0.66,151,0,0,0,0.64,156,0,0,0,0.92,52
\texttt{Chains100},\texttt{Dice3},0,0,0,0.66,301,1,0,0,0.65,306,0,0,0,0.96,102
\texttt{Chains200},\texttt{Dice3},0,0,0,0.66,601,1,0,0,0.66,606,0,0,0,0.98,202
\texttt{Chains300},\texttt{Dice3},0,0,0,0.66,901,2,0,0,0.66,906,0,0,0,0.99,302
\texttt{Chains400},\texttt{Dice3},1,0,0,0.67,1201,3,0,0,0.66,1206,1,0,0,0.99,402
\texttt{Chains500},\texttt{Dice3},1,0,0,0.67,1501,3,0,1,0.66,1506,1,0,0,0.99,502
\texttt{Chains3},\texttt{Dice4},0,0,0,0.50,10,0,0,0,0.40,15,0,0,0,0.20,5
\texttt{Chains10},\texttt{Dice4},0,0,0,0.61,31,0,0,0,0.56,36,0,0,0,0.67,12
\texttt{Chains20},\texttt{Dice4},0,0,0,0.64,61,0,0,0,0.61,66,0,0,0,0.82,22
\texttt{Chains50},\texttt{Dice4},0,0,0,0.66,151,0,0,0,0.64,156,0,0,0,0.92,52
\texttt{Chains100},\texttt{Dice4},0,0,0,0.66,301,1,0,0,0.65,306,0,0,0,0.96,102
\texttt{Chains200},\texttt{Dice4},1,0,0,0.66,601,1,0,0,0.66,606,1,0,0,0.98,202
\texttt{Chains500},\texttt{Dice4},2,0,0,0.67,1501,3,0,1,0.66,1506,2,0,0,0.99,502
\texttt{Chains3},\texttt{Dice5},0,0,0,0.50,20,0,0,0,0.44,25,0,0,0,0.20,5
\texttt{Chains10},\texttt{Dice5},0,0,0,0.56,54,0,0,0,0.53,60,0,0,0,0.67,12
\texttt{Chains20},\texttt{Dice5},0,0,0,0.58,104,1,0,0,0.56,110,0,0,0,0.82,22
\texttt{Chains50},\texttt{Dice5},1,0,0,0.59,254,2,0,0,0.56,364,1,0,0,0.92,52
\texttt{Chains100},\texttt{Dice5},1,0,0,0.60,504,4,0,0,0.57,714,1,0,0,0.96,102
\texttt{Chains200},\texttt{Dice5},2,0,0,0.60,1004,7,0,1,0.57,1414,2,0,0,0.98,202
\texttt{Chains500},\texttt{Dice5},5,0,0,0.60,2504,18,0,2,0.57,3514,5,0,1,0.99,502
\texttt{ChainsLoop3},\texttt{Dice3},0,0,0,0.72,107,0,0,0,0.78,185,0,0,0,0.85,65
\texttt{ChainsLoop10},\texttt{Dice3},0,0,0,0.80,320,1,0,0,0.81,402,0,0,0,0.93,142
\texttt{ChainsLoop20},\texttt{Dice3},0,0,0,0.82,630,1,0,0,0.82,712,0,0,0,0.96,252
\texttt{ChainsLoop50},\texttt{Dice3},1,0,0,0.83,1560,2,0,1,0.83,1642,1,0,0,0.98,582
\texttt{ChainsLoop100},\texttt{Dice3},1,0,0,0.83,3110,4,0,2,0.84,3192,1,0,1,0.99,1132
\texttt{ChainsLoop200},\texttt{Dice3},3,0,0,0.84,6210,9,0,3,0.84,6292,3,0,1,1.00,2232
\texttt{ChainsLoop300},\texttt{Dice3},4,0,0,0.84,9310,13,0,5,0.84,9392,4,0,2,1.00,3332
\texttt{ChainsLoop400},\texttt{Dice3},5,0,0,0.84,12410,17,0,6,0.84,12492,5,0,2,1.00,4432
\texttt{ChainsLoop500},\texttt{Dice3},6,0,0,0.84,15510,21,0,8,0.84,15592,6,0,3,1.00,5532
\texttt{ChainsLoop3},\texttt{Dice4},0,0,0,0.72,107,0,0,0,0.77,185,0,0,0,0.83,65
\texttt{ChainsLoop10},\texttt{Dice4},1,0,0,0.80,318,1,0,0,0.81,402,1,0,0,0.92,142
\texttt{ChainsLoop20},\texttt{Dice4},1,0,0,0.82,628,1,0,1,0.82,712,1,0,0,0.96,252
\texttt{ChainsLoop50},\texttt{Dice4},2,0,0,0.83,1558,3,0,1,0.83,1642,2,0,0,0.98,582
\texttt{ChainsLoop100},\texttt{Dice4},3,0,0,0.83,3108,5,0,3,0.83,3192,4,0,1,0.99,1132
\texttt{ChainsLoop200},\texttt{Dice4},5,0,0,0.84,6208,10,0,5,0.84,6292,9,0,2,1.00,2232
\texttt{ChainsLoop500},\texttt{Dice4},13,0,0,0.84,15508,25,0,13,0.84,15592,21,0,5,1.00,5532
\texttt{ChainsLoop3},\texttt{Dice5},0,0,0,0.72,107,1,0,0,0.77,185,1,0,0,0.83,65
\texttt{ChainsLoop10},\texttt{Dice5},1,0,0,0.80,321,1,0,0,0.81,402,TO,TO,TO,TO,TO
\texttt{ChainsLoop20},\texttt{Dice5},1,0,0,0.82,631,2,0,1,0.82,712,TO,TO,TO,TO,TO
\texttt{ChainsLoop50},\texttt{Dice5},2,0,0,0.83,1561,4,0,2,0.83,1642,TO,TO,TO,TO,TO
\texttt{ChainsLoop100},\texttt{Dice5},3,0,0,0.83,3111,8,0,4,0.83,3192,TO,TO,TO,TO,TO
\texttt{ChainsLoop200},\texttt{Dice5},6,0,0,0.84,6211,17,0,9,0.84,6292,TO,TO,TO,TO,TO
\texttt{ChainsLoop500},\texttt{Dice5},13,0,0,0.84,15511,47,0,26,0.84,15592,TO,TO,TO,TO,TO
\texttt{Chains10},\texttt{RmS},0,0,0,0.50,24,0,0,0,0.39,36,0,0,0,0.17,12
\texttt{Chains20},\texttt{RmS},0,0,0,0.50,44,0,0,0,0.48,66,0,0,0,0.45,22
\texttt{Chains50},\texttt{RmS},0,0,0,0.50,104,0,0,0,0.59,156,0,0,0,0.77,52
\texttt{Chains100},\texttt{RmS},0,0,0,0.50,204,0,0,0,0.61,306,0,0,0,0.83,102
\texttt{Chains200},\texttt{RmS},0,0,0,0.50,404,TO,TO,TO,TO,TO,0,0,0,0.94,202
\texttt{Chains500},\texttt{RmS},0,0,0,0.50,1004,TO,TO,TO,TO,TO,0,0,0,0.97,502
\texttt{Rooms10},\texttt{RmB},11,0,0,0.50,200,31,1,0,0.50,300,11,1,0,0.50,100
\texttt{Rooms20},\texttt{RmB},16,0,0,0.50,800,99,2,0,0.61,1200,16,2,0,0.84,400
\texttt{Rooms50},\texttt{RmB},TO,TO,TO,TO,TO,TO,TO,TO,TO,TO,24,2,1,0.97,2500
\texttt{Rooms100},\texttt{RmB},TO,TO,TO,TO,TO,TO,TO,TO,TO,TO,48,1,5,1.00,9999
\texttt{Rooms200},\texttt{RmB},TO,TO,TO,TO,TO,TO,TO,TO,TO,TO,213,1,19,1.00,39999
\texttt{Rooms250},\texttt{RmB},TO,TO,TO,TO,TO,TO,TO,TO,TO,TO,273,2,30,1.00,62499
\texttt{Rooms500},\texttt{RmB},TO,TO,TO,TO,TO,TO,TO,TO,TO,TO,TO,TO,TO,TO,TO
\texttt{Rooms10},\texttt{Dice2},1,0,0,0.50,600,TO,TO,TO,TO,TO,1,0,0,0.59,100
\texttt{Rooms20},\texttt{Dice2},2,0,0,0.50,5601,155,0,6,0.50,27600,2,0,0,0.84,400
\texttt{Rooms50},\texttt{Dice2},41,0,0,0.50,140013,TO,TO,TO,TO,TO,41,0,2,0.96,2500
\texttt{Rooms100},\texttt{Dice2},TO,TO,TO,TO,TO,TO,TO,TO,TO,TO,257,0,8,0.99,10000
\texttt{Rooms200},\texttt{Dice2},TO,TO,TO,TO,TO,TO,TO,TO,TO,TO,TO,TO,TO,TO,TO
\texttt{Rooms250},\texttt{Dice2},TO,TO,TO,TO,TO,TO,TO,TO,TO,TO,TO,TO,TO,TO,TO
\texttt{Rooms300},\texttt{Dice2},TO,TO,TO,TO,TO,TO,TO,TO,TO,TO,TO,TO,TO,TO,TO
\texttt{Rooms350},\texttt{Dice2},TO,TO,TO,TO,TO,TO,TO,TO,TO,TO,TO,TO,TO,TO,TO
\texttt{Rooms400},\texttt{Dice2},TO,TO,TO,TO,TO,TO,TO,TO,TO,TO,TO,TO,TO,TO,TO
\texttt{Rooms450},\texttt{Dice2},TO,TO,TO,TO,TO,TO,TO,TO,TO,TO,TO,TO,TO,TO,TO
\texttt{Rooms500},\texttt{Dice2},TO,TO,TO,TO,TO,TO,TO,TO,TO,TO,TO,TO,TO,TO,TO
\texttt{Rooms10},\texttt{RmS},0,0,0,0.50,200,TO,TO,TO,TO,TO,0,0,0,0.32,100
\texttt{Rooms20},\texttt{RmS},0,0,0,0.50,800,TO,TO,TO,TO,TO,6,0,0,0.70,400
\texttt{Rooms50},\texttt{RmS},1,0,0,0.50,5000,TO,TO,TO,TO,TO,50,0,2,0.94,2500
\texttt{Rooms100},\texttt{RmS},4,0,0,0.50,20000,TO,TO,TO,TO,TO,161,0,8,0.99,9999
\texttt{Rooms200},\texttt{RmS},15,0,0,0.50,80000,TO,TO,TO,TO,TO,758,0,32,1.00,39999
\texttt{Rooms500},\texttt{RmS},100,0,0,0.50,500000,TO,TO,TO,TO,TO,TO,TO,TO,TO,TO
\end{filecontents*}
\csvreader[late after line = \\, head to column names]{plots/table2full.csv}{}{\csvlinetotablerow}
  
\bottomrule
    \end{NiceTabular}
}
    \label{tab:cacheslong}
\end{table}

\section{Omitted Contents in~\cref{sec:prelinaries}}
\label{sec:details_VI}

We review the standard Bellman operator\footnote{More precisely, we generalise the Bellman operator marginally to a weighted variant. The standard Bellman operator is recovered by setting the weights of target states to one and of other states to zero.} and the (non-compositional) value iteration on MDPs.

\begin{mydefinition}[Bellman Operator]
\label{def:bellman_operator}
Let $\mdp{M}$ be an MDP, $T \subseteq S$ be target states, and $\convw{w}\defeq (w_t)_{t\in T}$ be a weight. The \emph{Bellman operator} $\wbellman{\mdp{M}}{T}{\convw{w}}\colon \probinterval^S \rightarrow \probinterval^S$ (wrt. $\convw{w}$) is given by 
\begin{align*}
    \wbellman{\mdp{M}}{T}{\convw{w}}(f)(s) \defeq \begin{cases}
        w_t   &\text{ if $s = t\in T$,}\\
        \max_{a\in A} \sum_{s'\in S} P(s, a, s')\cdot f(s') &\text{ if $s\in S \setminus T$.}\\
    \end{cases}
\end{align*}    
\end{mydefinition}
The set $\probinterval^S$ is a complete lattice with the functorial order and  $\wbellman{\mdp{M}}{T}{\convw{w}}$ is Scott-continuous~\cite{abramsky1994domain}. Thus, the Bellman operator has a least fixed point $\lfpoint{\wbellman{\mdp{M}}{T}{\convw{w}}}$.

The correctness of value iteration is ensured by Props~\ref{prop:charc_lfpoint} and \ref{prop:unique_bellmanOp}. In each iteration, the current value $f\colon S\rightarrow \probinterval$, which is an under approximation of $\lfpoint{\wbellman{\mdp{M}}{T}{\convw{w}}}$, is updated by $\wbellman{\mdp{M}}{T}{\convw{w}}(f)$. 

\begin{myproposition}[\!\!\cite{Puterman94,BaierK08}]
\label{prop:charc_lfpoint}
For a state $s\in S$, the least fixed point $\lfpoint{\wbellman{\mdp{M}}{T}{\convw{w}}}(s)$ coincides with the maximum weighted reachability probabilities $\MaxWReacha{\mdp{M}, T}{\convw{w}}{s}$.  
\end{myproposition}

\begin{myproposition}[\!\!\cite{cousot1979constructive,Baranga91}]
\label{prop:unique_bellmanOp}
Assume the setting of Def.~\ref{def:bellman_operator}.
The limit of the ascending $\omega$-chain $\big((\wbellman{\mdp{M}}{T}{\convw{w}})^n(\bot)\big)_{n\in \nat}$
is the least fixed point $\lfpoint{\wbellman{\mdp{M}}{T}{\convw{w}}}$ of $\wbellman{\mdp{M}}{T}{\convw{w}}$, where the order $\preceq$ on $\probinterval^S$ is induced by the ordinal total order in $\probinterval$, and $\bot$ is the least element in $\probinterval^S$.

\end{myproposition}

\section{Omitted Contents in~\cref{sec:pareto_curves}}
\label{sec:details_pareto_curves}

\subsection{Shortcut oMDP for String Diagram of MDPs}

Given a string diagram $\mathbb{D}$, and a weight $\convw{w}$, we first introduce the \emph{shortcut Bellman operator} $\compBellman{\mathbb{D}}{\convw{w}}$ as the standard Bellman operator on the operational semantics $\semantics{\shortsubst(\mathbb{D})}$ of the \emph{shortcut string diagram} $\shortsubst(\mathbb{D})$, which is a summary of $\mathbb{D}$ with only remaining open ends and setting actions as Pareto-optimal DM schedulers in each component oMDP $\mdp{A}\in \CP{\mathbb{D}}$. Recall that the set $\interface^{\mdp{A}}$ is the set of open ends in $\mdp{A}$.

\begin{mydefinition}[shortcut oMDP~\cite{WatanabeVHRJ24}]
\label{def:shortcut_MDP}
Let $\mdp{A}\defeq (\mdp{M}^{\mdp{A}}, \interface^{\mdp{A}})$ be an oMDP.  The \emph{shortcut oMDP} $\ecmdp{\mdp{A}}$ is an oMDP $\ecmdp{\mdp{A}} \defeq (\mdp{M}, \interface^{\mdp{A}})$ given by $\mdp{M} \defeq (S, A, P)$, where $S\defeq \interface^{\mdp{A}}\cup \{\star\}$, $A \defeq \Dmsched{\mdp{A}}$, and 
\begin{align*}
    P(s, \sigma, s') \defeq \begin{cases}
        \Reacha{\mdp{A},\sigma}{s}{s'} &\text{ if $s\in \en^{\mdp{A}}$, $\sigma\in \Dmsched{\mdp{A}}$ and $s'\in \ex^{\mdp{A}}$, }\\
        1-\sum_{o\in \ex^{\mdp{A}}}\Reacha{\mdp{A},\sigma}{s}{o}  &\text{ if $s\in \en^{\mdp{A}}$, $\sigma\in \Dmsched{\mdp{A}}$, and $s'=\star$, }\\
        0 &\text{ otherwise. }
    \end{cases}
\end{align*}
\end{mydefinition}
 
The shortcut oMDPs are similar to \emph{abstract MDPs} (or \emph{macro MDPs})~\cite{HauskrechtMKDB98,BarryKL11,JungesS22,GopalandLMSTWW17,JothimuruganBA21}; intuitively, we use them as summaries of oMDPs $\mdp{A}$ whose probabilistic transitions are the reachability probabilities that are induced by Pareto-optimal DM schedulers on $\mdp{A}$.

\begin{mydefinition}[shortcut string diagram]
Let $\mathbb{D}$ be a string diagram. The \emph{shortcut string diagram} $\shortsubst(\mathbb{D})$ is a string diagram inductively given as follows: 
\begin{itemize}
    \item if $\mathbb{D} \defeq\constMDP{\mdp{A}}$, then $\shortsubst(\mathbb{D}) $ is the shortcut oMDP $\ecmdp{\mdp{A}}$,
    \item if $\mathbb{D} \defeq\mathbb{D}_1\ast \mathbb{D}_2$ for $\ast\in \{\seqcomp, \oplus\}$, then $\shortsubst(\mathbb{D}) \defeq \shortsubst(\mathbb{D}_1)\ast \shortsubst(\mathbb{D}_2)$.
\end{itemize}
The \emph{shortcut oMDP} of the string diagram $\mathbb{D}$ is the operational semantics $\semantics{\shortsubst(\mathbb{D})}$ of the shortcut string diagram $\shortsubst(\mathbb{D})$. 
\end{mydefinition}

The construction is indeed correct in the following sense:  
\begin{myproposition}
\label{prop:correct_shortcut_construction}
Let $\mathbb{D}$ be a string diagram, and $\convw{w}$ be a weight. The equality $\MaxWReacha{\semantics{\mathbb{D}}}{\convw{w}}{\en} = \MaxWReacha{\semantics{\shortsubst(\mathbb{D})}}{\convw{w}}{\en}$ holds.

\end{myproposition}
\begin{proofs}
The proof relies on the theoretical development in~\cite{WatanabeVHRJ24}; this is a direct consequence of~\cite[Cors. 1, 2, and 3]{WatanabeVHRJ24}.  \qed
\end{proofs}

Then, we define a \emph{shortcut Bellman operator} $\compBellman{\mathbb{D}}{\convw{w}}$ for $\mathbb{D}$ as the standard Bellman operator for  $\semantics{\shortsubst(\mathbb{D})}$; we introduced standard Bellman operators in~\cref{sec:details_VI}. 
\begin{mydefinition}[shortcut Bellman operator]
\label{def:compBellmanOp}
Let $\mathbb{D}$ be a string diagram, and $\convw{w} \defeq (w_j)_{j\in \exGlobal{\mathbb{D}}}$ be a weight. The \emph{shortcut Bellman operator} $\compBellman{\mathbb{D}}{\convw{w}}$ is the (standard but weighted) Bellman operator $\wbellman{\semantics{\shortsubst(\mathbb{D})}}{\exGlobal{\mathbb{D}}}{\convw{w}}$ on $\semantics{\shortsubst(\mathbb{D})}$. Note that the exits in $\semantics{\shortsubst(\mathbb{D})}$ are $\exGlobal{\mathbb{D}}$. 
\end{mydefinition}

\subsection{Correctness of Opt-GSC }

We first recall the principle behind OVI (it is also closely related to the Knaster--Tarski theorem).
\begin{myproposition}[Park induction principle~\cite{park1969fixpoint}]
\label{prop:park_induction}
Let $V$ be a complete lattice, and $\Phi\colon V\rightarrow V$ be monotone. For any $v\in V$, if $\Phi(v) \leq v$, then $v$ is an upper bound of the least fixed point $\lfpoint{\Phi}$, that is, $ \lfpoint{\Phi}\leq v$. 
\end{myproposition}

Given an under-approximation $l$ and an error bound $\epsilon$, OVI and CVI with Opt-GSC construct a candidate $u$ of an over-approximation that satisfies $\|l - u\|\leq \epsilon$, and check the inequality $\Phi(u) \leq u$: if yes, they terminate and conclude that $l$ is an under-approximation that satisfies the error bound $\epsilon$. 

Secondly, we inductively characterize the shortcut Bellman operator $\compBellman{\mathbb{D}}{\convw{w}}$. The following lemma captures the base case $\compBellman{\constMDP{\mdp{A}}}{\convw{w}}$; this corresponds to a local VI in line~\ref{line:prototypeLocalVI} of~\cref{alg:CVIprototype}. Note that the set  $\interface^{\mdp{A}} (= \enLocal{\constMDP{\mdp{A}}}\uplus \exLocal{\constMDP{\mdp{A}}})$ is the set of states on $\shortsubst(\mdp{A})$ excluding the sink state $\star$. 
\begin{mylemma}
\label{lem:cha_bellman_oMDPs}
Let $\mdp{A}$ be an oMDP, and $\convw{w}\defeq (w_j)_{j\in \ex^{\mdp{A}}}$ be a weight. For each $f\in \probinterval^{\interface^{\mdp{A}}}$, it holds that: 
\begin{align*}
   \compBellman{\constMDP{\mdp{A}}}{\convw{w}}(f)(s)  &= \begin{cases}  
        \MaxWReacha{\mdp{A}}{\convw{w}}{s} &\text{ if $s\in \en^{\mdp{A}}$,}\\
        w_{s}&\text{ if $s \in \ex^{\mdp{A}}$.}\\
    \end{cases}
\end{align*}
\end{mylemma}
\begin{myproof}
    By Defs~\ref{def:shortcut_MDP} and \ref{def:compBellmanOp}, for any $i\in \en^{\mdp{A}}$, the following equation holds: 
    \begin{align*}
        \compBellman{\constMDP{\mdp{A}}}{\convw{w}}(f)(i) = \max_{\sigma\in \Dmsched{\cmdp{\mdp{A}}}}\sum_{j\in \ex^{\mdp{A}}} \Reacha{\cmdp{\mdp{A}},\sigma}{i}{j}\cdot w_j. 
    \end{align*}
    The RHS coincides with $\MaxWReacha{\mdp{A}}{\convw{w}}{i}$ because of the existence of optimal DM schedulers in $\cmdp{\mdp{A}}$ wrt. $\convw{w}$ by  Prop.~\ref{prop:correct_shortcut_construction} and Def.~\ref{def:compBellmanOp}. \qed
\end{myproof}

For the compositions, we use the following characterisations: this corresponds to the propagation (line~\ref{line:prototypePropag} in~\cref{alg:CVIprototype}). 
\begin{mylemma}
\label{lem:cha_bellman_comp}
Let $\ast\in \{\seqcomp, \oplus\}$, $\mathbb{D}\defeq \mathbb{E}\ast \mathbb{F}$ be a string diagram, and $\convw{w}\defeq (w_j)_{j\in \exGlobal{\mathbb{D}}}$ be a weight.  For each $f\in \probinterval^{\enLocal{\mathbb{D}}\uplus \exLocal{\mathbb{D}}}$, it holds that:
\begin{align*}
    \compBellman{\mathbb{D}}{\convw{w}}(f)(s) = \begin{cases}
        \compBellman{\mathbb{E}}{\convw{w}^{\mathbb{E}}_{\ast}}(f^{\mathbb{E}}_{\ast})(s) &\text{ if $s\in \enLocal{\mathbb{E}}\uplus \exLocal{\mathbb{E}}$,}\\
        \compBellman{\mathbb{F}}{\convw{w}^{\mathbb{F}}_{\ast}}(f^{\mathbb{F}}_{\ast})(s) &\text{ if $s\in \enLocal{\mathbb{F}}\uplus \exLocal{\mathbb{F}}$,}\\
    \end{cases}
\end{align*}    
 where weights $\convw{w}^{\mathbb{E}}_{\seqcomp} \defeq (w^{\mathbb{E}}_{\seqcomp, j})_{j\in \exGlobal{\mathbb{E}}}$ and $\convw{w}^{\mathbb{F}}_{\seqcomp} \defeq (w^{\mathbb{F}}_{\seqcomp, j})_{j\in \exGlobal{\mathbb{F}}}$ are given by 
\begin{align*}
    w^{\mathbb{E}}_{\seqcomp, j} &\defeq \begin{cases}
        w_j &\text{ if $j\in \exl^{\mathbb{E}}$,}\\
        f(\enrarg{i}^{\mathbb{F}}) &\text{ if $j= \exrarg{i}^{\mathbb{E}}$ for $i$,}\\
    \end{cases} 
    & w^{\mathbb{F}}_{\seqcomp, j} &\defeq \begin{cases}
        w_j &\text{ if $j\in \exr^{\mathbb{F}}$,}\\
        f(\enlarg{i}^{\mathbb{E}}) &\text{ if $j= \exlarg{i}^{\mathbb{F}}$ for $i$,}\\
    \end{cases} 
\end{align*}
and weights $\convw{w}^{\mathbb{E}}_{\oplus}$, $\convw{w}^{\mathbb{F}}_{\oplus}$,
and values $f^{\mathbb{E}}_{\ast}\colon \enLocal{\mathbb{E}}\uplus \exLocal{\mathbb{E}}\rightarrow \probinterval , f^{\mathbb{F}}_{\ast}\colon \enLocal{\mathbb{F}}\uplus \exLocal{\mathbb{F}}\rightarrow \probinterval$ are canonical restrictions. 

\end{mylemma}
\begin{proofs}
Let $\ast = \seqcomp$; the case $\ast = \oplus$ is easy. The only subtle thing is whether the equation holds for the states $s$ that have probabilistic transitions to the states $j\in \exr^{\mathbb{E}}\uplus \exl^{\mathbb{F}}$. This is indeed true because the LHS is based on the value $f(j)$, and the RHS correctly includes them in the weights $w^{\mathbb{E}}_{\seqcomp, j}$ and $w^{\mathbb{F}}_{\seqcomp, j}$. \qed 
\end{proofs}

Due to the characterisations (Prop.~\ref{prop:correct_shortcut_construction} and Lems.~\ref{lem:cha_bellman_oMDPs} and~\ref{lem:cha_bellman_comp}), we can define CVI with Opt-GSC as the standard OVI on the shortcut oMDP $\semantics{\shortsubst(\mathbb{D})}$ of $\mathbb{D}$ except using under-approximations for local computations (for under-approximations): this corresponds to replace the exact solution with an under-approximation in Lem.~\ref{lem:cha_bellman_oMDPs}. This is necessarily because of local VIs and Pareto caching. In order to apply the Park induction principle for a candidate of an over-approximation, we locally use exact solutions. 

\begin{myremark}[on-demand computation]
By Lems.~\ref{lem:cha_bellman_oMDPs} and~\ref{lem:cha_bellman_comp}, we can see that we do not have to explicitly construct the shortcut oMDP $\semantics{\shortsubst(\mathbb{D})}$ for running VIs. For each given weight $\convw{w}$ on component $\mdp{A}$, we only have to solve the weighted reachability probability problem on $\mdp{A}$ wrt. $\convw{w}$ for running VIs on the shortcut oMDP. 
    
\end{myremark}

Finally, we prove the $\epsilon$-soundness of CVI with Opt-GSC:

\begin{myproof}[Proof of Thm.~\ref{thm:epsilon_soundness} with Opt-GSC]
Let $\mathbb{D}$ be a string diagram. 
Prop.~\ref{prop:correct_shortcut_construction} and Lems.~\ref{lem:cha_bellman_oMDPs} and~\ref{lem:cha_bellman_comp} show that CVI with exact local solutions on $\mathbb{D}$ can be characterised as the standard VI on the shortcut oMDP $\semantics{\shortsubst(\mathbb{D})}$. Since the shortcut Bellman operator is monotone, the value $g$ in~\cref{alg:CVIprototype} is always a guaranteed under-approximation even when we replace exact local solutions with under-approximations, which CVI actually does. Since the computations for over-approximations are exact (due to the exact local solutions for over-approximations), we can directly apply the Park induction principle for the shortcut Bellman operator, and we can conclude that $l$ satisfies the desired result when CVI with Opt-GSC terminates. 
\qed
\end{myproof}

\subsection{Correctness of BU-GSC }
\begin{myproof}[Proof of Thm.~\ref{thm:epsilon_soundness} with BU-GSC]
It suffices to show that the obtained $U$ from a  Pareto cache $\paretoAssign$ is an over-approximation of the global Pareto curve. This is indeed true because of \cite[Props. 2, 4, and 5]{WatanabeVHRJ24}. 
\qed
\end{myproof}

\else
\fi
\end{document}